%% file: main.tex
\documentclass[fleqn,usenatbib]{mnras}

\usepackage{newtxtext,newtxmath}

\usepackage[T1]{fontenc}

\DeclareRobustCommand{\VAN}[3]{#2}
\let\VANthebibliography\thebibliography
\def\thebibliography{\DeclareRobustCommand{\VAN}[3]{##3}\VANthebibliography}

\usepackage{graphicx}	
\usepackage{amsmath}	

\usepackage{scalerel}
\usepackage{tikz}
\usetikzlibrary{svg.path}

\definecolor{orcidlogocol}{HTML}{A6CE39}
\tikzset{
  orcidlogo/.pic={
    \fill[orcidlogocol] svg{M256,128c0,70.7-57.3,128-128,128C57.3,256,0,198.7,0,128C0,57.3,57.3,0,128,0C198.7,0,256,57.3,256,128z};
    \fill[white] svg{M86.3,186.2H70.9V79.1h15.4v48.4V186.2z}
                 svg{M108.9,79.1h41.6c39.6,0,57,28.3,57,53.6c0,27.5-21.5,53.6-56.8,53.6h-41.8V79.1z M124.3,172.4h24.5c34.9,0,42.9-26.5,42.9-39.7c0-21.5-13.7-39.7-43.7-39.7h-23.7V172.4z}
                 svg{M88.7,56.8c0,5.5-4.5,10.1-10.1,10.1c-5.6,0-10.1-4.6-10.1-10.1c0-5.6,4.5-10.1,10.1-10.1C84.2,46.7,88.7,51.3,88.7,56.8z};
  }
}

\newcommand\orcidicon[1]{\href{https://orcid.org/#1}{\mbox{\scalerel*{
\begin{tikzpicture}[yscale=-1,transform shape]
\pic{orcidlogo};
\end{tikzpicture}
}{|}}}}

\usepackage{hyperref}
\usepackage{comment}

\title[kSZ + GGL]{Disentangling the Halo: Joint Model for Measurements of the
Kinetic Sunyaev-Zeldovich Effect and Galaxy-Galaxy Lensing}

\author[Sunseri et al.]{James Sunseri \orcidicon{0000-0003-4274-2662},$^{1}$\thanks{E-mail: jsunseri@princeton.edu (JS)}
Alexandra Amon \orcidicon{0000-0002-6445-0559},$^{1}$
Jo Dunkley \orcidicon{0000-0002-7450-2586},$^{1,2}$
Nicholas Battaglia \orcidicon{0000-0001-5846-0411},$^{3}$
Simone Ferraro \orcidicon{0000-0003-4992-7854},$^{4,5}$ 
\newauthor
Boryana Hadzhiyska \orcidicon{0000-0002-2312-3121},$^{4,5,6}$
Bernardita Ried Guachalla \orcidicon{0000-0002-0418-6258},$^{7,8,9}$
\& Emmanuel Schaan \orcidicon{0000-0002-4619-8927}$^{8,9}$
\\ \\
$^{1}$Department of Astrophysical Sciences, Princeton University, Princeton, NJ 08540, USA \\
$^{2}$Joseph Henry Laboratories of Physics, Jadwin Hall, Princeton University, Princeton, NJ 08544, USA \\
$^{3}$Department of Astronomy, Cornell University, Ithaca, NY 14853, USA \\
$^{4}$Physics Division, Lawrence Berkeley National Laboratory, Berkeley, California 94720, USA \\
$^{5}$Berkeley Center for Cosmological Physics, Department of Physics, University of California,
Berkeley, California 94720, USA \\
$^{6}$Miller Institute for Basic Research in Science, University of California, Berkeley, California 94720, USA \\
$^{7}$Department of Physics, Stanford University, Stanford, California 94305-4085, USA \\
$^{8}$Kavli Institute for Particle Astrophysics and Cosmology,
382 Via Pueblo Mall Stanford, California 94305-4060, USA \\
$^{9}$SLAC National Accelerator Laboratory, 2575 Sand Hill Road, Menlo Park, California 94025, USA
}

\date{Accepted XXX. Received YYY; in original form ZZZ}

\pubyear{2026}

\begin{document}
\label{firstpage}
\pagerange{\pageref{firstpage}--\pageref{lastpage}}
\maketitle

\begin{abstract}
We present the first joint analysis of the kinetic Sunyaev-Zeldovich (kSZ) effect with galaxy-galaxy lensing (GGL) for CMASS galaxies in the Baryon Oscillation Spectroscopic Survey (BOSS). We show these complementary probes can disentangle baryons from dark matter in the outskirts of galactic halos by alleviating model degeneracies that are present when fitting to kSZ or GGL measurements alone. In our joint kSZ+GGL analysis we show that the baryon density profile is well constrained on scales from 0.3 to 50 Mpc/$h$. With our well constrained profile of the baryon density, we provide direct comparisons to simulations. For our model we find an outer slope of the baryon distribution that is shallower than predicted by some hydrodynamical simulations, consistent with enhanced baryonic feedback. We also show that not including baryons in a model for GGL can shift halo mass estimates by $\sim 20\%$ compared to a model that includes baryons and is jointly fit to kSZ+GGL measurements. Our modelling code galaxy-galaxy lensing and kSZ (\texttt{glasz}) is publicly available at \url{https://github.com/James11222/glasz}. 
\end{abstract}




\section{Introduction}
\label{sec:intro}
\input{sections/01_introduction.tex}

\section{kSZ \& galaxy-galaxy lensing measurements}
\label{sec:measurements}
\input{sections/02_measurements.tex}

\section{Halo model}
\label{sec:model}
\input{sections/03_model.tex}

\section{Results: joint model constraints}
\label{sec:results}
\input{sections/04_results.tex}

\section{Conclusions}
\label{sec:conclusions}
\input{sections/05_conclusions.tex}

\section*{Acknowledgements}
We would like to thank Nick Kokron, Gabriela Sato-Polito, Dhayaa Anbajagane, Henry Schreiner, Adrian Bayer, and the members of the Amon research group for insightful discussions. We are pleased to acknowledge that the work reported on in this paper was substantially performed using the Princeton Research Computing resources at Princeton University which is a consortium of groups led by the Princeton Institute for Computational Science and Engineering (PICSciE) and Research Computing.
JS acknowledges and is thankful for the support from the Fannie \& John Hertz Foundation. JD acknowledges support from a Royal Society Wolfson Visiting Fellowship and from the Kavli Institute for Cosmology Cambridge and the Institute of Astronomy, Cambridge.

\section*{Data Availability}

The stacked kSZ profile measurements used in this work can be made available upon request and are from the original work of \cite{Schaan_2021_kSZ}. The GGL measurements used in this analysis are publicly available on this \href{https://github.com/aamon/Galaxy-galaxy-lensing-KiDS-DES-HSC}{github repository} and were originally made by \cite{Amon_&_Robertson_2023_GGL}.\footnote{\url{https://github.com/aamon/Galaxy-galaxy-lensing-KiDS-DES-HSC}} Our modelling code \texttt{glasz} is built upon the \texttt{pyccl} \cite{Chisari_2019_CCL} and \texttt{Mop-c-GT} \cite{Amodeo_2021_kSZ} packages and is publicly available at the \href{https://github.com/James11222/glasz}{\texttt{glasz} homepage}.\footnote{\url{https://github.com/James11222/glasz}} We request that any uses of the \texttt{glasz} package cite the DOI listed on the package homepage. 

\bibliographystyle{mnras}
\bibliography{biblio} 

\newpage

\appendix

\input{sections/06_appendix.tex}

\bsp	
\label{lastpage}
\end{document}

%% file: sections/01_introduction.tex
With large imaging surveys like the Dark Energy Survey (DES) \citep{DES_Y3_2022}, the 
Kilo-Degree Survey (KiDS) \citep{Heymans_2021_KiDS1000_S8},
and the Hyper Suprime-Cam Survey (HSC) \citep{HSC_3x2}, we have seen a rapid change in our ability to probe the large scale structure (LSS) and matter distribution in the non-linear regime using weak lensing (WL) and clustering. One of the biggest challenges standing in the way of using such measurements to test the dark components of the standard model is the modelling of baryons. In particular, baryonic feedback suppresses the matter power spectrum on non-linear scales, and therefore the weak lensing amplitude, and must be modelled accurately to avoid biasing cosmological inference \citep[e.g.,][]{VanDaalen_2011_Baryonic_Feedback, Chisari_2019_Baryonic_Feedback_Cosmology, Bigwood_2024_kSZ_WL}.

However, the complexity of baryonic feedback requires a large dynamic range in resolution and time that is difficult to simulate \citep[e.g.][]{Crain_2023_Galaxy_Population_Review}. This has led to the development of simplified `subgrid' prescription models in hydrodynamical cosmological simulations which are used to mimic the effects of baryonic feedback on LSS \citep{Steinborn_2015_AGN_Feedback_Subgrid, Chisari_2019_Baryonic_Feedback_Cosmology, Schaye_2023_FLAMINGO, Pakmor_2023_MTNG_Clusters}. Although the simulations are able to reproduce a range of observations, they give a wide spread of predictions for the amplitude and scale extent of the suppression of the matter power spectrum due to baryonic feedback \citep[e.g.,][]{ 
Chisari_2019_Baryonic_Feedback_Cosmology, Paco_2021_CAMELS, Sunseri_2023_Baryonic_Feedback}. Understanding and reducing this model uncertainty is the challenge. More recently, it has been possible to constrain the matter power suppression from a number of observables \citep[e.g.][]{Debackere_2020, Grandis_2024_Baryonic_Feedback, Bigwood_2024_kSZ_WL, LaPosta_2025_Baryonic_Feedback, Medlock_2025_FRB_Pk}, although these constraints are still not very precise.  

For this reason, WL and clustering cosmological analyses lose a substantial fraction of their constraining power to the uncertainty in modelling baryon feedback \citep[see, for example, Fig. 12 of][]{Amon_2022_DES_Y3_S8}.
Many analyses take the approach of throwing away measurements on small angular scales to circumvent modelling altogether \citep[e.g.][]{Prat_2022_GGL_modelling, Zacharegkas_2022_GGL_HOD, Amon_&_Robertson_2023_GGL, Lange_2023_2x2pt_GGL}. Even when attempting to model feedback, the uncertainty and uninformed parameter space is costly \citep[e.g.][]{huang_modelling_2019, Schneider_2022_Baryons_w_Cosmology, Arico_2023_DESY3_Baryons, Bigwood_2024_kSZ_WL}, unless informed by simulations \citep[][]{Asgari_2021_KiDS1000_S8,  
Bigwood_2024_kSZ_WL,
Zennaro_2024_GGL_w_Baryons}.
Another approach is to jointly model the WL and clustering data with probes of the baryon content \citep{Debackere_2020, Troster_2022_Joint_Cosmo_Baryons, Schneider_2022_Baryons_w_Cosmology, Bigwood_2024_kSZ_WL}.

The kinetic Sunyaev-Zeldovich (kSZ) effect is a secondary anisotropy in the CMB that comes from the Doppler boost of CMB photons by inverse Compton scattering of free electrons in the Intracluster Medium (ICM) and Circumgalactic Medium (CGM)
that have bulk motion along the line of sight relative to the CMB rest frame \citep{SZ_effect_OG, Mroczkowski_2019_kSZ_review}. The kSZ effect is small relative to other CMB secondaries (e.g. tSZ), follows a blackbody spectrum, and is directly proportional to the product of the density of electrons surrounding galaxies and their bulk velocity along the line of sight. The dependency on the bulk line of sight velocity and the small amplitude relative to other secondaries requires methods such as a velocity weighted stack using external velocity measurements from galaxy surveys to measure the signal. Stacked CMB measurements from the Atacama Cosmology Telescope (ACT) survey with a velocity reconstruction weighting have revealed that the electrons around galaxies in the Baryon Oscillation Spectroscopic Survey (BOSS) are significantly more extended than the dark matter distribution \citep{Schaan_Ferarro_2016_kSZ_detection,Schaan_2021_kSZ,Hadzhiyska_2024_ACT_DESI_kSZ, RG_2025_kSZ_ACT_DESI_Spec}. The kSZ effect is particularly useful for WL as it probes the density of electrons (and therefore baryons) in the outskirts of the halo ($\sim$1-4 virial radii) which lies in the range of  scales that baryons need to be better understood to improve WL cosmological constraints. Joint analyses of the kSZ effect with WL measurements are a promising way to constrain baryonic feedback and simultaneously improve cosmological constraints \citep[e.g.,][]{Bigwood_2024_kSZ_WL}.

In order to accurately interpret the kSZ effect, one must know the halo mass of the sample \citep{Moser_2021_SZ_choices} and have an appropriate forward model as was done in \cite{Amodeo_2021_kSZ}. By jointly analysing the kSZ effect with GGL measurements we can build a self-consistent framework for estimating the halo mass. Without GGL, analyses of the kSZ effect either rely on estimates of stellar mass \citep{Maraston_2011_stellar_pop, Maraston_2013_BOSS_stellar_mass} which are then used to estimate the halo mass through stellar-to-halo mass relations \citep{Miyatake_2015_BOSS_WL_SHMR, Kravtsov_2018_SHMR}, or fits to galaxy clustering \citep{Miyatake_2015_BOSS_WL_SHMR, Torres_2016_CMASS_Clustering}, both of which have greater uncertainties than galaxy-galaxy lensing estimates. \citet{Bigwood_2024_kSZ_WL} estimated that for the BOSS CMASS galaxy sample, these estimates span almost an order of magnitude. In this work we present the first joint fits of kSZ measurements and GGL measurements to disentangle baryons from dark matter in the outskirts of galactic halos, and more accurately constrain the mass of the sample. Accurate mass estimates where the baryon content is considered is also important for cluster cosmology \citep[e.g.][]{Cromer_2022_Baryons_WL_Mass}.

Measurements of the kSZ effect can be used to test baryon feedback models and hydrodynamical simulations. In this case, having an accurate halo mass of the sample is as important, in order to interpret the appropriate strength of feedback \citep{McCarthy_2024_kSZ_WL}. Previous analyses of the BOSS kSZ measurements of \citet{Schaan_2021_kSZ} are consistent with feedback that is stronger than most hydrodynamical simulations predict \citep{Bigwood_2024_kSZ_WL, McCarthy_2024_kSZ_WL}. This underestimation may be severe enough that it may even bias WL cosmological constraints \citep{Amon_2022_AMOD, Preston_2023_AMOD_2}. In this work we build a fast analytic modelling framework based on the halo model, complementary to previous approaches that rely on the baryonification method, or expensive hydrodynamical simulations. While out of the scope of this work, validation of our model with simulations will be necessary to use this model for cosmological analyses of GGL and clustering in the future.

This work is structured as follows: Section \ref{sec:measurements} describes the kSZ and GGL measurements that we jointly analyse, Section \ref{sec:model} details the model that we construct, Section \ref{sec:results} demonstrates the model in practice and presents 
the joint fits for BOSS galaxies and Section \ref{sec:conclusions} concludes the work.

%% file: sections/02_measurements.tex
\begin{figure*}
    \centering
    \includegraphics[width=\linewidth]{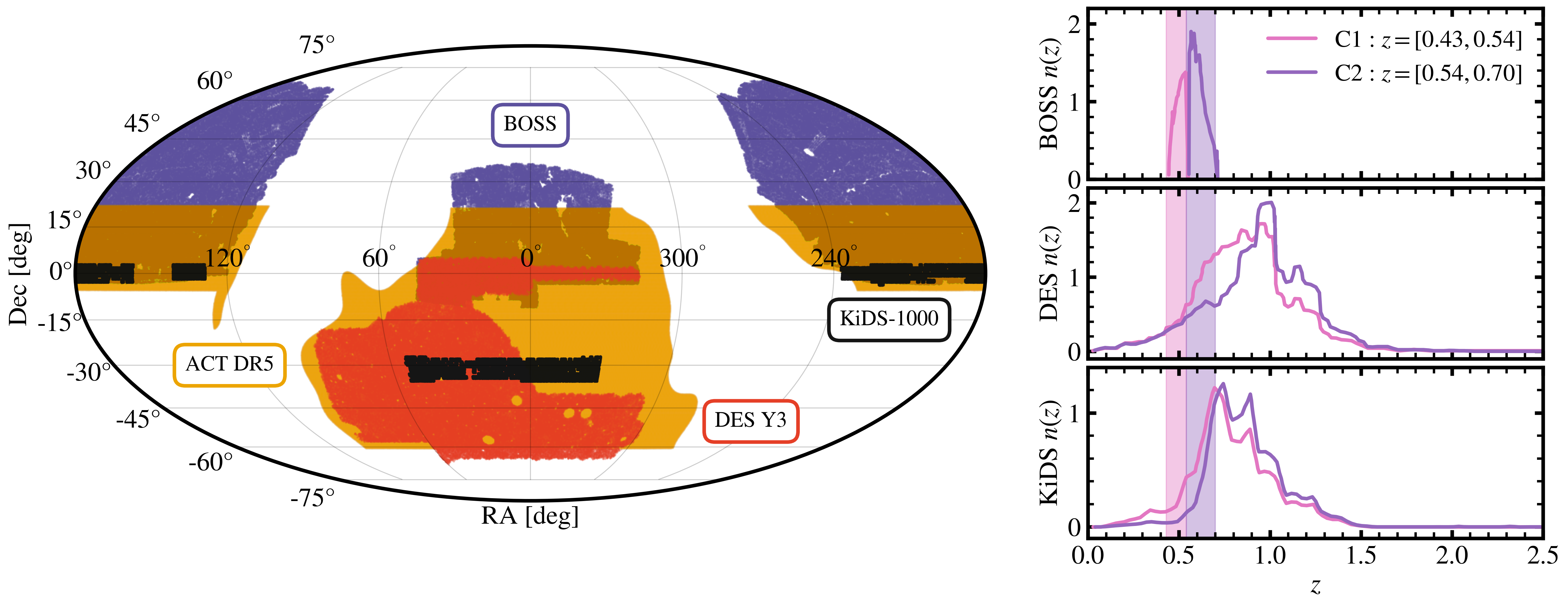}
    \caption{\textit{Left:} Sky footprints in equatorial coordinates of all surveys used in this work. Spectroscopic lenses are from the BOSS CMASS (purple) survey, photometric sources used for weak lensing come from DES Y3 (red) and KiDS-1000 (black), and the CMB backlight used for stacking the kSZ signal is from the ACT DR5 (orange) survey. We mask the full ACT footprint using the \textit{Planck} $f_{\rm sky} = 60 \%$ galactic mask from \citet{Planck_2020_Maps}. We estimate that DES Y3, KiDS-1000, and ACT have an overlap area of 777 deg$^2$, 409 deg$^2$, and 5737 deg$^2$ with BOSS respectively. \textit{Right:} Redshift distributions for the galaxy surveys used in this analysis from \citet{Amon_&_Robertson_2023_GGL} (BOSS scaled by $10^3$). The BOSS CMASS survey is divided into 2 samples C1 (pink) and C2 (purple) by redshift. DES Y3 and KiDS-1000 survey bins are coloured to indicate which lens bin they correspond to, such that the sources are sufficiently behind the lenses.}
    \label{fig:survey_footprints}
\end{figure*}

In this work, we build a model to consistently describe the excess surface mass density of halos, as measured by GGL \citep{Amon_&_Robertson_2023_GGL}, with the gas distribution, probed by measurements of the kSZ effect \citep{Schaan_2021_kSZ}. These measurements require data from a combination of several surveys (BOSS, DES Y3, KiDS-1000, ACT DR5), for clarity we summarize the sky footprints and redshift distributions of these surveys in Figure \ref{fig:survey_footprints}. We describe these measurements in more detail below.

\subsection{Galaxy-Galaxy Lensing}
\label{subsec:GGL}

In this work we use GGL measurements from 
\cite{Amon_&_Robertson_2023_GGL}\footnote{\url{https://github.com/aamon/Galaxy-galaxy-lensing-KiDS-DES-HSC}} to determine the excess surface mass density profile of BOSS DR12 CMASS galaxies. These are obtained from a cross-correlation of the spectroscopically determined positions of the BOSS galaxies with the photometric shapes of background source galaxies of the Dark Energy Survey Year 3 (DES Y3) and the Kilo-Degree Survey 1000 square degree (KiDS-1000, 4th KiDS Data Release) data. 

The DES Y3 data were collected with the
Dark Energy Camera on the 4-meter Blanco Telescope and spans 4143 $\rm deg^2$ of
5-band photometry (grizY) with an average number density of 5.59 arcmin$^{-2}$. The 
data include shapes of over 100 million galaxies across four tomographic 
redshift bins \citep{Myles_2021_DES_Y3}. The shapes that enabled these measurements were presented in \citet{Gatti_2021_DES_Y3}.

The KiDS-1000 weak lensing data were taken with
the 2.6 m VLT Survey Telescope (VST) \citep{Capaccioli_2011_VLT, Grado_2012_VLT} and span 777 $\rm deg^2$ of the sky \citep{Giblin_2021_KiDS} with a number density of 6.22 $\rm arcmin^{-2}$. 

The spectroscopic lenses from BOSS CMASS are subdivided into 2 lens redshift bins: C1 corresponding to $0.43 < z < 0.54$ and C2 corresponding to $0.54 < z < 0.70$. Each lensing survey (DES Y3 and KiDS-1000) providing background source galaxy shapes is also divided into 4 tomographic redshift bins, of which, we only use tomographic redshift bins sufficiently behind the foreground lens redshift bins (a GGL measurement of lens bin C1 uses both tomographic bins [3,4] while C2 uses only the [4] tomographic bin). The redshift distributions $n(z)$ of the lenses and sources are shown in the top-right panel of Figure \ref{fig:survey_footprints}. The tomographic bins used in DES Y3 and KiDS-1000 are calibrated using a tool from machine-learning known as the self-organizing map (SOM) \citep{Buchs_2019_DES, Wright_2020_KiDS} which requires spectroscopy to accurately estimate photometric redshifts in each tomographic bin.

Given spectroscopically measured redshifts of lenses and
photometric shape and redshift measurements of the sources, the quantity that is directly measured is the average tangential
shear $\langle \gamma_t \rangle$ as a function of angular 
separation $\theta$. We can then use this to determine the
differential surface mass density $\Delta \Sigma$
\begin{equation}
    \Delta \Sigma(R) = \frac{\langle \gamma_t(\theta) \rangle}{\overline{\Sigma_{\rm crit}^{-1}}} \; ,
\end{equation}
where the angular separation is related to the projected 
distance by $\theta = R / \chi(z_l)$. The denominator is 
referred to as the average inverse critical surface mass 
density, and is a function of the redshifts of both the lens
$z_l$ and the source $z_s$. For an individual lens-source 
pair, $\Sigma_{\rm crit}^{-1}$ is
\begin{equation}
    \Sigma_{\rm crit}^{-1} (z_l, z_s) = \frac{4\pi G}{c^2} \frac{\chi (z_l) [\chi (z_l) - \chi(z_s)]}{\chi (z_s) a(z_l)} \;,
\end{equation}
where $\chi$ is the line-of-sight comoving distance. In practice 
we measure $\Delta \Sigma$ using an estimator of the form
\begin{equation}
    \Delta\Sigma =
    \frac{\sum_{ls} w_{\mathrm{sys}, l} w_{ls} \Sigma_{\mathrm{crit}}
          (z_l, z_s) e_t}{\sum_{ls} w_{\mathrm{sys}, l} w_{ls}} \, .
\end{equation}
where $\sum_{ls}$ refers to the sum of all lens-source pairs, 
$w_{\mathrm{sys}, l}$ is the systematic weight of each lens 
provided by the lens survey to prevent selection bias, $e_t$ 
is the tangential ellipticity of a source, and $w_s$ is a 
source weight provided from the source survey to minimize 
systematic errors in the survey which we use to create a 
combined lens-source weight $w_{\rm ls}$,
\begin{equation}
    w_{\rm ls} = w_s \left(\Sigma_{\mathrm{crit}}^{-1} (z_l, z_s) \right)^2\;.
\end{equation}
This is a basic estimator, which is calibrated with additional corrections to account for
systematics such as photometric redshift uncertainties, shear 
response and calibration, overlapping lens and source populations (boost factors) and spurious systematic signals (random corrections). We refer to \cite{Amon_&_Robertson_2023_GGL} for a detailed description of 
all the relevant systematic corrections that are required to yield
a robust measurement of $\Delta \Sigma$.

\cite{Amon_&_Robertson_2023_GGL} create the DES Y3+KiDS-1000 combined data vector used in this analysis by taking the inverse-variance weighted average of GGL measurements made using DES Y3 and KiDS-1000 separately. They show that these measurements can be combined because the KiDS-BOSS and DES-BOSS on-sky footprints have no overlap (see Figure \ref{fig:survey_footprints}) and they are statistically consistent (see Section 5.1.5 of \cite{Amon_&_Robertson_2023_GGL} for more detail).

The error in the DES Y3 measurement is estimated using a Jackknife covariance. The lens galaxies and random points are split into 75 regions within the overlap area of 770 deg$^2$ using the \textsc{kmeans} algorithm, resulting in each square region spanning 3 deg on a side or $\sim 65 \; h^{-1} \rm Mpc$ (30 \% larger than the largest angular scale measured). The KiDS-1000 error estimation is computed using the bootstrap method with square regions of 4 deg on each side equivalent to $\sim 80 \; h^{-1} \; \rm Mpc$.

\subsection{kSZ Effect}
\label{subsec:kSZ}

The kinetic SZ effect refers to the contribution to the SZ effect that 
comes from the bulk motion of the electrons in galactic halos. For this reason, the 
kSZ effect is directly proportional to the number density of electrons and their bulk velocity along the line of sight (LoS).
The temperature fluctuations from the kSZ effect can be written as
\begin{equation}
    \frac{\delta T_{\rm kSZ}}{T_{\rm CMB}} = \frac{\sigma_T}{c} \int_{\rm LoS} \frac{d\chi}{1 + z} n_{\rm e}(\chi) v_{\rm p} e^{-\tau} \; ,
\end{equation}
where $\sigma_T$ is the Thompson cross-section, $c$ is the speed of light, $\chi$ is the comoving distance to redshift $z$, $n_{\rm e}$ is the physical (not comoving) electron number density, $v_{\rm p}$ is the peculiar velocity along the line of sight, and $\tau$ is the optical depth to Thompson 
scattering. 

The kSZ effect measurements used in this work were obtained by \cite{Schaan_2021_kSZ}
by stacking the CMB temperature fluctuations mapped by ACT (in 98 GHz and 150 GHz frequencies) around BOSS CMASS galaxies (DR10), weighted by their peculiar velocities. The covariance matrix is constructed using the bootstrap method.\footnote{Higher signal-to-noise kSZ measurements have recently been reported in \citep{Hadzhiyska_2024_ACT_DESI_kSZ, RG_2025_kSZ_ACT_DESI_Spec} using ACT DR6 data; our method could be extended in the future to incorporate these data.}

The coadded CMB temperature maps were 
created by combining the ACT DR5 day \& night maps \citep{Fowler_2007_ACT, Swetz_2011_ACT,  Thornton_2016_ACT, Henderson_2016_ACT} with data from the Planck satellite \citep{Planck_2020_Maps}. The beams used to 
create the f90 and f150 ACT DR5 maps are nearly Gaussian with FWHM
of 2.1 and 1.3 arcminutes respectively \citep{Lungu_2022_ACT_Beam}. The kSZ effect 
measurements of \cite{Schaan_2021_kSZ} can be extracted from a temperature map
$\delta T$ by using compensated aperture photometry (CAP) filtering with varying aperture radius 
$\theta_{\rm d}$. The CAP filter outputs an integrated temperature
$\mathcal{T}$ in the disk with radius $\theta_{\rm d}$ subtracted by the temperature
in a concentric ring of the same area around the original disk. Mathematically this is written as
\begin{equation}
    \mathcal{T}(\theta_{\rm d}) = \int d^2 \theta \delta T(\theta) W_{\theta_{\rm d}}(\theta) \; ,
\end{equation}
where the CAP filter $W_{\theta_{\rm d}}(\theta)$ is defined as
\begin{equation}
    W_{\theta_{\rm d}}(\theta) = 
    \begin{cases}
        1  & \theta < \theta_{\rm d} \\
        -1 & \theta_{\rm d} \leq \theta \leq \sqrt{2} \theta_{\rm d} \\
        0  & \text{else}
    \end{cases} \; .
\end{equation}
The measurements span a range of aperture radii $\theta_{\rm d}$ from 1 to 6 arcminutes which 
probes the kSZ profile in the range of 0.5 to 4 virial radii in the CMASS galaxy sample. 

These scales
were chosen to probe baryonic feedback while minimizing covariance in the measurements.  \cite{Schaan_2021_kSZ} stacked the measured temperatures 
$\mathcal{T}_i(\theta_{\rm d})$ around each galaxy in the sample. The stacking is done using a 
velocity weighted, inverse-variance weighted mean estimator given by
\begin{equation}
    \hat{T}_{\rm kSZ}(\theta_d) = -\frac{1}{r_v} \frac{v^{\rm rec}_{\rm rms}}{c} \frac{\sum_i \mathcal{T}_i (\theta_d) \left(\frac{v^{\rm rec}_i}{c} \right) \frac{1}{\sigma_i^2}}{\sum_i \left(\frac{v^{\rm rec}_i}{c} \right) \frac{1}{\sigma_i^2}} \; .
\label{eq: stacking estimator}
\end{equation}
We define $\sigma_i$ as the noise on the CAP filter for the $i$th galaxy. 
We note that the bias factor $r_v$, RMS reconstructed radial velocity $v^{\rm rec}_{\rm rms}$, and 
the reconstructed radial velocity of each galaxy $v^{\rm rec}_{i}$ are defined from the velocity 
reconstruction procedure outlined in \cite{Schaan_2021_kSZ} which is similar to the displacement reconstruction done in Baryon Acoustic Oscillation (BAO) studies \citep{Padmanabhan_2012_VR}. The velocity reconstruction
procedure involves solving the linearized continuity equation in redshift-space 
\begin{equation}
    \mathbf{\nabla} \cdot \mathbf{v} + f \mathbf{\nabla} \cdot [(\mathbf{v} \cdot \mathbf{\hat{n}})\mathbf{\hat{n}}] = -a H f \frac{\delta_{\rm g}}{b} \; ,
\end{equation}
where $\mathbf{v}$ is the velocity field, $f = d \ln \delta / d \ln a$ is the logarithmic 
linear growth rate, $\mathbf{\hat{n}}$ is the line-of-sight, $a$ is the scale factor, $H$ is the 
Hubble parameter, $b$ is the linear bias, and $\delta_{\rm g}$ is the galaxy overdensity. 

Solving the 
linearized continuity equation allows for the reconstruction of the radial velocity field which
can be assigned to each galaxy in the sample. The estimator is carefully chosen so that the radial 
direction of the velocity field does not cancel out thus leaving the kSZ signal intact. Lastly, the 
measurements of \cite{Schaan_2021_kSZ} perform a bootstrap resampling of the individual galaxies 
to estimate the covariance of the stacked kSZ profiles. Specifically, \cite{Schaan_2021_kSZ} repeatedly draw individual galaxies from the galaxy catalogue to generate a resampled galaxy catalogue of the same size. The kSZ profile is then measured on 10,000 realizations of the resampled galaxy catalogue which enable an estimation of the covariance matrix.
The measurements at smaller aperture sizes
have a dominant source of noise coming from the detector and foregrounds while the measurements at larger aperture
sizes are dominated by large-scale CMB fluctuations.

%% file: sections/03_model.tex
To model the kSZ and GGL signals we model the density profiles of both baryons and dark matter
around our lens galaxy sample. In this work we start by assuming the density profiles are axisymmetric. We can 
separate the matter density profile into two
components
\begin{equation}
    \rho_{\rm m}(r) = \rho_{\rm b}(r) + \rho_{\rm dm}(r) \; ,
\label{eq: matter density profile}
\end{equation}
where $\rho_{\rm b}(r)$ is the baryon density profile and $\rho_{\rm dm}(r)$ is the dark matter density profile. 
In this work, we are modelling the density profiles of matter at scales of order the virial radius and thus in our 
model we assume that the baryons are fully described by the gas such that $\rho_{\rm b}(r) \approx \rho_{\rm gas}(r)$ where $\rho_{\rm gas}(r)$ is the gas density profile. We can expect that the fraction of baryons in stars and cold non-ionized gas is $\lesssim 10 \%$ \citep{Kelly_2021_fb_discussion}, and since baryons make up $\sim 15 \%$ of the total matter, we estimate that this assumption should not lead to an error larger than $\sim 1.5 \%$ in the total matter density.
Each density profile in our model has two components
\begin{equation}
    \rho(r) = \rho^{\rm 1h}(r) + \rho^{\rm 2h}(r) \; .
\label{eq: density profile}
\end{equation}
The first term refers to the 1-halo density profile which corresponds to density 
from the effective central halo, and the second term refers to the 2-halo density profile 
which corresponds to contributions in the density coming from the surrounding 
neighbouring halos \citep{Asgari2023_Halo_Review}. We model the transition between the 1-halo and 2-halo terms as a sum rather than taking the maximum of the terms at any given scale. This simplistic treatment of the transition region should suffice for the current level of precision in our analysis, but future work that improves the precision of this model should explore the impact of this modelling choice. In the subsequent subsections below we will explain these 1-halo and 2-halo terms in more detail.

We use a fixed Planck flat $\Lambda$CDM cosmology with the following values for parameters: $\Omega_{\rm m} = 0.3111$, $\Omega_{\rm b} = 0.0490$,
$h = 0.6766$, $n_s = 0.9665$, $\sigma_8 = 0.8102$ \citep{Planck_2020a_S8}. We fix the redshift of our model to be $\langle z \rangle = 0.55$, the mean redshift of our lens sample (BOSS CMASS). This choice does not impact the results of our analysis when compared to a marginalization over the redshift distribution of the lens sample as shown in \cite{Amodeo_2021_kSZ}. We further verified this to be true while including GGL.
\subsection{The One Halo Term}
\label{subsec:1-halo}
The 1-halo term of the density profile is the contribution to the density from the central halo itself. For
this work we assume the host galaxy lies at the centre of a dark matter halo and that the gas is bound to the 
dark matter halo (i.e. no miscentering). In other words, we do not model the gas and dark matter around satellites and centrals with a Halo-Occupation Distribution (HOD) framework. Additionally, this model ignores the potential systematic effect of satellites or `miscentering',
where satellite galaxies offset the stacking procedures used on the central galaxies in both the kSZ effect measurements and the GGL measurements. 

The impacts of ignoring satellites on the kSZ and GGL signals have been shown in \citep{Hadzhiyska_2023_SZ_sims, McCarthy_2024_kSZ_WL, Siegel_2025_kSZ_GGL, Bigwood_2025_kSZ_GGL_benchmark}. For CMASS galaxies \cite{Siegel_2025_kSZ_GGL} showed that ignoring satellites shifts halo mass estimates at the $\sim10\%$ level, which is not significant enough to undermine the findings of this work. \cite{McCarthy_2024_kSZ_WL} and \cite{Bigwood_2025_kSZ_GGL_benchmark} showed with hydro simulations that for a CMASS-like sample in hydro simulations satellite fractions can be $\sim10-20\%$ and have a scale dependent influence on the amplitude of both the kSZ and GGL signals at the $\sim20\%$ level.

We also ignore an additional uncertainty on the reconstruction bias factor $r_v$ which is about $\sim 10 \%$ for the reconstruction used in this analysis \citep{Schaan_2021_kSZ}. We note that modelling of more precise measurements should include this effect \citep{Ried_Guachalla_2024_VR, Hadzhiyska_2024_ACT_DESI_kSZ, RG_2025_kSZ_ACT_DESI_Spec}. 

The central goal of this work is to demonstrate the complementarity of measurements of the kSZ effect and GGL. This goal does not require a level of precision that warrants the accounting of these additional systematic effects. For this reason, it is suitable to make these modelling simplifications in this first iteration of the model framework. In Sections \ref{subsec:halo mass} and \ref{subsec:feedback} we quantify how much these simplifications are expected to impact our parameter inference and by extension our analysis results. Future work designed for precise cosmological inference should account for these effects.
\subsubsection{Dark Matter}
We model the 1-halo dark matter density profile as a Navarro-Frank-White (NFW) profile \cite{NFW_1997} which follows the form
\begin{equation}
    \rho_{\rm NFW}(r) = \frac{\delta_{\rm c} \bar{\rho}_{\rm m} }{\left(\frac{r}{r_{\rm s}} \right) \left( 1 + \frac{r}{r_{\rm s}} \right)^2} \; ,
\end{equation}
where the scale radius $r_{\rm s}$ is related to the comoving spherical overdensity halo radius $r_{\Delta_{\rm m}}$ by the concentration parameter $c(M)$ 
by the relation $r_{\Delta_{\rm m}} = c(M,z) r_{\rm s}$ and $\bar{\rho}_{\rm m}$ is the comoving average cosmic matter density. 
We define the characteristic halo overdensity $\delta_{\rm c}$ as
\begin{equation}
    \delta_{\rm c} = \frac{\Delta c^3}{3[\log(1+c) - c/(1+c)]} \; ,
\end{equation}
where $\Delta$ is the spherical overdensity of the halo determining the mass definition of the halo. The comoving 
spherical overdensity halo radius $r_{\Delta}$ can be written as 
\begin{equation}
    r_{\Delta_{\rm m}} = \left( \frac{3 M_{\rm halo}}{4 \pi \Delta \bar{\rho}_{\rm m}} \right)^{1/3} \; .
\label{eq:r_delta}
\end{equation}
In our work we set the spherical overdensity $\Delta = 200$ and use a concentration-mass 
relation given by \cite{Duffy_2008_Concentration} 
\begin{equation}
    c(M_{\rm halo},z) = 5.71 \times \left( \frac{M_{\rm halo}}{2 \times 10^{12} \; \rm M_\odot}\right)^{-0.084} (1 + z)^{-0.47} \; ,
\end{equation}
to determine the concentration given a halo mass 
$c(M_{\rm halo})$ at $z = 0.55$. By using $\Delta = 200$, our definition of halo mass $M_{\rm halo}$ is the same as $M_{200}$. This concentration relation was determined from N-body simulations for halo masses $10^{11} - 10^{15} \; \rm M_\odot$ and redshifts $0 < z < 2$ which matches our selected sample. We use a comoving $\bar{\rho}_{\rm m}$ because the radius $r$ is in comoving units. With this
\begin{equation}
    \rho^{\rm 1h}_{\rm dm}(r ) = f_{\rm dm}\rho_{\rm NFW}(r ) \; ,
\end{equation}
where $f_{\rm dm}$ is the universal dark matter fraction of the universe defined as 
$f_{\rm dm} \equiv \Omega_{\rm dm} / \Omega_{\rm m}$. In our model, the matter in the universe 
is only composed of dark matter and baryons, so $f_{\rm dm} = 1 - f_{\rm b}$ where $f_{\rm b}$ is 
the universal baryon fraction defined as $f_{\rm b} \equiv \Omega_{\rm b} / \Omega_{\rm m}$. 

In this work, we don't model the potential effects of assembly bias from the surrounding environment in our CMASS sample. This effect could bias our results at the few percent level \citep{Yuan_2021_Assembly_Bias_CMASS, Amon_&_Robertson_2023_GGL} and is a necessary improvement for future work. We also choose to ignore the back-reaction of baryonic feedback on the dark matter distribution: a second-order effect which is not necessary to model for our current level of precision \citep{Val_Daalen_2011_BR, Springel_2018_BR_TNG}. Our model of the 1-halo term closely follows an approach from \cite{Cromer_2022_Baryons_WL_Mass} which modelled baryons in WL of clusters. This approach differs from HOD based halo models commonly used to model GGL \citep{Nishimichi_2019_DarkEmulator, Miyatake_2022_DarkEmulator, Zacharegkas_2022_GGL_HOD, Dvornik_2023_Halo_Model}, but it does allow for an explicit modelling of the baryon density profile which can be used to jointly analyse the kSZ measurements enabling constraints on baryon density parameters and dark matter parameters simultaneously.

\begin{figure*}
    \centering
    \includegraphics[width=\textwidth]{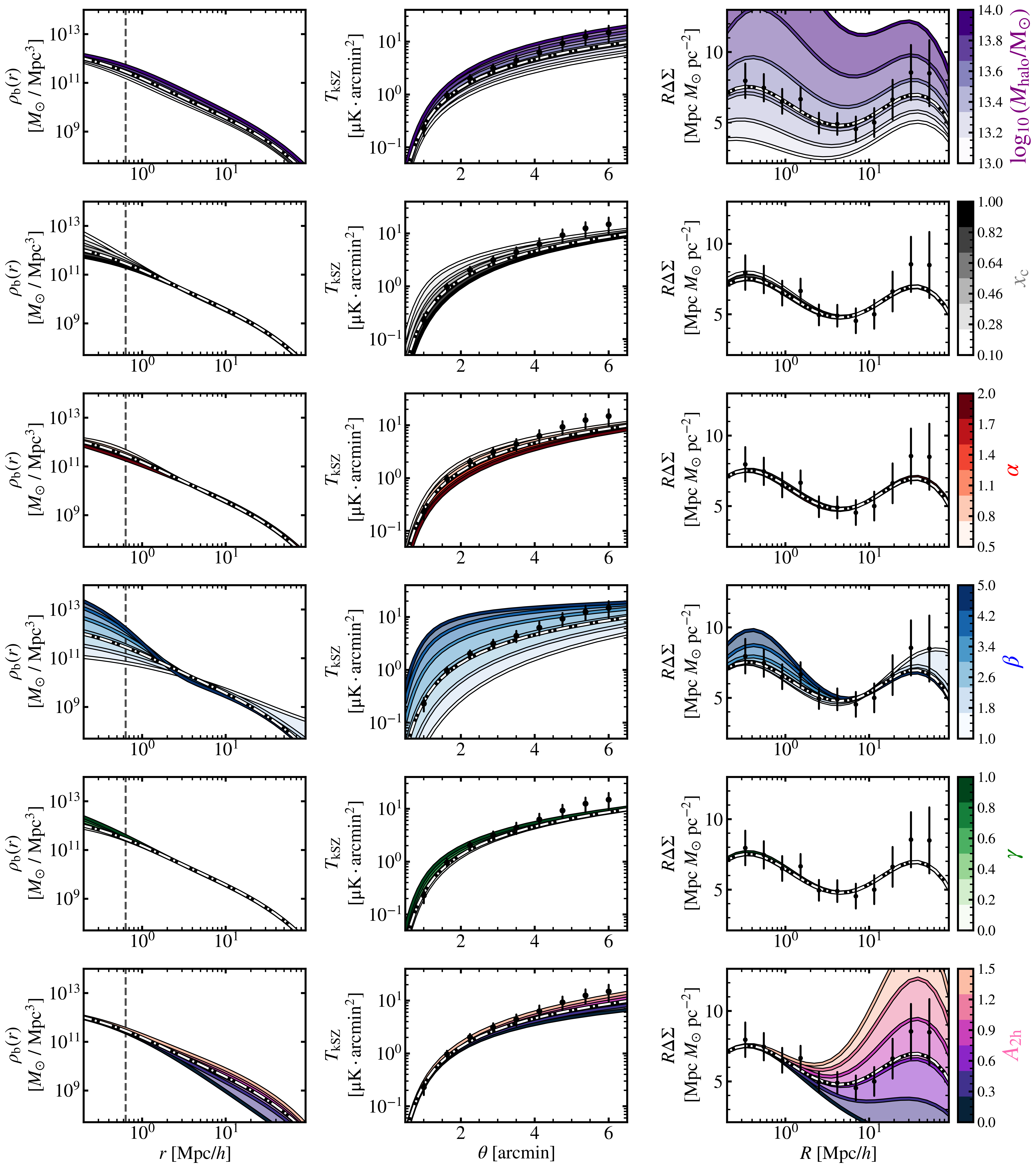}
    \caption{
    The impact of model parameters on the quantities of interest: 
    baryon density profile (left), kSZ temperature profile at 150 GHz (middle), and excess surface mass density $\Delta \Sigma$ (right). The model 
    contains 6 parameters varied from top to bottom: halo mass $M_{\rm halo}$, core scale fraction $x_{\rm c}$, intermediate 
    power law slope $\alpha$, outer power law slope $\beta$, inner power law slope $\gamma$, and the 2-halo amplitude factor 
    $A_{\rm 2h}$. For our fiducial model we hold several parameters fixed to alleviate degeneracy between parameters. 
    In the fiducial model we hold $\alpha = 1$ and $\gamma = 0.2$ fixed while using a concentration mass relation $c(M_{\rm halo})$. In the left most panel we show where $x = 1$ ($r = r_{\rm 200c}$).
    We found the halo mass $M_{\rm halo}$, the outer power law slope $\beta$, and the 2-halo amplitude $A_{\rm 2h}$ to be the most
    impactful parameters in the model given the scales we are probing with the kSZ effect and GGL. There are degeneracies between these parameters when only one measurement (kSZ or GGL) is being fit to which can be seen in either the $T_{\rm kSZ}$ or $R \Delta \Sigma$ panels. 
    }
    \label{fig: model}
\end{figure*}
\begin{table*}
    \begin{center}
    \begin{tabular} { l l c c c c}
          &  &  & \multicolumn{1}{c}{\bf kSZ Only} &  \multicolumn{1}{c}{\bf GGL Only} &  \multicolumn{1}{c}{\bf kSZ + GGL}\\
        \noalign{\vskip 3pt}\cline{4-6}\noalign{\vskip 3pt}
         Parameter & Description & Prior & 68\% limits &  68\% limits &  68\% limits\\
        \hline
        {\boldmath$\log_{10}(M_{\rm halo})$} & Halo Mass & $\mathcal{U}(12.5, 14.0)$ & $13.42^{+0.44}_{-0.27}    $ & $13.41^{+0.08}_{-0.07}   $ & $13.44^{+0.06}_{-0.05}          $\\
        {\boldmath$x_c            $} & Core scale fraction & $\mathcal{U}(0.0, 1.0)$ & $> 0.48                  $ & $> 0.46                   $ & $> 0.49   $\\
        {$\alpha          $}         & Intermediate slope & --- & --- & --- & --- \\
        {\boldmath$\beta          $} & Outer slope & $\mathcal{U}(1, 5)$  & $2.47^{+0.34}_{-0.46}             $ & $<3.51     $ & $2.45^{+0.30}_{-0.23}       $\\
        {$\gamma          $}         & Inner slope & --- & --- & --- & --- \\
        {\boldmath$A_{\rm 2h}     $} & Two halo term amplitude & $\mathcal{U}(0.0, 1.5)$ & $> 0.631         $ & $0.66\pm 0.16     $ & $0.64 \pm 0.16              $\\
        \hline
    \end{tabular}
    \caption{Summary of the halo model parameters, their priors, and the 68\% confidence intervals for the fit to only kSZ measurements, 
    only GGL measurements, and joint kSZ and GGL data. The bold font indicates the parameters that are allowed to vary in the fit. 
    In our fiducial model, we use a concentration-mass relation and we fix $\alpha = 1.0$ and $\gamma = 0.2$ to alleviate degeneracies in the GNFW model.}
    \label{tab:marginalized_parameters}
    \end{center}
\end{table*}

\subsubsection{Baryons}
We model the 1-halo density profile of baryons as a single Generalized NFW (GNFW) 
profile described in \citet{Zhao_1996_GNFW, Battaglia_2016_GNFW, Amodeo_2021_kSZ, Cromer_2022_Baryons_WL_Mass} which follows the form
\begin{equation}
    \rho_{\rm GNFW}(r) = \rho_0 \left( \frac{x}{x_{\rm c}} \right)^{-\gamma} \left( 1 + \left(\frac{x}{x_{\rm c}} \right)^{1/\alpha} \right)^{-(\beta - \gamma) \alpha} \; ,
\label{eq:GNFW}
\end{equation}
This profile is described by five parameters: $\rho_0$ the density amplitude of the gas, 
$\alpha$ the central power law slope ($x \sim 1$), $\beta$ the outer power law slope ($x \gg 1$), $\gamma$ the 
inner power law slope ($x \ll 1$), and $x_{\rm c}$ the core scale fraction.\footnote{We note that 
there are several slightly different versions of this profile in 
the literature, this implementation follows the original proposed definition of \cite{Zhao_1996_GNFW} where 
$\alpha,\beta,\gamma = 1, 3, 1$ corresponds to a standard NFW profile power law scaling.} We define 
\begin{equation}
    x \equiv \frac{r}{r_{\rm \Delta_{\rm c}}} = r \left( \frac{4 \pi \Delta \rho_{\rm crit}}{3M_{\rm halo}}\right)^{1/3} \; ,
\end{equation}
where we again use a spherical overdensity of $\Delta = 200$, but now we use the comoving critical density 
of the universe $\rho_{\rm crit}$ instead of the comoving average matter density $\bar{\rho}_{\rm m}$. We use 
a comoving $\rho_{\rm crit}$ because the radius $r$ is in comoving units. From here the 1-halo term of the baryon 
density profile is given by 
\begin{equation}
    \rho^{\rm 1h}_{\rm b}(r ) = f_{\rm b} \rho_{\rm GNFW}(r) \; .
\end{equation}
where $f_{\rm b}$ is the universal baryon fraction of the universe defined as 
$f_{\rm b} \equiv \Omega_{\rm b} / \Omega_{\rm m}$. It is important to note that $\rho_0$ 
is responsible for setting the amplitude of the baryon density profile, therefore we 
must be careful to ensure that the amplitude of the baryon density profile is normalized such that 
at sufficiently large scales (several virial radii) the mass fraction of baryons reaches the cosmic baryon fraction $f_{\rm b}$. This 
normalization can only be done numerically and is described in \cite{Cromer_2022_Baryons_WL_Mass}. We choose $\rho_0$ such that the mass ratio of baryons
to total matter enclosed in a sphere of radius $r_{\rm b}$ is equal to the cosmic baryon fraction
\begin{equation}
    f_{\rm b} = \frac{\int_0^{r_{\rm b}} 4 \pi r^2 dr \; \rho^{\rm 1h}_{\rm b}(r)}{\int_0^{r_{\rm b}} 4 \pi r^2 dr \; \rho^{\rm 1h}_{\rm m}(r)} \; .
\end{equation}
Using this condition, we solve for the baryon density amplitude $\rho_0$ to get 
\begin{equation}
    \rho_0 = \frac{\int_0^{r_{\rm b}} r^2 \rho_{\rm NFW}(r)}{\int_0^{r_{\rm b}} r^2 \left( \frac{x}{x_{\rm c}} \right)^{-\gamma} \left( 1 + \left(\frac{x}{x_{\rm c}} \right)^{1/\alpha} \right)^{-(\beta - \gamma) \alpha}} \; .
\label{eq: normalization}
\end{equation}
This normalization depends on the baryon radius $r_{\rm b}$ where we expect the mass fraction in baryons to be $f_{\rm b}$. In other words, 
$r_{\rm b}$ represents the lower-bound of where the baryons are expected to trace the dark matter. We choose $r_{\rm b} = 10 \times r_{200} \approx 6.5 \; \mathrm{Mpc}/h$ in comoving units. We note $r_{200}$ is the virial radius of the halo described in Eq. \ref{eq:r_delta}. The choice of $r_{\rm b}$ does impact $\rho_0$ if it is not 
chosen to be sufficiently large, for this reason we show it's impact in Figure \ref{fig:rb_impact} and discuss this choice further in Appendix \ref{sec:app_rb}. We note that this modelling choice does not explicitly preserve the mass of the halo (among other desirable properties) like the model framework of \cite{Bolliet_2023_kSZ}, but it was shown by \cite{Cromer_2022_Baryons_WL_Mass} that the inferred halo mass in galaxy cluster lensing simulations with baryons was unbiased with different choices of $r_{\rm b}$. In future work we anticipate modelling the entire profile using a halo model with galaxy-halo connection similar to works that model the imprint of the kSZ effect on the CMB power spectrum as a bispectrum \citep{Smith_2018_kSZ_Halo_Model, Munchmeyer_2019_kSZ, Cayuso_2023_kSZ, Roy_2023_kSZ, Bolliet_2023_kSZ}. When doing a full halo model treatment, the choice of truncation radius is crucial and must be done rigorously so as to accurately capture the amount of gas contained in each halo, but for our work we do not truncate the 1-halo term as we are modelling a single effective halo which smoothly transitions into the 2-halo term through summation. We do not truncate the 1-halo term since it's density amplitude will become negligible in comparison to the 2-halo density at sufficiently large radii $\sim 5 \; \mathrm{Mpc}/h$ in comoving units. 
\subsection{The Two Halo Term}
\label{subsec:2-halo}
The kSZ effect probes the gas on the outskirts of halos thus we have to model the contribution of gas coming 
from neighbouring halos. Similarly, the GGL signal probes the excess surface mass density around galaxies across a 
wide range of scales from the virial radius of the central halo to scales hundreds of times the virial radius 
of the central halo. We use an analytical halo model to calculate the 2-halo term of the density profiles. The 
contribution to the total matter density profile from neighbouring halos is given by 
\begin{equation}
    \rho^{\rm 2h}_{\rm m}(r) = \bar{\rho}_{\rm m} \xi^{\rm 2h}_{\rm hm}(r) A_{\rm 2h} \; ,
\end{equation} 
where $A_{\rm 2h}$ is an additional nuisance parameter to account for any potential systematic modelling errors in the 2-halo term (e.g. satellites, assembly bias, etc...), $\xi^{\rm 2h}_{\rm hm}(r)$ is the 2-halo term of the halo-matter 
correlation function, related to the matter correlation function by linear halo bias 
$\xi^{\rm 2h}_{\rm hm} = b(M) \xi^{\rm 2h}_{\rm mm}$, 
and $\bar{\rho}_{\rm m}$ is the comoving average cosmic matter density. We expect $A_{\rm 2h}$ to be less than unity to match the low amplitude of the 2-halo GGL signal \cite{Leauthaud_2017_lensing_is_low, Amon_&_Robertson_2023_GGL} compared to predictions from clustering in Planck $\Lambda$CDM. As can be seen in Figure \ref{fig:fit}, the fit to only the kSZ effect measurements prefers much larger values of $A_{\rm 2h}$ than unity which is not physically justifiable. We choose our upper prior bound for $A_{\rm 2h}$ to minimize the prior volume effect \citep[e.g.][]{RG_2024_Prior_Volume} and reduce potential bias on the physical parameters of interest. We note that $M$ denotes the mass of each halo. The correlation function is 
related to the power spectrum by the inverse Fourier transform
\begin{equation}
    \xi_{\rm mm}^{\rm 2h}(r) = \int \frac{k^2 dk}{2 \pi^2} P^{\rm 2h}_{\rm mm}(k)  j_0(kr)\; .
\end{equation}
The 2-halo term of the matter power spectrum $P^{\rm 2h}_{\rm mm}(k)$ is given by 
\begin{equation}
    P^{\rm 2h}_{\rm mm}(k) = P^{\rm lin}(k) \left[\int dM' \frac{dn}{dM'} b(M') \hat{\rho}_{\rm NFW}(k) \right]^2 \; ,
\label{eq:P_mm}
\end{equation}
where $P^{\rm lin}(k)$ is the linear matter power spectrum computed using the core cosmology library (\texttt{pyccl}) 
python package \cite{Chisari_2019_CCL}, $dn/dM$ is the halo mass function \cite{Tinker_2008_HMF}, $b(M)$ is the linear halo bias \cite{Tinker_2010_bias}, 
and $\hat{\rho}_{\rm NFW}(k)$ is the Fourier transform of the NFW profile. In this model we 
integrate over halo mass from $10^{10} M_\odot$ and $10^{15} M_\odot$. 
This completes the calculation of the 2-halo term of the total matter 
density profile. To compute the 2-halo contribution 
to the dark matter and baryon profiles we simply have
\begin{align}
\rho^{\rm 2h}_{\rm dm}(r) &= f_{\rm dm} \; \rho^{\rm 2h}_{\rm m}(r) \\
\rho^{\rm 2h}_{\rm b}(r) &= f_{\rm b} \; \rho^{\rm 2h}_{\rm m}(r) \; .
\end{align}
This completes the calculation of the 2-halo term of the density profiles of matter around galaxies. We note that 
we assume the baryons trace the dark matter on large scales
and that the gas is bound to the dark matter halos. For completeness, we write the complete density profiles for baryons and 
dark matter 
\begin{align}
    \rho_{\rm dm}(r) &= f_{\rm dm} \left( \rho_{\rm NFW}(r) +\bar{\rho}_{\rm m} \xi^{\rm 2h}_{\rm hm}(r) A_{\rm 2h} \right) \\
    \rho_{\rm b}(r)  &= f_{\rm b}  \left(\rho_{\rm GNFW}(r) +\bar{\rho}_{\rm m} \xi^{\rm 2h}_{\rm hm}(r) A_{\rm 2h} \right) \; .
\end{align}
In summary, the dark matter is described by an NFW profile and the baryons are described by a GNFW profile out to a radius of 
$r_{\rm b} = 10 \times r_{\rm 200} \approx 6.5 \; \mathrm{Mpc}/h$ in comoving units. Beyond this radius the baryons trace the dark matter with cosmic mass fractions.
\subsection{Modelling the kSZ Effect}

For the galaxies in our lens 
sample which have a redshift range of $0.4 \lesssim z \lesssim 0.7$, the optical 
depth is at the sub-percent level so $e^{-\tau} \approx 1$, which allows us to write
\begin{equation}
    \frac{\delta T_{\rm kSZ}}{T_{\rm CMB}} \approx \tau_{\rm gal}(\theta) \left( \frac{v_{\rm rms}^{\rm true}}{c} \right) \; .
\end{equation}
From cosmology and linear theory we know the RMS velocity of our lens sample is $v_{\rm rms}^{\rm true} \approx 1.06 \times 10^{-3} c$ \citep{Schaan_2021_kSZ, Amodeo_2021_kSZ}, and $\tau_{\rm gal}$
is the optical depth to electrons of the galaxy group which can be written as
\begin{equation}
    \tau_{\rm gal}(\theta) = \sigma_T \int_{\rm LoS} \frac{d\chi}{1 + z} n_{\rm e} \left( \sqrt{\left(\frac{\chi}{1 + z}\right)^2 + d_{\rm A}(z)^2 |\theta^2| } \right)  \; .
\label{eq: tau}
\end{equation}
Here we denote $d_{\rm A}(z)$ as the angular diameter distance of the galaxy group. With the kSZ effect we can probe extended gas in the outskirts 
of the dark matter halo. These measurements probe the electron density in the range of $\sim$1-4 virial radii and were first modelled by \cite{Amodeo_2021_kSZ}. The minimum scale $\theta_{\rm min} \approx 1'$ is dictated by the resolution of the stacked CMB maps while the maximum scale $\theta_{\rm max} \approx 6'$ is chosen to minimize noise contamination from CMB primary anisotropies. At these scales the contribution to baryons from stars is negligible, therefore we say that $\rho_{\rm gas}(r) \approx \rho_{\rm b}(r)$. 
Furthermore, the ICM and CGM are broadly understood to be mostly ionized \citep{McQuinn_2016_IGM, Tumlinson_2017_CGM}, for this reason we assume that the gas is fully ionized with primordial abundances 
at these scales so that
\begin{equation}
    n_{\rm e} = \frac{X_{\rm H} + 1}{2} \frac{\rho_{\rm gas}}{m_{\rm amu}} \; .
\end{equation}
We denote $X_{H}$ as the primordial mass fraction of hydrogen and $m_{\rm amu}$ to be an
atomic mass unit. Combining all of this together we can say that given $\rho_{\rm gas}(r)$ 
we can model the kSZ temperature fluctuations to the CMB as
\begin{multline}
    \frac{\delta T_{\rm kSZ}(\theta)}{T_{\rm CMB}} = \left(\frac{v_{\rm rms}^{\rm true}}{c}\right) \frac{\sigma_{T}}{m_{\rm amu}} \frac{X_H + 1}{2} \\ 
    \times \underbrace{\int_{\rm LoS} \frac{d\chi}{1 + z} \rho_{\rm gas}\left(\sqrt{\left(\frac{\chi}{1 + z}\right)^2 + d_{\rm A}(z)^2 |\theta|^2} \right)}_{\rho^{\rm 2D}_{\rm gas}(\theta)} \; .
\end{multline}
The integral is a projection of the gas density profile along the line-of-sight. The temperature 
fluctuation maps $\delta T$ that are CAP filtered are observed through the ACT instrument beam, therefore
we must convolve the projected density field with the real space beam profile $B(\theta)$ to make a 
direct comparison to the data. To perform this numerical convolution quickly and accurately, we use the Fast-Hankel Transform method described in \cite{Moser_2023_SZ_systematics}. The end result becomes
\begin{equation}
    \delta T_{\rm kSZ}(\theta) = T_{\rm CMB} \left(\frac{v_{\rm rms}^{\rm true}}{c}\right) \frac{\sigma_{T}}{m_{\rm amu}} \frac{X_H + 1}{2} \left( \rho^{\rm 2D}_{\rm gas}(\theta) * B(\theta) \right)\; .
\end{equation}
With this expression for the kSZ effect temperature fluctuations of a model average galaxy of our
sample, we can then do the same CAP filtering procedure described in Section \ref{subsec:kSZ} to 
compare to the measurements. This step is important for comparing the model to observations because there is some loss in signal caused by the CAP filter.

\subsection{Modelling Galaxy-Galaxy Lensing}
In GGL the measurement is of the excess surface mass density $\Delta \Sigma(R)$ which is fully described by the 
matter density profile $\rho_{\rm m}(r)$ previously described in Sections \ref{subsec:1-halo} and \ref{subsec:2-halo}. The 
excess surface mass density $\Delta \Sigma(R)$ is a function of the comoving surface mass density $\Sigma(R)$ which we define as 
\begin{equation}
    \Sigma(R) =  \int_{\rm LoS} \rho_{\rm m}\left(\sqrt{\chi^2 + R^2} \right) d \chi \; ,
\label{eq: sigma theory}
\end{equation}
where $\chi$ is the comoving line-of-sight distance to a redshift $z$ and $R$ refers to the 
projected comoving radius in the plane of the sky.\footnote{Throughout this work $R$ refers to projected radii and $r$ refers to non-projected radii.} The excess surface mass density 
$\Delta \Sigma$ is then defined as 
\begin{equation}
    \Delta \Sigma(R) \equiv \overline{\Sigma}(\leq R) - \Sigma(R) \;,
\end{equation}
where the first term corresponds to the average comoving surface mass density enclosed within $R$ and is given by
\begin{equation}
    \overline{\Sigma}(<R)= \frac{2}{R^2} \int dR' \; R' \; \Sigma(R') \;.
\label{eq: ccl cumul2d}
\end{equation}
We compute the total matter excess surface mass density using the FFTLog algorithm implemented in \texttt{pyccl} \cite{Chisari_2019_CCL}. With this result we have a complete consistent framework for modelling the both the kSZ effect and GGL measurements.

%% file: sections/04_results.tex
\begin{figure}
    \centering
    \includegraphics[width=\columnwidth]{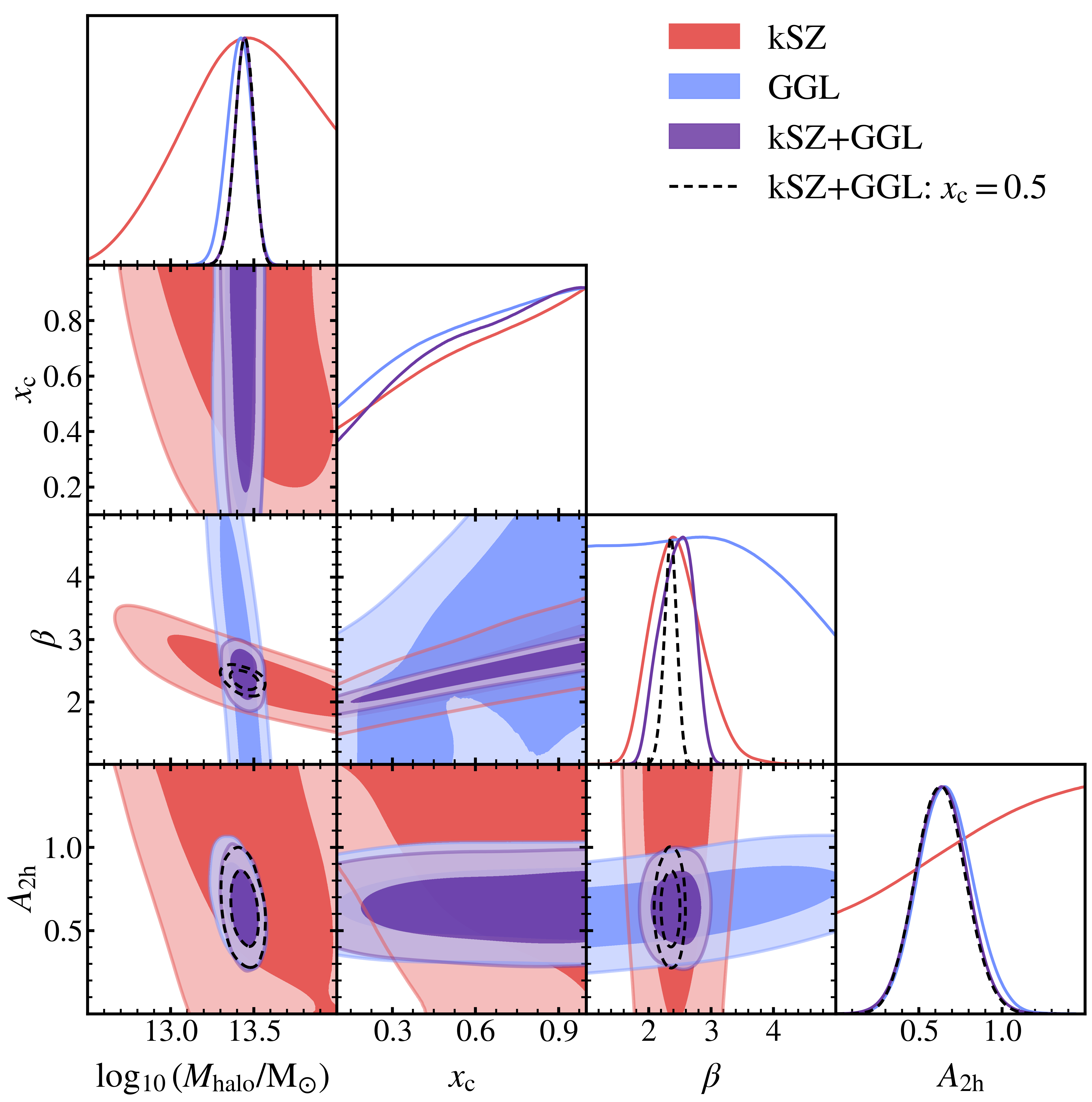}
    \caption{68\% and 95\% constraints on the halo model parameters (halo mass $M_{\rm halo}$, core scale fraction $x_{\rm c}$, outer power law slope $\beta$, and 2-halo amplitude $A_{\rm 2h}$) for the analysis of the BOSS kSZ temperature profile measurements (kSZ, red),  GGL excess surface 
    mass density profile measurements (GGL, blue) and joint observations (kSZ+GGL, purple). 
    Jointly fitting both the kSZ and GGL measurements allows for tighter constraints on dark matter and gas parameters compared to fits using only one of the datasets, effectively "disentangling" the dark matter from the gas. The GGL measurements tightly constrain the mass of the sample, and probe the matter density out to much larger 
    scales than the kSZ measurements, which effectively pins the amplitude of the 
    2-halo term $A_{\rm 2h}$ alleviating degeneracies in the other free parameters.
    }
    \label{fig:fit}
\end{figure}

\begin{figure*}
    \centering
    \includegraphics[width=2\columnwidth]{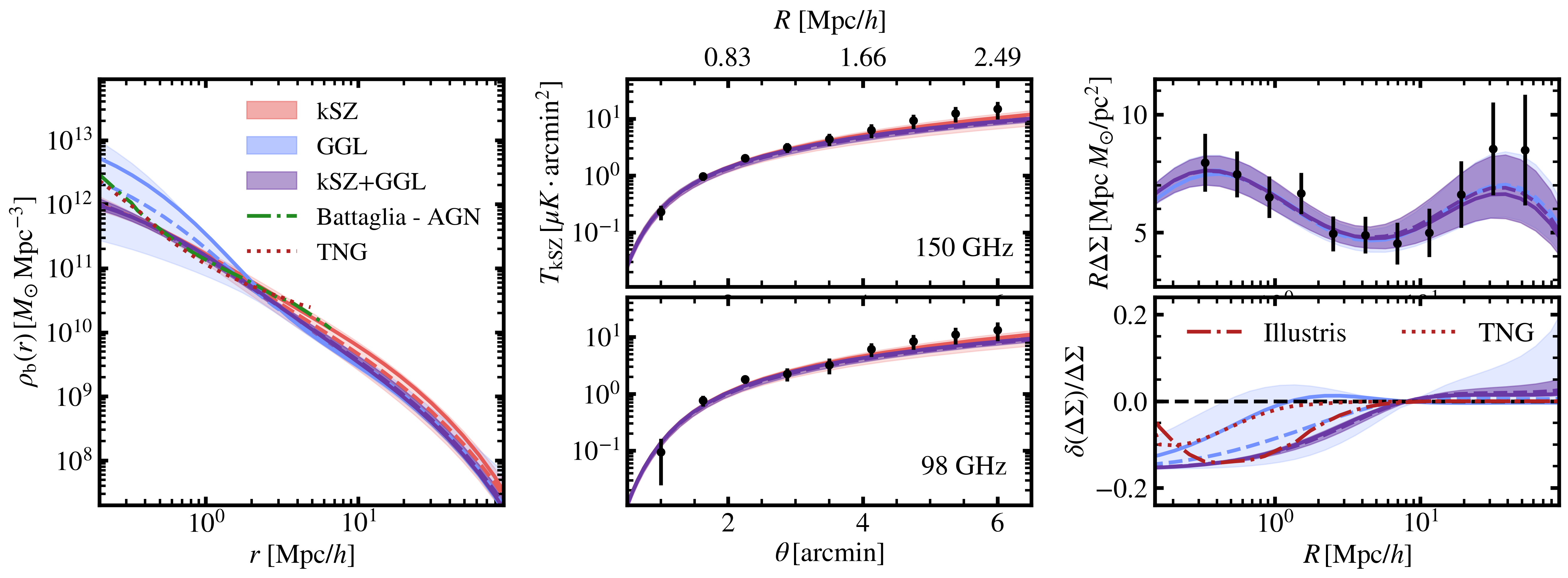}
    \caption{The baryon density profile (left), kSZ temperature profile at 
    150 GHz (top center) and 98 GHz (bottom center), and the GGL excess surface mass density profile (top right). We additionally 
    show the fractional difference in $\Delta \Sigma$ relative to a model without any baryons (bottom right). 
    The black points correspond to observations, the red corresponds to our fiducial model being fit to only the 
    kSZ measurements, the blue corresponds to our fiducial model being fit to only the GGL measurements, and the purple 
    corresponds to our joint fits to both the kSZ and GGL data. Dashed curves refer to the median (50th percentile) model of 
    the posterior distribution and the bands denote the $\pm 1\sigma$ (32nd-68th percentiles) of the distribution of 
    model curves. Lastly, solid curves denote the best-fit to the model. We show that the profile of the baryon density is tightly constrained by the joint fit relative to our kSZ or GGL only fits and 
    qualitatively different compared to the data limited fits. We show that the joint fit of kSZ with GGL tightly constrains the baryon 
    suppression in $\Delta \Sigma$ at small scales.}
    \label{fig: band_plot}
\end{figure*}
Here, we present a joint kSZ+GGL analysis of CMASS galaxies in 
the BOSS survey to disentangle the baryons and dark matter in the outskirts of the galactic halo. In our parameter fitting we assume a 
Gaussian likelihood for both measurements. The joint 
likelihood can be written as
\begin{multline}
    \ln \mathcal{L}\left[ \vec{d}_{\rm kSZ}, \vec{d}_{\rm GGL} \big| \vec{m}(\theta)\right] = -\frac{1}{2} \left(\vec{d}_{\rm kSZ} - \vec{m}(\theta)\right)^T \mathbf{C}^{-1}_{\rm kSZ} \left(\vec{d}_{\rm kSZ} - \vec{m}(\theta)\right) \\
     -\frac{1}{2}\left(\vec{d}_{\rm GGL} - \vec{m}(\theta)\right)^T \mathbf{C}^{-1}_{\rm GGL} \left( \vec{d}_{\rm GGL} - \vec{m}(\theta)\right) \; .
\end{multline}
We refer to $\vec{d}_{\rm kSZ}$, $\vec{d}_{\rm GGL}$ as the measurements for the kSZ effect and GGL, and $\mathbf{C}_{\rm kSZ}$, $\mathbf{C}_{\rm GGL}$ as their
respective covariance matrices, as described in Sections \ref{subsec:GGL} and \ref{subsec:kSZ}. One might expect that there would be non-negligible covariance between the two measurements because the kSZ probes the optical depth around galaxies while GGL probes the matter overdensity field around galaxies. We suspect that the optical depth could be correlated to the matter overdensity field and thus there could be non-negligible covariance for regions of the sky with significant overlap. In this analysis we assume the covariance between the kSZ and GGL measurements to be negligible due to a lack of overlap between the measurements on the sky (see Figure 4 of \cite{Schaan_2021_kSZ} and Figure 1 of \cite{Amon_&_Robertson_2023_GGL} for survey footprints of the kSZ and GGL measurements respectively).

We can write an expression for the posterior $\mathcal{P}$ as
\begin{equation}
    \ln \mathcal{P}(\theta \big| \vec{d}_{\rm kSZ}, \vec{d}_{\rm GGL}) = \ln \left[ \mathcal{L} \left(\vec{d}_{\rm kSZ}, \vec{d}_{\rm GGL} \big| \vec{m}(\theta)\right) \mathrm{Pr}(\theta) \right] \; .
\end{equation}
We refer to $\vec{m}(\theta)$ as our model function for both $T_{\rm kSZ}$ and $\Delta \Sigma$ given our model parameters $\theta$. We define $\mathrm{Pr}(\theta)$ to be the priors on our model parameters $\theta$ which can be found in Table \ref{tab:marginalized_parameters}. We sample the posterior using an affine-invariant ensemble Markov Chain Monte-Carlo (MCMC) algorithm implemented in the package \textsc{emcee} \cite{FM_2013_emcee}. To ensure convergence we ran 
the chains sufficiently long such that the 
Gelman-Rubin convergence parameter $\mathcal{R}$ 
was less than 1.01 for all model parameters \cite{Gelman_1992_Stat}.

\subsection{The complementarity of kSZ and GGL}
\label{subsec:complimentarity}

We use the model that describes the two likelihoods to jointly fit for the mass of the dark matter halo $M_{\rm halo}$, 
the core scale fraction in the baryon gas profile $x_{\rm c}$, the outer power law slope for the baryon gas profile $\beta$, and the overall amplitude of the 2-halo term $A_{\rm 2h}$, which we include as a nuisance parameter. Figure \ref{fig: model} shows that the GGL measurements will be most sensitive to $M_{\rm halo}$ and $A_{\rm 2h}$ while the kSZ effect measurements will be most sensitive to the baryon gas parameters $x_{\rm c}$ and $\beta$. We also see that nearly all baryon gas parameters are either degenerate with $\beta$ or have negligible effects when varied. To alleviate some degeneracy we fix the inner power law slope $\gamma = 0.2$, the intermediate power law slope $\alpha = 1$.

Figure \ref{fig:fit} shows the posteriors for our model parameters obtained from fitting to 
the stacked kSZ profile data (red). The best-fit (minimum $\chi^2$) for this model is $\chi^2_{\rm min}/ \mathrm{d.o.f} = 21.8/16$ (PTE $=0.15$). As one would expect, the halo mass and amplitude of the 2-halo term are largely unconstrained when the model only has access to the kSZ profile.\footnote{Note that we tested the impact of using a substantially flexible prior on the amplitude of the 2-halo term,  $A_{\rm 2h} = \mathcal{U}(0.0, 20.0)$. In this case, we do constrain the parameter, $A_{\rm 2h} = 3.2^{+1.4}_{-2.4}$. For our main analysis, we keep the physically-motivated prior to avoid exploring large parts of unphysical parameter space. } Complementary are the blue contours, which show the posterior from the GGL data. The GGL measurements offer a lot of constraining power on $M_{\rm halo}$ and  $A_{\rm 2h}$ but don't have much constraining power for $\beta$ or $x_{\rm c}$. This fit to the data gave a best-fit $\chi^2_{\rm min}/ \mathrm{d.o.f} = 2.5/9$ (PTE $=0.98$) implying that the GGL measurements were over-fit by our model. Furthermore, the GGL measurements are a probe of the total matter distribution which is comprised of mostly dark matter making it difficult to infer substantial information of the baryon density profile. 

The joint fit to the data is shown in purple. This fit yielded a best-fit $\chi^2_{\rm min}/ \mathrm{d.o.f} = 24.8/26$ (PTE $=0.53$) implying a good fit to the data. In these purple contours we can see the overlapping regions of unconstrained red and blue contours leading to relatively tight constraints on all parameters except $x_{\rm c}$. If we fix $x_{\rm c} = 0.5$, then we can see that $M_{\rm halo}$, $\beta$, and $A_{\rm 2h}$ are all tightly constrained (denoted by dashed black line in Figure \ref{fig:fit}) with little degeneracy. By jointly fitting to both kSZ and GGL data we have `disentangled' baryons from dark matter. 

Figure \ref{fig: band_plot} shows the best-fit models from the kSZ, GGL, and kSZ+GGL data for the baryon density profile $\rho_{\rm b}(r)$ as a function of radius (left), the kSZ profile (middle) and the GGL (right). The kSZ measurements constrain the density on small scales near the transition from the 1-halo to 2-halo term and the GGL measurements are good at constraining $\rho_{\rm b}(r)$ at larger scales in the 2-halo term. When combined we see a tight constraint on the density profile across all scales, within our model where $\alpha=1$ and $\gamma=0.2$ are fixed. We explored allowing $\alpha$ to be a free parameter within the model and found that the density profile is still well constrained across all scales despite $\alpha$ being unconstrained. Having tight constraints on the baryon density profile across such a wide range of scales enables more direct comparisons to simulations.

In the bottom right panel of Figure \ref{fig: band_plot} we show the fractional difference in the GGL signal due to baryons $\delta (\Delta \Sigma)/\Delta \Sigma = (\Delta \Sigma - \Delta \Sigma_{\rm nb}) / \Delta \Sigma_{\rm nb}$, where $\Delta \Sigma_{\rm nb}$ refers to the same fiducial model except we exclude baryons ($f_{\rm b} = 0 \; \rightarrow \; f_{\rm cdm} = 1$). When the model only has access to the GGL measurements, the suppression in $\Delta \Sigma$ at small scales is unconstrained but when we include the kSZ profile measurements into the fit we can clearly see that suppression in $\Delta \Sigma$ at small scales is well constrained. 

\begin{figure}
    \centering
    \includegraphics[width=\columnwidth]{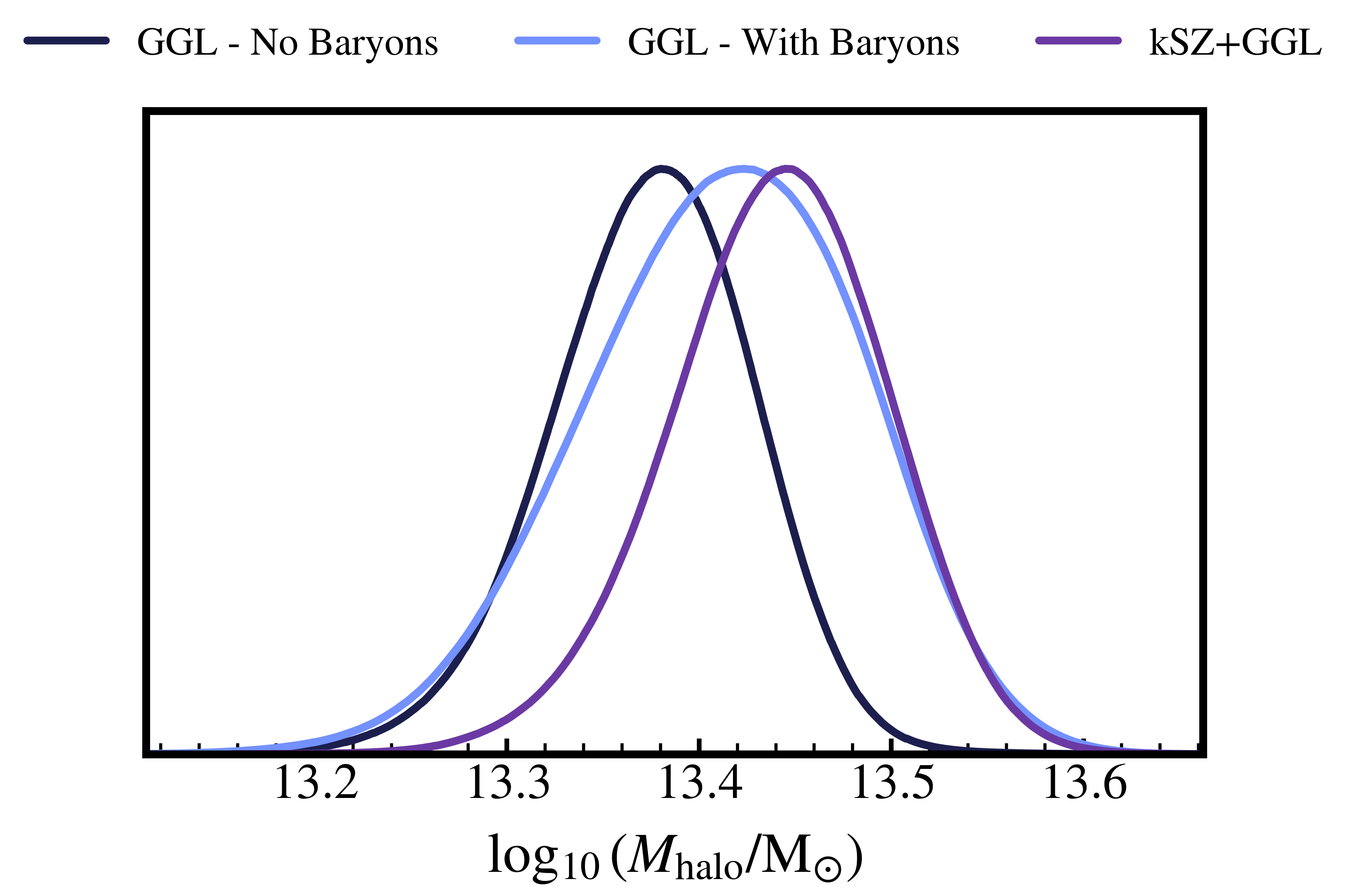}
    \caption{Constraints on the halo mass 
    from GGL where baryons are neglected (dark blue, $f_{\rm cdm} = 1$), where the baryon contribution and its uncertainty is modelled (blue, $f_{\rm cdm} = 1 - f_{\rm b}$) and where kSZ is jointly fit with the GGL to constrain the baryon parameters (purple). The prior range for the halo mass adopted by \citep{Bigwood_2024_kSZ_WL} to encompass the uncertainty on the stellar masses and the SHMR spans beyond the extent of the plot.
    }
    \label{fig: b_vs_nb}
\end{figure}

\subsection{Halo Mass}
\label{subsec:halo mass}

As a consequence of the suppression of $\Delta \Sigma$ caused by baryons, as seen in Figure \ref{fig: band_plot} (lower right), neglecting to account for baryons can bias halo mass estimates $M_{\rm halo}$, compared to the true mass \citep[e.g.][]{Debackere_2020, Cromer_2022_Baryons_WL_Mass}. This effect is demonstrated in Figure~\ref{fig: b_vs_nb} where we show the posteriors of the halo mass for three analyses of the GGL data: fitting a model that neglects baryons ($f_{\rm b} = 0 \; \rightarrow \; f_{\rm cdm} = 1$) to the GGL data, fitting our model which includes baryons to the same GGL data, and the same, where we jointly analyse GGL with kSZ profile measurements. We find that neglecting the impact of baryons lowers our estimate of the halo mass by $\sim 20\%$. Including the impact of baryons in the model but only using GGL data inflates our posterior, which is improved by the inclusion of the kSZ measurements. That is, a joint analysis of GGL and kSZ can account for the gas parameters, shrinking the width of the posterior back down, while the halo mass estimate undergoes a $\sim 20\%$ shift compared to the no-baryon model. 

Our GGL+Baryons and kSZ+GGL constraints on $M_{\rm halo}$ are consistent with the mean halo mass $\langle M_{\rm halo} \rangle = 3 \times 10^{13} \; \rm M_\odot$ obtained with the SHMR  \cite{Schaan_2021_kSZ} at the $1\sigma$ level but our GGL with no baryons constraint on halo mass shifts downward to agree at the $2\sigma$ level. We note that our mass estimate could be underestimated by $\lesssim 10 \%$  due to the exclusion of miscentering in our model \citep{Skibba_2011_miscenter, Hoshino_2015_miscenter}. In future work we will improve our model by including satellites and accounting for miscentering which will yield an unbiased estimate of the halo mass.

A crucial step in using kSZ profile measurements as a benchmark for feedback models and simulations is an accurate constraint on the halo mass. The halo mass can change both the shape and the amplitude of the kSZ signal as seen in Figure \ref{fig: model}. The ratio of kSZ signals modelled with halo masses of $10^{14} \; \rm M_\odot$ and $10^{13} \; \rm M_\odot$ is $\approx 2.1$ at $\theta = 1'$ and grows with increasing $\theta$, plateauing to $\approx 3.3$ at $\theta = 6'$. Assuming a certain halo mass estimated from stellar masses is potentially dangerous as stellar masses and SHMR can be biased \citep{Yuan_2021_Assembly_Bias_CMASS, Oyarzun_2024_SHMR_bias}. 

These kSZ profile measurements were jointly analysed with cosmic shear for the first time by \cite{Bigwood_2024_kSZ_WL} to constrain the power spectrum and they found a best-fitting model with stronger baryonic feedback than hydrodynamical simulations predict. As discussed in \cite{Bigwood_2024_kSZ_WL} (their Appendix B3), a major source of uncertainty in constraining the impact of baryonic feedback from kSZ and cosmic shear measurements, was the halo mass of the sample, which spans almost an order of magnitude: $M_{\rm halo} = 0.8-30 \times 10^{13} \; M_{\odot}$ (significantly larger than the range of masses plotted in Figure \ref{fig: b_vs_nb}). For this reason, \cite{Bigwood_2024_kSZ_WL} had to use a large prior for $M_{\rm halo}: \mathcal{U}[5\times10^{12},7\times10^{13}]$. Having such a large uncertainty on halo masses makes it difficult to interpret the strength of baryonic feedback from the stacked kSZ profiles measured by \cite{Schaan_2021_kSZ, Hadzhiyska_2024_ACT_DESI_kSZ} and future measurements to come. The methodology in this work can provide a more precise prior on the effective halo mass of the sample. 

We compare our ($2 \sigma$) posteriors on the halo mass $\log_{10}(M_{\rm halo, 200m} / \rm M_\odot) = 13.44 \pm 0.12$ to the halo mass found by \cite{McCarthy_2024_kSZ_WL} where they jointly analysed the kSZ and GGL measurements with the FLAMINGO simulation suite \citep{Schaye_2023_FLAMINGO}. In this work we've used a mass definition of $M_{\rm 200m}$, to match \cite{McCarthy_2024_kSZ_WL} we transform to a mass definition of $M_{\rm 500c}$ using an NFW profile \citep{NFW_1997} and the concentration mass relation of \cite{Duffy_2008_Concentration}. Our ($2 \sigma$) posteriors on the halo mass are $\log_{10}(M_{\rm halo, 500c} / \rm M_\odot) = 13.22 \pm 0.12$ and the ($2\sigma$) posteriors on the halo mass from \cite{McCarthy_2024_kSZ_WL} are $\log_{10}(M_{\rm 500c} / \rm M_\odot) = 13.34 \pm 0.04$. We find that our estimate of the halo mass is slightly lower than what was found by \cite{McCarthy_2024_kSZ_WL} but still statistically consistent at the $2 \sigma$ level. We note that there is additional uncertainty introduced by our method of converting between mass definitions due to the fact that the baryons are not described by an NFW profile, but the baryon fraction at $r_{\rm 500c}$ is lower than the cosmic baryon fraction and should only introduce uncertainty at the few percent level.

\subsection{Baryonic Feedback}
\label{subsec:feedback}

The kSZ effect probes the distribution of gas in the outskirts of halos. Only recently have direct measurements of the stacked kSZ effect profile had high enough SNR to be a powerful probe of the gas \citep{Schaan_Ferarro_2016_kSZ_detection, Schaan_2021_kSZ, Hadzhiyska_2024_ACT_DESI_kSZ, RG_2025_kSZ_ACT_DESI_Spec}. The ACT-BOSS kSZ measurements used in this analysis from \cite{Schaan_2021_kSZ} were first analysed with their corresponding tSZ measurements to study the thermodynamic properties of the halo by \cite{Amodeo_2021_kSZ} and have recently been jointly analysed with cosmic shear for the first time by \cite{Bigwood_2024_kSZ_WL} to constrain the matter power spectrum. The joint analysis of \cite{Bigwood_2024_kSZ_WL} found a best-fitting model that preferred a more extreme baryonic feedback scenario than hydrodynamical simulations predict. These same kSZ and GGL measurements were then analysed using the FLAMINGO simulations directly by \cite{McCarthy_2024_kSZ_WL}. Their analysis found that the kSZ measurements required more extreme baryonic feedback than the FLAMINGO simulations predicted by $7 \sigma$. With this context in mind, we compare our joint kSZ+GGL constraints to the GNFW gas profile scaling relations derived from hydrodynamical simulations in \cite{Battaglia_2016_GNFW}.

\begin{figure}
    \centering
    \includegraphics[width=\columnwidth]{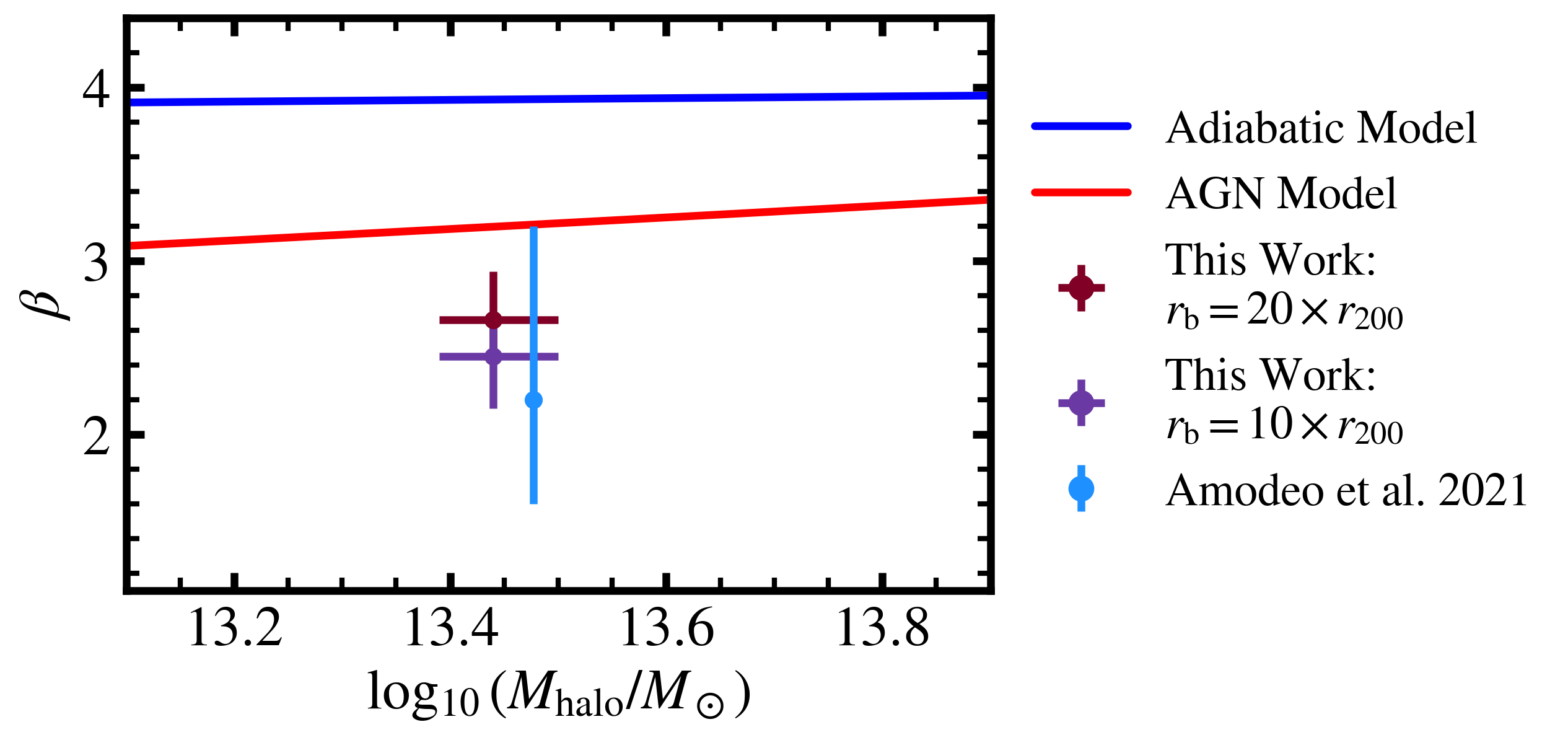}
    \caption{
    Scaling relations derived from \citet{Battaglia_2016_GNFW} hydrodynamical simulations  for the outer power law slope, $\beta$, of the baryon density profile in our model as a function of halo mass, $M_{\rm halo}$. The blue line corresponds to simulations with only gravitational heating and the red line corresponds to simulations with additional radiative cooling, star formation, supernova feedback, cosmic rays, and AGN feedback. Lower values of the outer power law slope 
    $\beta$ indicate the baryons are more diffusely spread out in the density profile to the outermost regions of
    the halo. This suggests that the data prefers stronger baryonic feedback relative to the hydrodynamical simulations of \citet{Battaglia_2016_GNFW}. In our model we use a fiducial value for the baryon radius $r_{\rm b} = 10 \times r_{200}$ (purple), but we include an additional data point with $r_{\rm b} = 20 \times r_{200}$ to show the range of values $\beta$ could be with this model depending on the choice of $r_{\rm b}$. All fits plotted from this work allow $x_{\rm c}$ to be a free parameter.
    }
    \label{fig: scalings}
\end{figure}

It can be seen in Figure \ref{fig: model} that the outer power law slope $\beta$ is the most important parameter in our model for describing the distribution of baryons in the halo outskirts. Since we are not modelling the entire distribution of galaxies with an HOD, we are measuring an effective value of $\beta$ which inherently assumes that all halos in our sample have the same mass. This effective $\beta$ could be different than an HOD-based model would find and we leave this as a future improvement to the model. In Figure \ref{fig: scalings} we compare our constraints on $\beta$ to scaling relations $\beta(M, z)$ derived from the cosmological hydrodynamic simulations of \cite{Battaglia_2016_GNFW} (corrected for difference in GNFW profile definition, see Appendix \ref{sec:app_amodeo}). The $\beta(M, z)$ relation was derived from gas profiles of resolved halos with masses of $M_{500} = 10^{14} - 10^{15} \; \rm M_\odot$, for this reason we acknowledge that our comparisons are to an extrapolation of the $\beta(M, z)$ relation of \cite{Battaglia_2016_GNFW}. 

We show our $\pm 1 \sigma$ estimates on $\beta$ and $M_{\rm halo}$ compared to 2 models from \cite{Battaglia_2016_GNFW}. The blue curve in Figure \ref{fig: scalings} is the \textit{Adiabatic} model which corresponds to a simulation with only gravitational heating, the red curve is the \textit{AGN} model which includes radiative cooling, star formation, supernova feedback, cosmic rays, and AGN feedback (more extreme feedback). A lower value of $\beta$ implies that the baryons are more spread out to larger radii implying stronger baryonic feedback. We find that our posteriors on halo mass and $\beta$ are consistent with the assumptions and findings of \cite{Amodeo_2021_kSZ} where they found $\beta = 2.6^{+1.0}_{-0.6}$ (at fixed mean halo mass $\langle M_{\rm halo}\rangle = 3\times10^{13} \; \rm M_\odot$: light blue in Figure \ref{fig: scalings} with necessary correction discussed in Appendix \ref{sec:app_amodeo}) when fitting a GNFW gas profile to the same kSZ measurements.\footnote{We note that the modelling done in \cite{Amodeo_2021_kSZ} includes a mass-weighted average over the halo masses in their sample, for comparison we plot their constraints on $\beta$ at the mean halo mass in their sample in Figure \ref{fig: scalings}.} We find our constraints on $\beta$ are tighter than  \cite{Amodeo_2021_kSZ} by a factor of 2 when fitting our model to the same data (kSZ Only fit in Table \ref{tab:marginalized_parameters}). This improvement in constraining $\beta$ is due to the way we normalize the baryon (gas) density profile (see Eq. \ref{eq: normalization}) which effectively increases the importance of $\beta$ in the model. When we include GGL in the kSZ+GGL joint analysis our constraints on $\beta$ tighten further by 50\% yielding a total factor of 3 improvement compared to \cite{Amodeo_2021_kSZ} while also giving a direct estimate on the halo mass $M_{\rm halo}$.

When fitting our fiducial model to kSZ+GGL measurements, we find that our constraints on $\beta$ are lower than the \textit{AGN} cosmological hydrodynamic simulation feedback model of \cite{Battaglia_2016_GNFW}. We note that our constraints on $\beta$ do change with respect to our choice of baryon radius $r_{\rm b}$ described in Appendix \ref{sec:app_rb}. In Figure \ref{fig: scalings}, we also show our fiducial choice of $r_{\rm b} = 10 \times r_{200}$ compared to an extreme choice of $r_{\rm b} = 20 \times r_{200}$. The baryon radius $r_{\rm b}$ represents the radius at which the mass fraction of baryons to matter is equal to the cosmic baryon fraction $f_{\rm b}$ and it is crucial for normalizing the amplitude of the baryon density profile $\rho_0$. As $r_{\rm b}$ increases $\beta$ does too, but if the radius is too large the applicability of the NFW profile used to describe the dark matter and the GNFW profile used to describe the baryons is no longer appropriate. We show the impact of our choice of $r_{\rm b}$ on kSZ+GGL posteriors in Figure \ref{fig:rb_impact}, but in all cases our joint kSZ+GGL analysis implies that baryonic feedback is stronger than the cosmological hydrodynamic simulations of \cite{Battaglia_2016_GNFW} with AGN feedback predict. 

In Figure \ref{fig: band_plot} we compare our results to hydrodynamical simulations where available. In the rightmost panel we show the suppression of $\Delta \Sigma$ in prominent hydrodynamical simulations: Illustris \citep{Vogelsberger_2014_Illustrist} (dark red, dot-dashed) and IllustrisTNG-300 \citep{Pillepich_2018_TNG} (dark red, dotted). These curves are from \cite{Lange_2019_BOSS_baryon_curves} where they show the impact of baryonic feedback on the GGL signal around CMASS-like halos with a stellar mass range of ($\log_{10}(M_* / \rm M_\odot) = 10.5 - 12$) in Illustris and IllustrisTNG. They select halos from snapshots matching the mean redshift of the CMASS sample $\langle z \rangle = 0.55$ and compute the suppression in $\Delta \Sigma$ from baryons for 4 separate stellar mass bins. For a given simulation (Illustris or IllustrisTNG), the difference in suppression within $R \lesssim 1 \; \mathrm{Mpc}/h$ (in comoving units) is within a few percent. In Figure \ref{fig: band_plot} we plot the average of the $11 \leq \log_{10}(M_* / \rm M_\odot) \leq 11.5$ bin and the $11.5 \leq \log_{10}(M_* / \rm M_\odot) \leq 12.0$ to most closely represent the expected suppression of our sample.

The difference in the suppression of $\Delta \Sigma$ between the older Illustris simulation and the TNG-300 simulation can be explained by the differences in subgrid model implementations for feedback. The Illustris model is known to have more extreme feedback than the TNG-300 model but is also known to have shortcomings in predicting observations such as gas fractions, galaxy color and morphologies, and several cluster properties \citep{Nelson_2019_TNG_50, Kauffman_2019_Illustris, Rodriguez_Gomez_2019_Illustris}. Our kSZ+GGL best-fit suggests a suppression of $\Delta \Sigma$ that gradually increases from 0\% at 10 Mpc/$h$ to $15 \%$ for $R \lesssim 1 \; \mathrm{Mpc}/h$ (in comoving units) which appears to more closely reflect the Illustris feedback model than the IllustrisTNG-300 feedback model. Newer measurements of the kSZ effect profile in DESI-Y1 data also appear to have a shape that agrees with the older Illustris feedback model more than the IllustrisTNG-300 feedback model \citep{Hadzhiyska_2024_ACT_DESI_kSZ, RG_2025_kSZ_ACT_DESI_Spec}. 

Our last point of comparison to simulations is directly to the baryon (gas) density profile shown in the leftmost panel of Figure \ref{fig: band_plot}. Alongside our fits, we show curves from \cite{Amodeo_2021_kSZ} for the \textit{AGN} feedback model of \cite{Battaglia_2016_GNFW} (green, dash-dot) also depicted in Figure \ref{fig: scalings} (red, solid) and the IllustrisTNG simulations (dark red, dotted). To produce the density profile for the \textit{AGN} feedback simulation, \cite{Amodeo_2021_kSZ} used the best-fit relationships from \cite{Battaglia_2016_GNFW} to compute a mass-weighted density profile with weights chosen to match the CMASS halo mass distribution. They chose the redshift to be the mean redshift of the CMASS sample $\langle z \rangle = 0.55$. For the IllustrisTNG density profile, they modelled the profile by selecting ``red'' galaxies with colours \texttt{sdss\_g - sdss\_r} $\geq$ 0.6 \citep{Nelson_2018_Color_Bimodality} and weighting each galaxy by stellar mass to match the corresponding stellar mass distribution of the CMASS sample. We see that the two simulations are similar in shape and amplitude and appear to be comparable to our fits to the measurements. A deeper exploration of modelling choices on sample selection in the simulations was done by \cite{Moser_2021_SZ_choices}, and they found for the IllustrisTNG simulations with the same mock selection as done in \cite{Amodeo_2021_kSZ} that $\beta = 3.20 \pm 0.11$ ($\beta = 2.8 \pm 0.11$ when corrected for difference in profile definition, see Appendix \ref{sec:app_amodeo}). This value of $\beta$ sits between the \textit{AGN} model of \cite{Battaglia_2016_GNFW} and our kSZ+GGL fits with $r_{\rm b} = 10 \times r_{200}$. It would be informative to have similar GNFW density profile parameter fits found in \cite{Battaglia_2016_GNFW} from different hydrodynamical simulations to get a better understanding of how our measurements compare to other feedback schemes. 

We note that our model neglects satellites, for this reason we are neglecting the potential systematic of 'miscentering'. We expect that not accounting for this effect causes an underestimation of the kSZ amplitude in simulations by up to $\sim 20\%$ for CMASS-like galaxies ($11.5 < \log M_* < 12.1$) as shown in Figure 6 of \cite{Hadzhiyska_2023_SZ_sims}, at the current level of statistical power, this effect would not significantly change the conclusions of this work. 
We note that this effect is dependent on the satellite fraction of the sample which means that miscentering is more significant for lower mass galaxies ($10.9 < \log M_* < 11.5$) and less significant for more massive galaxies ($12.1 < \log M_* < 12.7$). 
Considering Figure~\ref{fig: model}, we can see that ignoring miscentering from satellites in our model would most likely mean our values for $\beta$ are overestimated by $\sim 0.2-0.4$ which would not significantly impact the conclusions from this study. There is of course, a possibility that miscentering could alter the shape of the kSZ profile further altering the value of $\beta$ changing our results. Including this effect in the model can become computationally expensive, but it has been modelled for observations of the tSZ effect \citep{Popik_tSZ_HOD_Model}. We leave this improvement to our model of kSZ+GGL as future work.

%% file: sections/05_conclusions.tex
In this work we present an analytic framework to model GGL measurements jointly with stacked kSZ profile measurements to disentangle the halo. We jointly analyse combined GGL measurements from DESY3 and KiDS-1000 to probe the matter density around BOSS CMASS galaxies \cite{Amon_&_Robertson_2023_GGL}, while our stacked kSZ profile measurements from ACT DR5 and Planck maps probe the gas density around the same galaxies \cite{Schaan_2021_kSZ}. Jointly modelling these two signals offers great potential for 
disentangling baryons from dark matter in the outskirts of dark matter halos at the expense of requiring a complex model to consistently describe both signals which (to our knowledge) has never been done before. We model the signals using an analytical halo model to fit for both the 
halo mass and the gas density profile parameters simultaneously. 
Our most important findings from our analysis are:
\begin{itemize}
    \item[\textbf{I.}] the kSZ effect serves as a complementary probe to GGL. By doing a joint fit to kSZ+GGL data, we can alleviate degeneracies in the model that exist when fitting to a single dataset (kSZ or GGL) as seen in Figures \ref{fig: model} and \ref{fig:fit}. When our fiducial model is fit to both kSZ and GGL measurements, the baryon density profile is tightly constrained at all relevant scales $r \approx 0.3 - 50$ Mpc/$h$ (in comoving units) as seen in the left panel of Figure \ref{fig: band_plot}. By having a tight constraint on the baryon density profile around the outskirts of the galactic halo, we determine empirically how baryons suppress $\Delta \Sigma$ at small scales as seen in the right panel of Figure \ref{fig: band_plot}. This work opens an avenue for using small scales in future cosmological weak lensing analyses. Additionally, not including baryons in a model of GGL can shift halo mass estimates by $\sim 20 \%$ compared to a model that includes baryons fit to kSZ+GGL measurements as seen in Figure \ref{fig: b_vs_nb}.
    \newline
    \item [\textbf{II.}] Given our modelling choices and assumptions, we find our model is consistent with enhanced baryonic feedback compared to the hydrodynamical simulations from \cite{Battaglia_2016_GNFW}. This comes from comparing our kSZ+GGL constraints of the outer power law slope $\beta$ on the baryon density profile to relationships $\beta(M,z)$ derived from \citet {Battaglia_2016_GNFW} hydrodynamical simulations, as seen in Figure \ref{fig: scalings}. Furthermore, we find that our kSZ+GGL constraints on the suppression in $\Delta \Sigma$ due to baryons is better represented by the Illustris simulation than the IllustrisTNG simulation (see Figure \ref{fig: band_plot}).
\end{itemize}

Our model neglects the modelling of satellites which can have a significant impact on the kSZ profile through miscentering and could bias our inferred effective value of $\beta$. The next step forward in improving the model would be to account for the galaxy-halo connection using an HOD framework similar to the work done by \cite{Popik_tSZ_HOD_Model} for the tSZ effect. This improvement would enable us to account for the presence of satellites and it enables a self-consistent halo model treatment of the 1-halo and 2-halo terms for an accurate model of the matter spectrum. This would allow the kSZ measurements to be used a cosmological analysis of galaxy-galaxy lensing and clustering to model the impact of baryon feedback on smaller scale measurements. 

Going forward, there are many avenues to build upon this analysis. First, new measurements of the stacked kSZ profile from ACT-DR6 and DESI-Y1 data using photometric \citep{Hadzhiyska_2024_ACT_DESI_kSZ, Ried_Guachalla_2024_VR, Hadzhiyska_2024_VR} and spectroscopic \citep{RG_2025_kSZ_ACT_DESI_Spec} redshifts have significantly higher SNR than the measurements from \cite{Schaan_2021_kSZ} (analysed in this work) with the ability to measure the signal in a more narrow range of redshift and mass. In the near future, Simons Observatory (SO) will bring better CMB measurements to accompany the DESI spectroscopic survey and yield better kSZ profile measurements. At the same time, upcoming WL surveys such as Euclid, LSST, and Roman will provide substantially more powerful WL measurements. Additionally, when combined, the kSZ effect and the tSZ effect can probe the thermodynamic profiles around galaxies \citep{Amodeo_2021_kSZ}, when combined with additional information from GGL one can have unprecedented constraining power for testing baryonic feedback models. Together, these measurements will enable a better understanding of baryonic feedback, halo masses, and cosmology.

%% file: sections/06_appendix.tex
\section{Impact of Baryon Radius}
\label{sec:app_rb}
\begin{figure}
    \centering
    \includegraphics[width=\columnwidth]{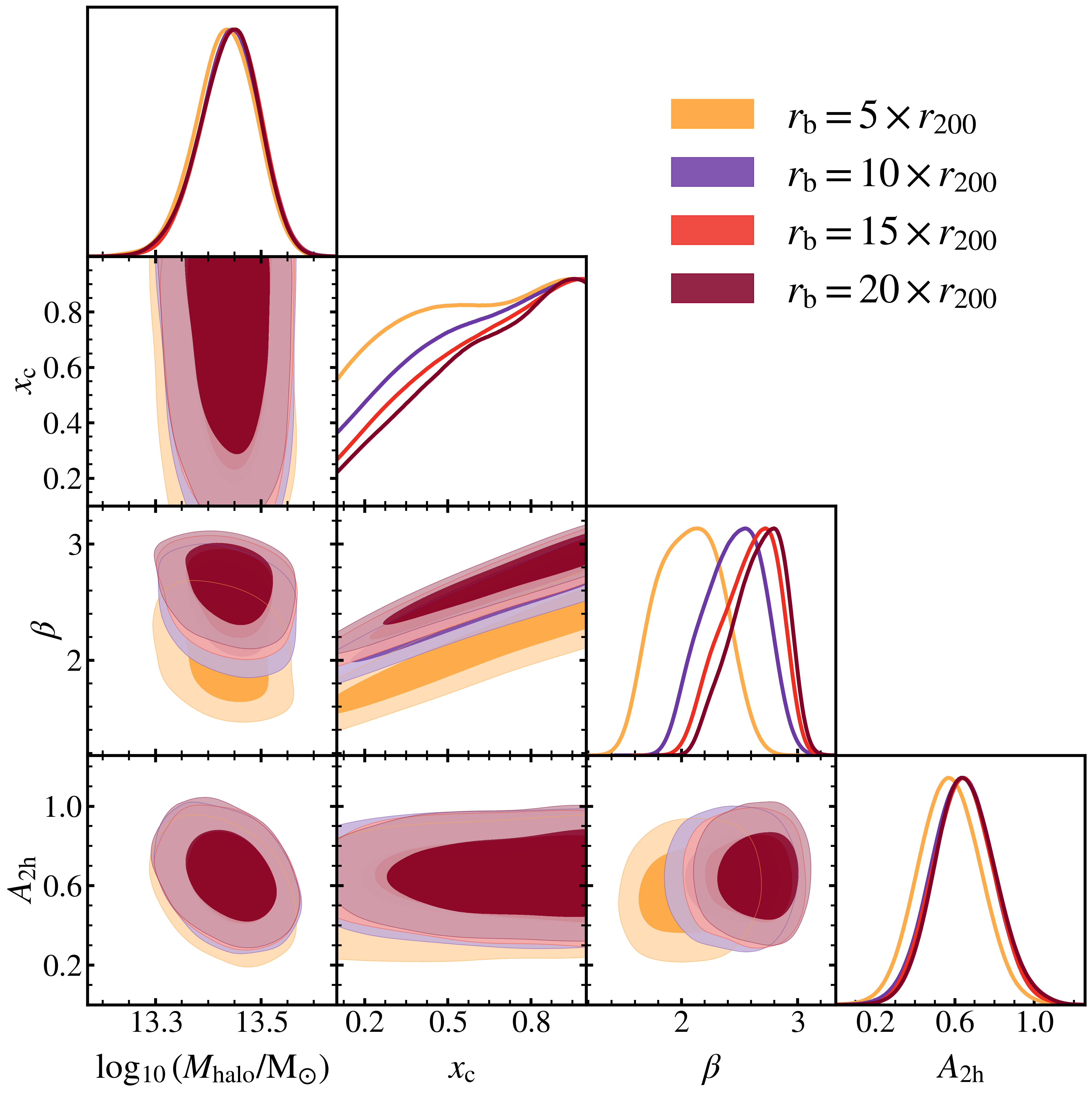}
    \caption{We show 4 different choices of $r_{\rm b}$ ranging from $r_{\rm b} = 5 \times r_{200}$ to $r_{\rm b} = 20 \times r_{200}$ to demonstrate the impact of the choice of $r_{\rm b}$ on our posterior for dark matter and baryon profile parameters. As $r_{\rm b}$ increases we note that the outer power law slope $\beta$ increases while the halo mass $M_{\rm halo}$ and 2-halo amplitude $A_{\rm 2h}$ remain relatively unchanged. We note that as $r_{\rm b}$ becomes sufficiently large $r_{\rm b} \gtrsim 10 \times r_{200}$, it's affect on $\beta$ saturates in our fits. If $r_{\rm b}$ is chosen to be too large, then the validity of the single halo radial profiles used in our model breaks down. If $r_{\rm b}$ is chosen to be too small (i.e $r_{\rm b} = 5 \times r_{200}$) then we add risk of assuming universal mass fraction prematurely which implies lower values of $\beta$ potentially implying stronger unphysical baryonic feedback. For these reasons, our fiducial model uses $r_{\rm b} = 10 \times r_{200} \approx 6.5 \; \mathrm{Mpc}/h$ in comoving units.}
    \label{fig:rb_impact}
\end{figure}

We define the baryon radius $r_{\rm b}$ to be the radius where we expect the mass fraction in baryons to be the cosmic fraction $f_{\rm b}$. In our analysis, we chose $r_{\rm b} = 10 \times r_{200}$, but the choice of this radius does shift the posteriors of our model parameters. One can see the exact behaviour in Figure \ref{fig:rb_impact} where we show what posteriors look like for four choices of $r_{\rm b}$ with increasing size. The most notable effect of increasing $r_{\rm b}$ is that the outer power law slope $\beta$ of the baryon density profile also increases. We note that this shift in $\beta$ does begin to saturate at sufficiently large $r_{\rm b} \gtrsim 10 \times r_{200}$. In all cases, our posteriors of $\beta$ still suggest values less than to be expected by the cosmological hydrodynamic simulations with AGN feedback used in \cite{Battaglia_2016_GNFW} as mentioned in Section \ref{sec:results}. 

If $r_{\rm b}$ is chosen to be too small, then we see a bias in $A_{\rm 2h}$. This bias occurs because we don't enforce normalization of the baryon density profile amplitude $\rho_0$ sufficiently far out causing $\beta$ to compensate creating an outer power law slope that is comparable to or shallower than an NFW profile. If the baryon profile is shallower than the dark matter profile then one would expect the baryon density 1-halo term to be non-physically large at large scales forcing $A_{\rm 2h}$ be biased low to compensate for this. 
\section{Differences from Amodeo Halo Model}
\label{sec:app_amodeo}
In this section we unpack a few important differences from our model and pre-existing work from \cite{Amodeo_2021_kSZ} that directly precedes ours. 
\begin{itemize}
    \item We use a slightly different GNFW form for the density profile than Eq. 16 of \cite{Amodeo_2021_kSZ}. Our version follows the original form of \cite{Zhao_1996_GNFW}. As mentioned in \cite{Smith_2018_kSZ_Halo_Model}, the GNFW form used in \citep{Battaglia_2016_GNFW, Amodeo_2021_kSZ} must be corrected to be consistent with the universal NFW profile proposed in \cite{Zhao_1996_GNFW}. The difference between the two profiles amounts to $\Delta \beta = 2 |\gamma|$ assuming fixed $\alpha = 1$. More specifically, any $\beta$ which uses the GNFW profile of \citep{Battaglia_2016_GNFW, Amodeo_2021_kSZ} will be larger by $\Delta \beta = 0.4$ assuming a fixed $|\gamma| = 0.2$ where the sign is chosen so that all radial terms in the profile are in the denominator. While correcting for this does bring our joint fits closer to the simulated models for $\beta(M,z)$ from \cite{Battaglia_2016_GNFW}, it does not substantially change the results of this work. In all comparisons of $\beta$ to other works \citep[e.g.][]{Battaglia_2016_GNFW, Amodeo_2021_kSZ, Moser_2021_SZ_choices} we account for this necessary correction.
    \item Our framework removes $\rho_0$ as a free parameter from the model of the baryon density profile by forcing a normalization of the profile at $r_{\rm b}$. This normalization procedure was inspired by \cite{Cromer_2022_Baryons_WL_Mass}. Without this normalization the model prefers the amplitude of baryons to be an order of magnitude lower than required by cosmological abundance.  
    \item The \href{https://github.com/samodeo/Mop-c-GT}{\texttt{Mop-c-GT}} code used in \cite{Amodeo_2021_kSZ} has dependencies on two cosmology codes \href{https://hmf.readthedocs.io/en/latest/}{\texttt{HMF}} and \href{https://bdiemer.bitbucket.io/colossus/}{\texttt{colossus}} which have different assumptions of units. These different assumptions do not appear to be properly accounted for in the 2-halo term module of \href{https://github.com/samodeo/Mop-c-GT}{\texttt{Mop-c-GT}}. For this reason we found that the 2-halo density curve used in \cite{Amodeo_2021_kSZ} was a factor of $1/h^2$ larger than it should have been. While $A_{\rm k2h}$ was a nuisance parameter and was marginalised over, this finding implies that their posteriors for $A_{\rm k2h}$ should be $1/h^2$ larger. Our implementation of this model in \href{https://github.com/James11222/glasz}{\texttt{glasz}} uses \href{https://ccl.readthedocs.io/en/latest/}{\texttt{pyccl}} for all halo model computations rather than combining multiple cosmological codes.
\end{itemize}

\section{Degeneracy in Baryon Profile Parameters}
\label{sec:degeneracy}

The 2 baryon density profile parameters $x_{\rm c}$ and $\beta$ are highly degenerate with each other. While we have defined the core scale fraction $x_{\rm c}$ to span $[0, 1]$ based off of the definition proposed previously in precursory work \citep{Battaglia_2016_GNFW, Amodeo_2021_kSZ}, this is simply a choice in the model. In \cite{Smith_2018_kSZ_Halo_Model} the core scale fraction is fixed to $x_{\rm c} = 2$ rather than 0.5. A higher value of $x_{\rm c}$ implies that the turnover in the density profile occurs further out into the halo which requires a higher (steeper) value $\beta$ for the baryon density to remain equivalent in the outskirts of the halo.

To better understand this degeneracy between $x_{\rm c}$ and $\beta$, we fit the data with expanded priors on the $x_{\rm c}$ parameter. We find that only one of the two parameters can be constrained at a time. More specifically, when the MCMC chains reach a prior wall for one of the two parameters, the other parameter will be constrained. If the upper bound on the prior for $x_{\rm c}$ is 5, then $\beta$ will remain unconstrained. If the upper bound on the prior for $\beta$ is expanded to 10 in this case, then $\beta$ will be constrained while $x_{\rm c}$ remains unconstrained (the posteriors look similar to Figure \ref{fig:fit} only shifted to higher values). Fixing $x_{\rm c} = 2.0$ necessarily shifts the marginalized constraint on the outer power law slope to $\beta = 3.60 \pm 0.19$ while leaving $A_{\rm 2h}$ and $M_{\rm halo}$ unchanged. Ultimately, in order to make a fair comparison to the simulations of \cite{Battaglia_2016_GNFW}, we choose to keep the fiducial prior $\mathcal{U}(0.0, 1.0)$ for $x_{\rm c}$ which is consistent with the model definition.

%% file: biblio.bib
@ARTICLE{Debackere_2020,
       author = {{Debackere}, Stijn N.~B. and {Schaye}, Joop and {Hoekstra}, Henk},
        title = "{The impact of the observed baryon distribution in haloes on the total matter power spectrum}",
      journal = {\mnras},
         year = 2020,
        month = feb,
       volume = {492},
       number = {2},
        pages = {2285-2307},
          doi = {10.1093/mnras/stz3446},
archivePrefix = {arXiv},
       eprint = {1908.05765},
 primaryClass = {astro-ph.CO},
       adsurl = {https://ui.adsabs.harvard.edu/abs/2020MNRAS.492.2285D}
}

@ARTICLE{HSC_3x2,
       author = {{Sugiyama}, Sunao and {Miyatake}, Hironao and {More}, Surhud and {Li}, Xiangchong and {Shirasaki}, Masato and {Takada}, Masahiro and {Kobayashi}, Yosuke and {Takahashi}, Ryuichi and {Nishimichi}, Takahiro and {Nishizawa}, Atsushi J. and {Rau}, Markus M. and {Zhang}, Tianqing and {Dalal}, Roohi and {Mandelbaum}, Rachel and {Strauss}, Michael A. and {Hamana}, Takashi and {Oguri}, Masamune and {Osato}, Ken and {Kannawadi}, Arun and {Hsieh}, Bau-Ching and {Luo}, Wentao and {Armstrong}, Robert and {Bosch}, James and {Komiyama}, Yutaka and {Lupton}, Robert H. and {Lust}, Nate B. and {Miyazaki}, Satoshi and {Murayama}, Hitoshi and {Okura}, Yuki and {Price}, Paul A. and {Tait}, Philip J. and {Tanaka}, Masayuki and {Wang}, Shiang-Yu},
        title = "{Hyper Suprime-Cam Year 3 results: Cosmology from galaxy clustering and weak lensing with HSC and SDSS using the minimal bias model}",
      journal = {\prd},
         year = 2023,
        month = dec,
       volume = {108},
       number = {12},
          eid = {123521},
        pages = {123521},
          doi = {10.1103/PhysRevD.108.123521},
archivePrefix = {arXiv},
       eprint = {2304.00705},
 primaryClass = {astro-ph.CO},
       adsurl = {https://ui.adsabs.harvard.edu/abs/2023PhRvD.108l3521S}
}

@ARTICLE{Asgari2023_Halo_Review,
       author = {{Asgari}, Marika and {Mead}, Alexander J. and {Heymans}, Catherine},
        title = "{The halo model for cosmology: a pedagogical review}",
      journal = {The Open Journal of Astrophysics},
         year = 2023,
        month = nov,
       volume = {6},
          eid = {39},
        pages = {39},
          doi = {10.21105/astro.2303.08752},
archivePrefix = {arXiv},
       eprint = {2303.08752},
 primaryClass = {astro-ph.CO},
       adsurl = {https://ui.adsabs.harvard.edu/abs/2023OJAp....6E..39A}
}

@ARTICLE{FM_2013_emcee,
       author = {{Foreman-Mackey}, Daniel and {Hogg}, David W. and {Lang}, Dustin and {Goodman}, Jonathan},
        title = "{emcee: The MCMC Hammer}",
      journal = {\pasp},
         year = 2013,
        month = mar,
       volume = {125},
       number = {925},
        pages = {306},
          doi = {10.1086/670067},
archivePrefix = {arXiv},
       eprint = {1202.3665},
 primaryClass = {astro-ph.IM},
       adsurl = {https://ui.adsabs.harvard.edu/abs/2013PASP..125..306F}
}

@ARTICLE{Gelman_1992_Stat,
       author = {{Gelman}, Andrew and {Rubin}, Donald B.},
        title = "{Inference from Iterative Simulation Using Multiple Sequences}",
      journal = {Statistical Science},
         year = 1992,
        month = jan,
       volume = {7},
        pages = {457-472},
          doi = {10.1214/ss/1177011136},
       adsurl = {https://ui.adsabs.harvard.edu/abs/1992StaSc...7..457G}
}

@ARTICLE{Hoshino_2015_miscenter,
       author = {{Hoshino}, Hanako and {Leauthaud}, Alexie and {Lackner}, Claire and {Hikage}, Chiaki and {Rozo}, Eduardo and {Rykoff}, Eli and {Mandelbaum}, Rachel and {More}, Surhud and {More}, Anupreeta and {Saito}, Shun and {Vulcani}, Benedetta},
        title = "{Luminous red galaxies in clusters: central occupation, spatial distributions and miscentring}",
      journal = {\mnras},
         year = 2015,
        month = sep,
       volume = {452},
       number = {1},
        pages = {998-1013},
          doi = {10.1093/mnras/stv1271},
archivePrefix = {arXiv},
       eprint = {1503.05200},
 primaryClass = {astro-ph.CO},
       adsurl = {https://ui.adsabs.harvard.edu/abs/2015MNRAS.452..998H}
}

@ARTICLE{Skibba_2011_miscenter,
       author = {{Skibba}, Ramin A. and {van den Bosch}, Frank C. and {Yang}, Xiaohu and {More}, Surhud and {Mo}, Houjun and {Fontanot}, Fabio},
        title = "{Are brightest halo galaxies central galaxies?}",
      journal = {\mnras},
         year = 2011,
        month = jan,
       volume = {410},
       number = {1},
        pages = {417-431},
          doi = {10.1111/j.1365-2966.2010.17452.x},
archivePrefix = {arXiv},
       eprint = {1001.4533},
 primaryClass = {astro-ph.CO},
       adsurl = {https://ui.adsabs.harvard.edu/abs/2011MNRAS.410..417S}
}

@ARTICLE{Amodeo_2021_kSZ,
       author = {{Amodeo}, Stefania and {Battaglia}, Nicholas and {Schaan}, Emmanuel and {Ferraro}, Simone and {Moser}, Emily and {Aiola}, Simone and {Austermann}, Jason E. and {Beall}, James A. and {Bean}, Rachel and {Becker}, Daniel T. and {Bond}, Richard J. and {Calabrese}, Erminia and {Calafut}, Victoria and {Choi}, Steve K. and {Denison}, Edward V. and {Devlin}, Mark and {Duff}, Shannon M. and {Duivenvoorden}, Adriaan J. and {Dunkley}, Jo and {D{\"u}nner}, Rolando and {Gallardo}, Patricio A. and {Hall}, Kirsten R. and {Han}, Dongwon and {Hill}, J. Colin and {Hilton}, Gene C. and {Hilton}, Matt and {Hlo{\v{z}}ek}, Ren{\'e}e and {Hubmayr}, Johannes and {Huffenberger}, Kevin M. and {Hughes}, John P. and {Koopman}, Brian J. and {MacInnis}, Amanda and {McMahon}, Jeff and {Madhavacheril}, Mathew S. and {Moodley}, Kavilan and {Mroczkowski}, Tony and {Naess}, Sigurd and {Nati}, Federico and {Newburgh}, Laura B. and {Niemack}, Michael D. and {Page}, Lyman A. and {Partridge}, Bruce and {Schillaci}, Alessandro and {Sehgal}, Neelima and {Sif{\'o}n}, Crist{\'o}bal and {Spergel}, David N. and {Staggs}, Suzanne and {Storer}, Emilie R. and {Ullom}, Joel N. and {Vale}, Leila R. and {van Engelen}, Alexander and {Van Lanen}, Jeff and {Vavagiakis}, Eve M. and {Wollack}, Edward J. and {Xu}, Zhilei},
        title = "{Atacama Cosmology Telescope: Modeling the gas thermodynamics in BOSS CMASS galaxies from kinematic and thermal Sunyaev-Zel'dovich measurements}",
      journal = {\prd},
         year = 2021,
        month = mar,
       volume = {103},
       number = {6},
          eid = {063514},
        pages = {063514},
          doi = {10.1103/PhysRevD.103.063514},
archivePrefix = {arXiv},
       eprint = {2009.05558},
 primaryClass = {astro-ph.CO},
       adsurl = {https://ui.adsabs.harvard.edu/abs/2021PhRvD.103f3514A}
}

@ARTICLE{Schaan_2021_kSZ,
       author = {{Schaan}, Emmanuel and {Ferraro}, Simone and {Amodeo}, Stefania and {Battaglia}, Nicholas and {Aiola}, Simone and {Austermann}, Jason E. and {Beall}, James A. and {Bean}, Rachel and {Becker}, Daniel T. and {Bond}, Richard J. and {Calabrese}, Erminia and {Calafut}, Victoria and {Choi}, Steve K. and {Denison}, Edward V. and {Devlin}, Mark J. and {Duff}, Shannon M. and {Duivenvoorden}, Adriaan J. and {Dunkley}, Jo and {D{\"u}nner}, Rolando and {Gallardo}, Patricio A. and {Guan}, Yilun and {Han}, Dongwon and {Hill}, J. Colin and {Hilton}, Gene C. and {Hilton}, Matt and {Hlo{\v{z}}ek}, Ren{\'e}e and {Hubmayr}, Johannes and {Huffenberger}, Kevin M. and {Hughes}, John P. and {Koopman}, Brian J. and {MacInnis}, Amanda and {McMahon}, Jeff and {Madhavacheril}, Mathew S. and {Moodley}, Kavilan and {Mroczkowski}, Tony and {Naess}, Sigurd and {Nati}, Federico and {Newburgh}, Laura B. and {Niemack}, Michael D. and {Page}, Lyman A. and {Partridge}, Bruce and {Salatino}, Maria and {Sehgal}, Neelima and {Schillaci}, Alessandro and {Sif{\'o}n}, Crist{\'o}bal and {Smith}, Kendrick M. and {Spergel}, David N. and {Staggs}, Suzanne and {Storer}, Emilie R. and {Trac}, Hy and {Ullom}, Joel N. and {Van Lanen}, Jeff and {Vale}, Leila R. and {van Engelen}, Alexander and {Maga{\~n}a}, Mariana Vargas and {Vavagiakis}, Eve M. and {Wollack}, Edward J. and {Xu}, Zhilei and {Atacama Cosmology Telescope Collaboration}},
        title = "{Atacama Cosmology Telescope: Combined kinematic and thermal Sunyaev-Zel'dovich measurements from BOSS CMASS and LOWZ halos}",
      journal = {\prd},
         year = 2021,
        month = mar,
       volume = {103},
       number = {6},
          eid = {063513},
        pages = {063513},
          doi = {10.1103/PhysRevD.103.063513},
archivePrefix = {arXiv},
       eprint = {2009.05557},
 primaryClass = {astro-ph.CO},
       adsurl = {https://ui.adsabs.harvard.edu/abs/2021PhRvD.103f3513S}
}

@ARTICLE{Smith_2018_kSZ_Halo_Model,
       author = {{Smith}, Kendrick M. and {Madhavacheril}, Mathew S. and {M{\"u}nchmeyer}, Moritz and {Ferraro}, Simone and {Giri}, Utkarsh and {Johnson}, Matthew C.},
        title = "{KSZ tomography and the bispectrum}",
      journal = {arXiv e-prints},
         year = 2018,
        month = oct,
          eid = {arXiv:1810.13423},
        pages = {arXiv:1810.13423},
          doi = {10.48550/arXiv.1810.13423},
archivePrefix = {arXiv},
       eprint = {1810.13423},
 primaryClass = {astro-ph.CO},
       adsurl = {https://ui.adsabs.harvard.edu/abs/2018arXiv181013423S}
}

@ARTICLE{Amon_&_Robertson_2023_GGL,
       author = {{Amon}, A. and {Robertson}, N.~C. and {Miyatake}, H. and {Heymans}, C. and {White}, M. and {DeRose}, J. and {Yuan}, S. and {Wechsler}, R.~H. and {Varga}, T.~N. and {Bocquet}, S. and {Dvornik}, A. and {More}, S. and {Ross}, A.~J. and {Hoekstra}, H. and {Alarcon}, A. and {Asgari}, M. and {Blazek}, J. and {Campos}, A. and {Chen}, R. and {Choi}, A. and {Crocce}, M. and {Diehl}, H.~T. and {Doux}, C. and {Eckert}, K. and {Elvin-Poole}, J. and {Everett}, S. and {Fert{\'e}}, A. and {Gatti}, M. and {Giannini}, G. and {Gruen}, D. and {Gruendl}, R.~A. and {Hartley}, W.~G. and {Herner}, K. and {Hildebrandt}, H. and {Huang}, S. and {Huff}, E.~M. and {Joachimi}, B. and {Lee}, S. and {MacCrann}, N. and {Myles}, J. and {Navarro-Alsina}, A. and {Nishimichi}, T. and {Prat}, J. and {Secco}, L.~F. and {Sevilla-Noarbe}, I. and {Sheldon}, E. and {Shin}, T. and {Tr{\"o}ster}, T. and {Troxel}, M.~A. and {Tutusaus}, I. and {Wright}, A.~H. and {Yin}, B. and {Aguena}, M. and {Allam}, S. and {Annis}, J. and {Bacon}, D. and {Bilicki}, M. and {Brooks}, D. and {Burke}, D.~L. and {Carnero Rosell}, A. and {Carretero}, J. and {Castander}, F.~J. and {Cawthon}, R. and {Costanzi}, M. and {da Costa}, L.~N. and {Pereira}, M.~E.~S. and {de Jong}, J. and {De Vicente}, J. and {Desai}, S. and {Dietrich}, J.~P. and {Doel}, P. and {Ferrero}, I. and {Frieman}, J. and {Garc{\'\i}a-Bellido}, J. and {Gerdes}, D.~W. and {Gschwend}, J. and {Gutierrez}, G. and {Hinton}, S.~R. and {Hollowood}, D.~L. and {Honscheid}, K. and {Huterer}, D. and {Kannawadi}, A. and {Kuehn}, K. and {Kuropatkin}, N. and {Lahav}, O. and {Lima}, M. and {Maia}, M.~A.~G. and {Marshall}, J.~L. and {Menanteau}, F. and {Miquel}, R. and {Mohr}, J.~J. and {Morgan}, R. and {Muir}, J. and {Paz-Chinch{\'o}n}, F. and {Pieres}, A. and {Plazas Malag{\'o}n}, A.~A. and {Porredon}, A. and {Rodriguez-Monroy}, M. and {Roodman}, A. and {Sanchez}, E. and {Serrano}, S. and {Shan}, H. and {Suchyta}, E. and {Swanson}, M.~E.~C. and {Tarle}, G. and {Thomas}, D. and {To}, C. and {Zhang}, Y.},
        title = "{Consistent lensing and clustering in a low-S$_{8}$ Universe with BOSS, DES Year 3, HSC Year 1, and KiDS-1000}",
      journal = {\mnras},
         year = 2023,
        month = jan,
       volume = {518},
       number = {1},
        pages = {477-503},
          doi = {10.1093/mnras/stac2938},
archivePrefix = {arXiv},
       eprint = {2202.07440},
 primaryClass = {astro-ph.CO},
       adsurl = {https://ui.adsabs.harvard.edu/abs/2023MNRAS.518..477A}
}

@ARTICLE{Duffy_2008_Concentration,
       author = {{Duffy}, Alan R. and {Schaye}, Joop and {Kay}, Scott T. and {Dalla Vecchia}, Claudio},
        title = "{Dark matter halo concentrations in the Wilkinson Microwave Anisotropy Probe year 5 cosmology}",
      journal = {\mnras},
         year = 2008,
        month = oct,
       volume = {390},
       number = {1},
        pages = {L64-L68},
          doi = {10.1111/j.1745-3933.2008.00537.x},
archivePrefix = {arXiv},
       eprint = {0804.2486},
 primaryClass = {astro-ph},
       adsurl = {https://ui.adsabs.harvard.edu/abs/2008MNRAS.390L..64D}
}

@ARTICLE{Tinker_2008_HMF,
       author = {{Tinker}, Jeremy and {Kravtsov}, Andrey V. and {Klypin}, Anatoly and {Abazajian}, Kevork and {Warren}, Michael and {Yepes}, Gustavo and {Gottl{\"o}ber}, Stefan and {Holz}, Daniel E.},
        title = "{Toward a Halo Mass Function for Precision Cosmology: The Limits of Universality}",
      journal = {\apj},
         year = 2008,
        month = dec,
       volume = {688},
       number = {2},
        pages = {709-728},
          doi = {10.1086/591439},
archivePrefix = {arXiv},
       eprint = {0803.2706},
 primaryClass = {astro-ph},
       adsurl = {https://ui.adsabs.harvard.edu/abs/2008ApJ...688..709T}
}

@ARTICLE{Tinker_2010_bias,
       author = {{Tinker}, Jeremy L. and {Robertson}, Brant E. and {Kravtsov}, Andrey V. and {Klypin}, Anatoly and {Warren}, Michael S. and {Yepes}, Gustavo and {Gottl{\"o}ber}, Stefan},
        title = "{The Large-scale Bias of Dark Matter Halos: Numerical Calibration and Model Tests}",
      journal = {\apj},
         year = 2010,
        month = dec,
       volume = {724},
       number = {2},
        pages = {878-886},
          doi = {10.1088/0004-637X/724/2/878},
archivePrefix = {arXiv},
       eprint = {1001.3162},
 primaryClass = {astro-ph.CO},
       adsurl = {https://ui.adsabs.harvard.edu/abs/2010ApJ...724..878T}
}

@ARTICLE{Battaglia_2016_GNFW,
       author = {{Battaglia}, N.},
        title = "{The tau of galaxy clusters}",
      journal = {\jcap},
         year = 2016,
        month = aug,
       volume = {2016},
       number = {8},
          eid = {058},
        pages = {058},
          doi = {10.1088/1475-7516/2016/08/058},
archivePrefix = {arXiv},
       eprint = {1607.02442},
 primaryClass = {astro-ph.CO},
       adsurl = {https://ui.adsabs.harvard.edu/abs/2016JCAP...08..058B}
}

@ARTICLE{NFW_1997,
       author = {{Navarro}, Julio F. and {Frenk}, Carlos S. and {White}, Simon D.~M.},
        title = "{A Universal Density Profile from Hierarchical Clustering}",
      journal = {\apj},
         year = 1997,
        month = dec,
       volume = {490},
       number = {2},
        pages = {493-508},
          doi = {10.1086/304888},
archivePrefix = {arXiv},
       eprint = {astro-ph/9611107},
 primaryClass = {astro-ph},
       adsurl = {https://ui.adsabs.harvard.edu/abs/1997ApJ...490..493N}
}

@ARTICLE{Lungu_2022_ACT_Beam,
       author = {{Lungu}, Marius and {Storer}, Emilie R. and {Hasselfield}, Matthew and {Duivenvoorden}, Adriaan J. and {Calabrese}, Erminia and {Chesmore}, Grace E. and {Choi}, Steve K. and {Dunkley}, Jo and {D{\"u}nner}, Rolando and {Gallardo}, Patricio A. and {Golec}, Joseph E. and {Guan}, Yilun and {Hill}, J. Colin and {Hincks}, Adam D. and {Hubmayr}, Johannes and {Madhavacheril}, Mathew S. and {Mallaby-Kay}, Maya and {McMahon}, Jeff and {Moodley}, Kavilan and {Naess}, Sigurd and {Nati}, Federico and {Niemack}, Michael D. and {Page}, Lyman A. and {Partridge}, Bruce and {Puddu}, Roberto and {Schillaci}, Alessandro and {Sif{\'o}n}, Crist{\'o}bal and {Staggs}, Suzanne and {Sunder}, Dhaneshwar D. and {Wollack}, Edward J. and {Xu}, Zhilei},
        title = "{The Atacama Cosmology Telescope: measurement and analysis of 1D beams for DR4}",
      journal = {\jcap},
         year = 2022,
        month = may,
       volume = {2022},
       number = {5},
          eid = {044},
        pages = {044},
          doi = {10.1088/1475-7516/2022/05/044},
archivePrefix = {arXiv},
       eprint = {2112.12226},
 primaryClass = {astro-ph.IM},
       adsurl = {https://ui.adsabs.harvard.edu/abs/2022JCAP...05..044L}
}

@ARTICLE{McCarthy_2024_kSZ_WL,
       author = {{McCarthy}, Ian G. and {Amon}, Alexandra and {Schaye}, Joop and {Schaan}, Emmanuel and {Angulo}, Raul E. and {Salcido}, Jaime and {Schaller}, Matthieu and {Bigwood}, Leah and {Elbers}, Willem and {Kugel}, Roi and {Helly}, John C. and {Forouhar Moreno}, Victor J. and {Frenk}, Carlos S. and {McGibbon}, Robert J. and {Ondaro-Mallea}, Lurdes and {van Daalen}, Marcel P.},
        title = "{FLAMINGO: combining kinetic SZ effect and galaxy-galaxy lensing measurements to gauge the impact of feedback on large-scale structure}",
      journal = {arXiv e-prints},
         year = 2024,
        month = oct,
          eid = {arXiv:2410.19905},
        pages = {arXiv:2410.19905},
          doi = {10.48550/arXiv.2410.19905},
archivePrefix = {arXiv},
       eprint = {2410.19905},
 primaryClass = {astro-ph.CO},
       adsurl = {https://ui.adsabs.harvard.edu/abs/2024arXiv241019905M}
}

@ARTICLE{Bigwood_2024_kSZ_WL,
       author = {{Bigwood}, L. and {Amon}, A. and {Schneider}, A. and {Salcido}, J. and {McCarthy}, I.~G. and {Preston}, C. and {Sanchez}, D. and {Sijacki}, D. and {Schaan}, E. and {Ferraro}, S. and {Battaglia}, N. and {Chen}, A. and {Dodelson}, S. and {Roodman}, A. and {Pieres}, A. and {Fert{\'e}}, A. and {Alarcon}, A. and {Drlica-Wagner}, A. and {Choi}, A. and {Navarro-Alsina}, A. and {Campos}, A. and {Ross}, A.~J. and {Carnero Rosell}, A. and {Yin}, B. and {Yanny}, B. and {S{\'a}nchez}, C. and {Chang}, C. and {Davis}, C. and {Doux}, C. and {Gruen}, D. and {Rykoff}, E.~S. and {Huff}, E.~M. and {Sheldon}, E. and {Tarsitano}, F. and {Andrade-Oliveira}, F. and {Bernstein}, G.~M. and {Giannini}, G. and {Diehl}, H.~T. and {Huang}, H. and {Harrison}, I. and {Sevilla-Noarbe}, I. and {Tutusaus}, I. and {Elvin-Poole}, J. and {McCullough}, J. and {Zuntz}, J. and {Blazek}, J. and {DeRose}, J. and {Cordero}, J. and {Prat}, J. and {Myles}, J. and {Eckert}, K. and {Bechtol}, K. and {Herner}, K. and {Secco}, L.~F. and {Gatti}, M. and {Raveri}, M. and {Kind}, M. Carrasco and {Becker}, M.~R. and {Troxel}, M.~A. and {Jarvis}, M. and {MacCrann}, N. and {Friedrich}, O. and {Alves}, O. and {Leget}, P. -F. and {Chen}, R. and {Rollins}, R.~P. and {Wechsler}, R.~H. and {Gruendl}, R.~A. and {Cawthon}, R. and {Allam}, S. and {Bridle}, S.~L. and {Pandey}, S. and {Everett}, S. and {Shin}, T. and {Hartley}, W.~G. and {Fang}, X. and {Zhang}, Y. and {Aguena}, M. and {Annis}, J. and {Bacon}, D. and {Bertin}, E. and {Bocquet}, S. and {Brooks}, D. and {Carretero}, J. and {Castander}, F.~J. and {da Costa}, L.~N. and {Pereira}, M.~E.~S. and {De Vicente}, J. and {Desai}, S. and {Doel}, P. and {Ferrero}, I. and {Flaugher}, B. and {Frieman}, J. and {Garc{\'\i}a-Bellido}, J. and {Gaztanaga}, E. and {Gutierrez}, G. and {Hinton}, S.~R. and {Hollowood}, D.~L. and {Honscheid}, K. and {Huterer}, D. and {James}, D.~J. and {Kuehn}, K. and {Lahav}, O. and {Lee}, S. and {Marshall}, J.~L. and {Mena-Fern{\'a}ndez}, J. and {Miquel}, R. and {Muir}, J. and {Paterno}, M. and {Plazas Malag{\'o}n}, A.~A. and {Porredon}, A. and {Romer}, A.~K. and {Samuroff}, S. and {Sanchez}, E. and {Sanchez Cid}, D. and {Smith}, M. and {Soares-Santos}, M. and {Suchyta}, E. and {Swanson}, M.~E.~C. and {Tarle}, G. and {To}, C. and {Weaverdyck}, N. and {Weller}, J. and {Wiseman}, P. and {Yamamoto}, M.},
        title = "{Weak lensing combined with the kinetic Sunyaev-Zel'dovich effect: a study of baryonic feedback}",
      journal = {\mnras},
         year = 2024,
        month = oct,
       volume = {534},
       number = {1},
        pages = {655-682},
          doi = {10.1093/mnras/stae2100},
archivePrefix = {arXiv},
       eprint = {2404.06098},
 primaryClass = {astro-ph.CO},
       adsurl = {https://ui.adsabs.harvard.edu/abs/2024MNRAS.534..655B}
}

@ARTICLE{Chisari_2019_CCL,
       author = {{Chisari}, Nora Elisa and {Alonso}, David and {Krause}, Elisabeth and {Leonard}, C. Danielle and {Bull}, Philip and {Neveu}, J{\'e}r{\'e}my and {Villarreal}, Antonia Sierra and {Singh}, Sukhdeep and {McClintock}, Thomas and {Ellison}, John and {Du}, Zilong and {Zuntz}, Joe and {Mead}, Alexander and {Joudaki}, Shahab and {Lorenz}, Christiane S. and {Tr{\"o}ster}, Tilman and {Sanchez}, Javier and {Lanusse}, Francois and {Ishak}, Mustapha and {Hlozek}, Ren{\'e}e and {Blazek}, Jonathan and {Campagne}, Jean-Eric and {Almoubayyed}, Husni and {Eifler}, Tim and {Kirby}, Matthew and {Kirkby}, David and {Plaszczynski}, St{\'e}phane and {Slosar}, An{\v{z}}e and {Vrastil}, Michal and {Wagoner}, Erika L. and {LSST Dark Energy Science Collaboration}},
        title = "{Core Cosmology Library: Precision Cosmological Predictions for LSST}",
      journal = {\apjs},
         year = 2019,
        month = may,
       volume = {242},
       number = {1},
          eid = {2},
        pages = {2},
          doi = {10.3847/1538-4365/ab1658},
archivePrefix = {arXiv},
       eprint = {1812.05995},
 primaryClass = {astro-ph.CO},
       adsurl = {https://ui.adsabs.harvard.edu/abs/2019ApJS..242....2C}
}

@ARTICLE{Moser_2021_SZ_choices,
       author = {{Moser}, Emily and {Amodeo}, Stefania and {Battaglia}, Nicholas and {Alvarez}, Marcelo A. and {Ferraro}, Simone and {Schaan}, Emmanuel},
        title = "{The Impacts of Modeling Choices on the Inference of Circumgalactic Medium Properties from Sunyaev-Zeldovich Observations}",
      journal = {\apj},
         year = 2021,
        month = sep,
       volume = {919},
       number = {1},
          eid = {2},
        pages = {2},
          doi = {10.3847/1538-4357/ac0cea},
archivePrefix = {arXiv},
       eprint = {2103.02469},
 primaryClass = {astro-ph.GA},
       adsurl = {https://ui.adsabs.harvard.edu/abs/2021ApJ...919....2M}
}

@ARTICLE{Moser_2023_SZ_systematics,
       author = {{Moser}, Emily and {Battaglia}, Nicholas and {Amodeo}, Stefania},
        title = "{Searching for Systematics in Forward Modeling Sunyaev-Zeldovich Profiles}",
      journal = {arXiv e-prints},
         year = 2023,
        month = jul,
          eid = {arXiv:2307.10919},
        pages = {arXiv:2307.10919},
          doi = {10.48550/arXiv.2307.10919},
archivePrefix = {arXiv},
       eprint = {2307.10919},
 primaryClass = {astro-ph.CO},
       adsurl = {https://ui.adsabs.harvard.edu/abs/2023arXiv230710919M}
}

@ARTICLE{Lange_2023_2x2pt_GGL,
       author = {{Lange}, Johannes U. and {Hearin}, Andrew P. and {Leauthaud}, Alexie and {van den Bosch}, Frank C. and {Xhakaj}, Enia and {Guo}, Hong and {Wechsler}, Risa H. and {DeRose}, Joseph},
        title = "{Constraints on S$_{8}$ from a full-scale and full-shape analysis of redshift-space clustering and galaxy-galaxy lensing in BOSS}",
      journal = {\mnras},
         year = 2023,
        month = apr,
       volume = {520},
       number = {4},
        pages = {5373-5393},
          doi = {10.1093/mnras/stad473},
archivePrefix = {arXiv},
       eprint = {2301.08692},
 primaryClass = {astro-ph.CO},
       adsurl = {https://ui.adsabs.harvard.edu/abs/2023MNRAS.520.5373L}
}

@ARTICLE{Amon_2022_AMOD,
       author = {{Amon}, Alexandra and {Efstathiou}, George},
        title = "{A non-linear solution to the S$_{8}$ tension?}",
      journal = {\mnras},
         year = 2022,
        month = nov,
       volume = {516},
       number = {4},
        pages = {5355-5366},
          doi = {10.1093/mnras/stac2429},
archivePrefix = {arXiv},
       eprint = {2206.11794},
 primaryClass = {astro-ph.CO},
       adsurl = {https://ui.adsabs.harvard.edu/abs/2022MNRAS.516.5355A}
}

@ARTICLE{Preston_2023_AMOD_2,
       author = {{Preston}, Calvin and {Amon}, Alexandra and {Efstathiou}, George},
        title = "{A non-linear solution to the S$_{8}$ tension - II. Analysis of DES Year 3 cosmic shear}",
      journal = {\mnras},
         year = 2023,
        month = nov,
       volume = {525},
       number = {4},
        pages = {5554-5564},
          doi = {10.1093/mnras/stad2573},
archivePrefix = {arXiv},
       eprint = {2305.09827},
 primaryClass = {astro-ph.CO},
       adsurl = {https://ui.adsabs.harvard.edu/abs/2023MNRAS.525.5554P}
}

@ARTICLE{Heymans_2021_KiDS1000_S8,
       author = {{Heymans}, Catherine and {Tr{\"o}ster}, Tilman and {Asgari}, Marika and {Blake}, Chris and {Hildebrandt}, Hendrik and {Joachimi}, Benjamin and {Kuijken}, Konrad and {Lin}, Chieh-An and {S{\'a}nchez}, Ariel G. and {van den Busch}, Jan Luca and {Wright}, Angus H. and {Amon}, Alexandra and {Bilicki}, Maciej and {de Jong}, Jelte and {Crocce}, Martin and {Dvornik}, Andrej and {Erben}, Thomas and {Fortuna}, Maria Cristina and {Getman}, Fedor and {Giblin}, Benjamin and {Glazebrook}, Karl and {Hoekstra}, Henk and {Joudaki}, Shahab and {Kannawadi}, Arun and {K{\"o}hlinger}, Fabian and {Lidman}, Chris and {Miller}, Lance and {Napolitano}, Nicola R. and {Parkinson}, David and {Schneider}, Peter and {Shan}, HuanYuan and {Valentijn}, Edwin A. and {Verdoes Kleijn}, Gijs and {Wolf}, Christian},
        title = "{KiDS-1000 Cosmology: Multi-probe weak gravitational lensing and spectroscopic galaxy clustering constraints}",
      journal = {\aap},
         year = 2021,
        month = feb,
       volume = {646},
          eid = {A140},
        pages = {A140},
          doi = {10.1051/0004-6361/202039063},
archivePrefix = {arXiv},
       eprint = {2007.15632},
 primaryClass = {astro-ph.CO},
       adsurl = {https://ui.adsabs.harvard.edu/abs/2021A&A...646A.140H}
}

@ARTICLE{Asgari_2021_KiDS1000_S8,
       author = {{Asgari}, Marika and {Lin}, Chieh-An and {Joachimi}, Benjamin and {Giblin}, Benjamin and {Heymans}, Catherine and {Hildebrandt}, Hendrik and {Kannawadi}, Arun and {St{\"o}lzner}, Benjamin and {Tr{\"o}ster}, Tilman and {van den Busch}, Jan Luca and {Wright}, Angus H. and {Bilicki}, Maciej and {Blake}, Chris and {de Jong}, Jelte and {Dvornik}, Andrej and {Erben}, Thomas and {Getman}, Fedor and {Hoekstra}, Henk and {K{\"o}hlinger}, Fabian and {Kuijken}, Konrad and {Miller}, Lance and {Radovich}, Mario and {Schneider}, Peter and {Shan}, HuanYuan and {Valentijn}, Edwin},
        title = "{KiDS-1000 cosmology: Cosmic shear constraints and comparison between two point statistics}",
      journal = {\aap},
         year = 2021,
        month = jan,
       volume = {645},
          eid = {A104},
        pages = {A104},
          doi = {10.1051/0004-6361/202039070},
archivePrefix = {arXiv},
       eprint = {2007.15633},
 primaryClass = {astro-ph.CO},
       adsurl = {https://ui.adsabs.harvard.edu/abs/2021A&A...645A.104A}
}

@ARTICLE{Amon_2022_DES_Y3_S8,
       author = {{Amon}, A. and {Gruen}, D. and {Troxel}, M.~A. and {MacCrann}, N. and {Dodelson}, S. and {Choi}, A. and {Doux}, C. and {Secco}, L.~F. and {Samuroff}, S. and {Krause}, E. and {Cordero}, J. and {Myles}, J. and {DeRose}, J. and {Wechsler}, R.~H. and {Gatti}, M. and {Navarro-Alsina}, A. and {Bernstein}, G.~M. and {Jain}, B. and {Blazek}, J. and {Alarcon}, A. and {Fert{\'e}}, A. and {Lemos}, P. and {Raveri}, M. and {Campos}, A. and {Prat}, J. and {S{\'a}nchez}, C. and {Jarvis}, M. and {Alves}, O. and {Andrade-Oliveira}, F. and {Baxter}, E. and {Bechtol}, K. and {Becker}, M.~R. and {Bridle}, S.~L. and {Camacho}, H. and {Carnero Rosell}, A. and {Carrasco Kind}, M. and {Cawthon}, R. and {Chang}, C. and {Chen}, R. and {Chintalapati}, P. and {Crocce}, M. and {Davis}, C. and {Diehl}, H.~T. and {Drlica-Wagner}, A. and {Eckert}, K. and {Eifler}, T.~F. and {Elvin-Poole}, J. and {Everett}, S. and {Fang}, X. and {Fosalba}, P. and {Friedrich}, O. and {Gaztanaga}, E. and {Giannini}, G. and {Gruendl}, R.~A. and {Harrison}, I. and {Hartley}, W.~G. and {Herner}, K. and {Huang}, H. and {Huff}, E.~M. and {Huterer}, D. and {Kuropatkin}, N. and {Leget}, P. and {Liddle}, A.~R. and {McCullough}, J. and {Muir}, J. and {Pandey}, S. and {Park}, Y. and {Porredon}, A. and {Refregier}, A. and {Rollins}, R.~P. and {Roodman}, A. and {Rosenfeld}, R. and {Ross}, A.~J. and {Rykoff}, E.~S. and {Sanchez}, J. and {Sevilla-Noarbe}, I. and {Sheldon}, E. and {Shin}, T. and {Troja}, A. and {Tutusaus}, I. and {Tutusaus}, I. and {Varga}, T.~N. and {Weaverdyck}, N. and {Yanny}, B. and {Yin}, B. and {Zhang}, Y. and {Zuntz}, J. and {Aguena}, M. and {Allam}, S. and {Annis}, J. and {Bacon}, D. and {Bertin}, E. and {Bhargava}, S. and {Brooks}, D. and {Buckley-Geer}, E. and {Burke}, D.~L. and {Carretero}, J. and {Costanzi}, M. and {da Costa}, L.~N. and {Pereira}, M.~E.~S. and {De Vicente}, J. and {Desai}, S. and {Dietrich}, J.~P. and {Doel}, P. and {Ferrero}, I. and {Flaugher}, B. and {Frieman}, J. and {Garc{\'\i}a-Bellido}, J. and {Gaztanaga}, E. and {Gerdes}, D.~W. and {Giannantonio}, T. and {Gschwend}, J. and {Gutierrez}, G. and {Hinton}, S.~R. and {Hollowood}, D.~L. and {Honscheid}, K. and {Hoyle}, B. and {James}, D.~J. and {Kron}, R. and {Kuehn}, K. and {Lahav}, O. and {Lima}, M. and {Lin}, H. and {Maia}, M.~A.~G. and {Marshall}, J.~L. and {Martini}, P. and {Melchior}, P. and {Menanteau}, F. and {Miquel}, R. and {Mohr}, J.~J. and {Morgan}, R. and {Ogando}, R.~L.~C. and {Palmese}, A. and {Paz-Chinch{\'o}n}, F. and {Petravick}, D. and {Pieres}, A. and {Romer}, A.~K. and {Sanchez}, E. and {Scarpine}, V. and {Schubnell}, M. and {Serrano}, S. and {Smith}, M. and {Soares-Santos}, M. and {Tarle}, G. and {Thomas}, D. and {To}, C. and {Weller}, J. and {DES Collaboration}},
        title = "{Dark Energy Survey Year 3 results: Cosmology from cosmic shear and robustness to data calibration}",
      journal = {\prd},
         year = 2022,
        month = jan,
       volume = {105},
       number = {2},
          eid = {023514},
        pages = {023514},
          doi = {10.1103/PhysRevD.105.023514},
archivePrefix = {arXiv},
       eprint = {2105.13543},
 primaryClass = {astro-ph.CO},
       adsurl = {https://ui.adsabs.harvard.edu/abs/2022PhRvD.105b3514A}
}

@ARTICLE{Planck_2020a_S8,
       author = {{Planck Collaboration} and {Aghanim}, N. and {Akrami}, Y. and {Ashdown}, M. and {Aumont}, J. and {Baccigalupi}, C. and {Ballardini}, M. and {Banday}, A.~J. and {Barreiro}, R.~B. and {Bartolo}, N. and {Basak}, S. and {Battye}, R. and {Benabed}, K. and {Bernard}, J. -P. and {Bersanelli}, M. and {Bielewicz}, P. and {Bock}, J.~J. and {Bond}, J.~R. and {Borrill}, J. and {Bouchet}, F.~R. and {Boulanger}, F. and {Bucher}, M. and {Burigana}, C. and {Butler}, R.~C. and {Calabrese}, E. and {Cardoso}, J. -F. and {Carron}, J. and {Challinor}, A. and {Chiang}, H.~C. and {Chluba}, J. and {Colombo}, L.~P.~L. and {Combet}, C. and {Contreras}, D. and {Crill}, B.~P. and {Cuttaia}, F. and {de Bernardis}, P. and {de Zotti}, G. and {Delabrouille}, J. and {Delouis}, J. -M. and {Di Valentino}, E. and {Diego}, J.~M. and {Dor{\'e}}, O. and {Douspis}, M. and {Ducout}, A. and {Dupac}, X. and {Dusini}, S. and {Efstathiou}, G. and {Elsner}, F. and {En{\ss}lin}, T.~A. and {Eriksen}, H.~K. and {Fantaye}, Y. and {Farhang}, M. and {Fergusson}, J. and {Fernandez-Cobos}, R. and {Finelli}, F. and {Forastieri}, F. and {Frailis}, M. and {Fraisse}, A.~A. and {Franceschi}, E. and {Frolov}, A. and {Galeotta}, S. and {Galli}, S. and {Ganga}, K. and {G{\'e}nova-Santos}, R.~T. and {Gerbino}, M. and {Ghosh}, T. and {Gonz{\'a}lez-Nuevo}, J. and {G{\'o}rski}, K.~M. and {Gratton}, S. and {Gruppuso}, A. and {Gudmundsson}, J.~E. and {Hamann}, J. and {Handley}, W. and {Hansen}, F.~K. and {Herranz}, D. and {Hildebrandt}, S.~R. and {Hivon}, E. and {Huang}, Z. and {Jaffe}, A.~H. and {Jones}, W.~C. and {Karakci}, A. and {Keih{\"a}nen}, E. and {Keskitalo}, R. and {Kiiveri}, K. and {Kim}, J. and {Kisner}, T.~S. and {Knox}, L. and {Krachmalnicoff}, N. and {Kunz}, M. and {Kurki-Suonio}, H. and {Lagache}, G. and {Lamarre}, J. -M. and {Lasenby}, A. and {Lattanzi}, M. and {Lawrence}, C.~R. and {Le Jeune}, M. and {Lemos}, P. and {Lesgourgues}, J. and {Levrier}, F. and {Lewis}, A. and {Liguori}, M. and {Lilje}, P.~B. and {Lilley}, M. and {Lindholm}, V. and {L{\'o}pez-Caniego}, M. and {Lubin}, P.~M. and {Ma}, Y. -Z. and {Mac{\'\i}as-P{\'e}rez}, J.~F. and {Maggio}, G. and {Maino}, D. and {Mandolesi}, N. and {Mangilli}, A. and {Marcos-Caballero}, A. and {Maris}, M. and {Martin}, P.~G. and {Martinelli}, M. and {Mart{\'\i}nez-Gonz{\'a}lez}, E. and {Matarrese}, S. and {Mauri}, N. and {McEwen}, J.~D. and {Meinhold}, P.~R. and {Melchiorri}, A. and {Mennella}, A. and {Migliaccio}, M. and {Millea}, M. and {Mitra}, S. and {Miville-Desch{\^e}nes}, M. -A. and {Molinari}, D. and {Montier}, L. and {Morgante}, G. and {Moss}, A. and {Natoli}, P. and {N{\o}rgaard-Nielsen}, H.~U. and {Pagano}, L. and {Paoletti}, D. and {Partridge}, B. and {Patanchon}, G. and {Peiris}, H.~V. and {Perrotta}, F. and {Pettorino}, V. and {Piacentini}, F. and {Polastri}, L. and {Polenta}, G. and {Puget}, J. -L. and {Rachen}, J.~P. and {Reinecke}, M. and {Remazeilles}, M. and {Renzi}, A. and {Rocha}, G. and {Rosset}, C. and {Roudier}, G. and {Rubi{\~n}o-Mart{\'\i}n}, J.~A. and {Ruiz-Granados}, B. and {Salvati}, L. and {Sandri}, M. and {Savelainen}, M. and {Scott}, D. and {Shellard}, E.~P.~S. and {Sirignano}, C. and {Sirri}, G. and {Spencer}, L.~D. and {Sunyaev}, R. and {Suur-Uski}, A. -S. and {Tauber}, J.~A. and {Tavagnacco}, D. and {Tenti}, M. and {Toffolatti}, L. and {Tomasi}, M. and {Trombetti}, T. and {Valenziano}, L. and {Valiviita}, J. and {Van Tent}, B. and {Vibert}, L. and {Vielva}, P. and {Villa}, F. and {Vittorio}, N. and {Wandelt}, B.~D. and {Wehus}, I.~K. and {White}, M. and {White}, S.~D.~M. and {Zacchei}, A. and {Zonca}, A.},
        title = "{Planck 2018 results. VI. Cosmological parameters}",
      journal = {\aap},
         year = 2020,
        month = sep,
       volume = {641},
          eid = {A6},
        pages = {A6},
          doi = {10.1051/0004-6361/201833910},
archivePrefix = {arXiv},
       eprint = {1807.06209},
 primaryClass = {astro-ph.CO},
       adsurl = {https://ui.adsabs.harvard.edu/abs/2020A&A...641A...6P}
}

@ARTICLE{Prat_2022_GGL_modelling,
       author = {{Prat}, J. and {Blazek}, J. and {S{\'a}nchez}, C. and {Tutusaus}, I. and {Pandey}, S. and {Elvin-Poole}, J. and {Krause}, E. and {Troxel}, M.~A. and {Secco}, L.~F. and {Amon}, A. and {DeRose}, J. and {Zacharegkas}, G. and {Chang}, C. and {Jain}, B. and {MacCrann}, N. and {Park}, Y. and {Sheldon}, E. and {Giannini}, G. and {Bocquet}, S. and {To}, C. and {Alarcon}, A. and {Alves}, O. and {Andrade-Oliveira}, F. and {Baxter}, E. and {Bechtol}, K. and {Becker}, M.~R. and {Bernstein}, G.~M. and {Camacho}, H. and {Campos}, A. and {Carnero Rosell}, A. and {Carrasco Kind}, M. and {Cawthon}, R. and {Chen}, R. and {Choi}, A. and {Cordero}, J. and {Crocce}, M. and {Davis}, C. and {De Vicente}, J. and {Diehl}, H.~T. and {Dodelson}, S. and {Doux}, C. and {Drlica-Wagner}, A. and {Eckert}, K. and {Eifler}, T.~F. and {Elsner}, F. and {Everett}, S. and {Fang}, X. and {Farahi}, A. and {Fert{\'e}}, A. and {Fosalba}, P. and {Friedrich}, O. and {Gatti}, M. and {Gruen}, D. and {Gruendl}, R.~A. and {Harrison}, I. and {Hartley}, W.~G. and {Herner}, K. and {Huang}, H. and {Huff}, E.~M. and {Huterer}, D. and {Jarvis}, M. and {Kuropatkin}, N. and {Leget}, P. -F. and {Lemos}, P. and {Liddle}, A.~R. and {McCullough}, J. and {Muir}, J. and {Myles}, J. and {Navarro-Alsina}, A. and {Porredon}, A. and {Raveri}, M. and {Rodriguez-Monroy}, M. and {Rollins}, R.~P. and {Roodman}, A. and {Rosenfeld}, R. and {Ross}, A.~J. and {Rykoff}, E.~S. and {Sanchez}, J. and {Sevilla-Noarbe}, I. and {Shin}, T. and {Troja}, A. and {Varga}, T.~N. and {Weaverdyck}, N. and {Wechsler}, R.~H. and {Yanny}, B. and {Yin}, B. and {Zuntz}, J. and {Abbott}, T.~M.~C. and {Aguena}, M. and {Allam}, S. and {Annis}, J. and {Bacon}, D. and {Brooks}, D. and {Burke}, D.~L. and {Carretero}, J. and {Conselice}, C. and {Costanzi}, M. and {da Costa}, L.~N. and {Pereira}, M.~E.~S. and {Desai}, S. and {Dietrich}, J.~P. and {Doel}, P. and {Evrard}, A.~E. and {Ferrero}, I. and {Flaugher}, B. and {Frieman}, J. and {Garc{\'\i}a-Bellido}, J. and {Gaztanaga}, E. and {Gerdes}, D.~W. and {Giannantonio}, T. and {Gschwend}, J. and {Gutierrez}, G. and {Hinton}, S.~R. and {Hollowood}, D.~L. and {Honscheid}, K. and {James}, D.~J. and {Kuehn}, K. and {Lahav}, O. and {Lin}, H. and {Maia}, M.~A.~G. and {Marshall}, J.~L. and {Martini}, P. and {Melchior}, P. and {Menanteau}, F. and {Miller}, C.~J. and {Miquel}, R. and {Mohr}, J.~J. and {Morgan}, R. and {Ogando}, R.~L.~C. and {Palmese}, A. and {Paz-Chinch{\'o}n}, F. and {Petravick}, D. and {Plazas Malag{\'o}n}, A.~A. and {Sanchez}, E. and {Serrano}, S. and {Smith}, M. and {Soares-Santos}, M. and {Suchyta}, E. and {Tarle}, G. and {Thomas}, D. and {Weller}, J. and {DES Collaboration}},
        title = "{Dark energy survey year 3 results: High-precision measurement and modeling of galaxy-galaxy lensing}",
      journal = {\prd},
         year = 2022,
        month = apr,
       volume = {105},
       number = {8},
          eid = {083528},
        pages = {083528},
          doi = {10.1103/PhysRevD.105.083528},
archivePrefix = {arXiv},
       eprint = {2105.13541},
 primaryClass = {astro-ph.CO},
       adsurl = {https://ui.adsabs.harvard.edu/abs/2022PhRvD.105h3528P}
}

@ARTICLE{Miyatake_2015_BOSS_WL_SHMR,
       author = {{Miyatake}, Hironao and {More}, Surhud and {Mandelbaum}, Rachel and {Takada}, Masahiro and {Spergel}, David N. and {Kneib}, Jean-Paul and {Schneider}, Donald P. and {Brinkmann}, J. and {Brownstein}, Joel R.},
        title = "{The Weak Lensing Signal and the Clustering of BOSS Galaxies. I. Measurements}",
      journal = {\apj},
         year = 2015,
        month = jun,
       volume = {806},
       number = {1},
          eid = {1},
        pages = {1},
          doi = {10.1088/0004-637X/806/1/1},
archivePrefix = {arXiv},
       eprint = {1311.1480},
 primaryClass = {astro-ph.CO},
       adsurl = {https://ui.adsabs.harvard.edu/abs/2015ApJ...806....1M}
}

@ARTICLE{Kravtsov_2018_SHMR,
       author = {{Kravtsov}, A.~V. and {Vikhlinin}, A.~A. and {Meshcheryakov}, A.~V.},
        title = "{Stellar Mass{\textemdash}Halo Mass Relation and Star Formation Efficiency in High-Mass Halos}",
      journal = {Astronomy Letters},
         year = 2018,
        month = jan,
       volume = {44},
       number = {1},
        pages = {8-34},
          doi = {10.1134/S1063773717120015},
archivePrefix = {arXiv},
       eprint = {1401.7329},
 primaryClass = {astro-ph.CO},
       adsurl = {https://ui.adsabs.harvard.edu/abs/2018AstL...44....8K}
}

@ARTICLE{Maraston_2013_BOSS_stellar_mass,
       author = {{Maraston}, Claudia and {Pforr}, Janine and {Henriques}, Bruno M. and {Thomas}, Daniel and {Wake}, David and {Brownstein}, Joel R. and {Capozzi}, Diego and {Tinker}, Jeremy and {Bundy}, Kevin and {Skibba}, Ramin A. and {Beifiori}, Alessandra and {Nichol}, Robert C. and {Edmondson}, Edd and {Schneider}, Donald P. and {Chen}, Yanmei and {Masters}, Karen L. and {Steele}, Oliver and {Bolton}, Adam S. and {York}, Donald G. and {Weaver}, Benjamin A. and {Higgs}, Tim and {Bizyaev}, Dmitry and {Brewington}, Howard and {Malanushenko}, Elena and {Malanushenko}, Viktor and {Snedden}, Stephanie and {Oravetz}, Daniel and {Pan}, Kaike and {Shelden}, Alaina and {Simmons}, Audrey},
        title = "{Stellar masses of SDSS-III/BOSS galaxies at z {\ensuremath{\sim}} 0.5 and constraints to galaxy formation models}",
      journal = {\mnras},
         year = 2013,
        month = nov,
       volume = {435},
       number = {4},
        pages = {2764-2792},
          doi = {10.1093/mnras/stt1424},
archivePrefix = {arXiv},
       eprint = {1207.6114},
 primaryClass = {astro-ph.CO},
       adsurl = {https://ui.adsabs.harvard.edu/abs/2013MNRAS.435.2764M}
}

@ARTICLE{Maraston_2011_stellar_pop,
       author = {{Maraston}, C. and {Str{\"o}mb{\"a}ck}, G.},
        title = "{Stellar population models at high spectral resolution}",
      journal = {\mnras},
         year = 2011,
        month = dec,
       volume = {418},
       number = {4},
        pages = {2785-2811},
          doi = {10.1111/j.1365-2966.2011.19738.x},
archivePrefix = {arXiv},
       eprint = {1109.0543},
 primaryClass = {astro-ph.CO},
       adsurl = {https://ui.adsabs.harvard.edu/abs/2011MNRAS.418.2785M}
}

@ARTICLE{Kelly_2021_fb_discussion,
       author = {{Kelly}, Ashley J. and {Jenkins}, Adrian and {Frenk}, Carlos S.},
        title = "{The origin of X-ray coronae around simulated disc galaxies}",
      journal = {\mnras},
         year = 2021,
        month = apr,
       volume = {502},
       number = {2},
        pages = {2934-2951},
          doi = {10.1093/mnras/stab255},
archivePrefix = {arXiv},
       eprint = {2005.12926},
 primaryClass = {astro-ph.GA},
       adsurl = {https://ui.adsabs.harvard.edu/abs/2021MNRAS.502.2934K}
}

@ARTICLE{RG_2024_Prior_Volume,
       author = {{Ried Guachalla}, Bernardita and {Britt}, Dylan and {Gruen}, Daniel and {Friedrich}, Oliver},
        title = "{Informed total-error-minimizing priors: Interpretable cosmological parameter constraints despite complex nuisance effects}",
      journal = {\aap},
         year = 2025,
        month = jan,
       volume = {693},
          eid = {A178},
        pages = {A178},
          doi = {10.1051/0004-6361/202450575},
archivePrefix = {arXiv},
       eprint = {2405.00261},
 primaryClass = {astro-ph.CO},
       adsurl = {https://ui.adsabs.harvard.edu/abs/2025A&A...693A.178R}
}

@ARTICLE{Leauthaud_2017_lensing_is_low,
       author = {{Leauthaud}, Alexie and {Saito}, Shun and {Hilbert}, Stefan and {Barreira}, Alexandre and {More}, Surhud and {White}, Martin and {Alam}, Shadab and {Behroozi}, Peter and {Bundy}, Kevin and {Coupon}, Jean and {Erben}, Thomas and {Heymans}, Catherine and {Hildebrandt}, Hendrik and {Mandelbaum}, Rachel and {Miller}, Lance and {Moraes}, Bruno and {Pereira}, Maria E.~S. and {Rodr{\'\i}guez-Torres}, Sergio A. and {Schmidt}, Fabian and {Shan}, Huan-Yuan and {Viel}, Matteo and {Villaescusa-Navarro}, Francisco},
        title = "{Lensing is low: cosmology, galaxy formation or new physics?}",
      journal = {\mnras},
         year = 2017,
        month = may,
       volume = {467},
       number = {3},
        pages = {3024-3047},
          doi = {10.1093/mnras/stx258},
archivePrefix = {arXiv},
       eprint = {1611.08606},
 primaryClass = {astro-ph.CO},
       adsurl = {https://ui.adsabs.harvard.edu/abs/2017MNRAS.467.3024L}
}

@ARTICLE{Hadzhiyska_2024_ACT_DESI_kSZ,
       author = {{Hadzhiyska}, B. and {Ferraro}, S. and {Ried Guachalla}, B. and {Schaan}, E. and {Aguilar}, J. and {Battaglia}, N. and {Bond}, J.~R. and {Brooks}, D. and {Calabrese}, E. and {Choi}, S.~K. and {Claybaugh}, T. and {Coulton}, W.~R. and {Dawson}, K. and {Devlin}, M. and {Dey}, B. and {Doel}, P. and {Duivenvoorden}, A.~J. and {Dunkley}, J. and {Farren}, G.~S. and {Font-Ribera}, A. and {Forero-Romero}, J.~E. and {Gallardo}, P.~A. and {Gazta{\~n}aga}, E. and {Gontcho Gontcho}, S. and {Gralla}, M. and {Le Guillou}, L. and {Gutierrez}, G. and {Guy}, J. and {Hill}, J.~C. and {Hlo{\v{z}}ek}, R. and {Honscheid}, K. and {Juneau}, S. and {Kisner}, T. and {Kremin}, A. and {Landriau}, M. and {Liu}, R.~H. and {Louis}, T. and {MacCrann}, N. and {de Macorra}, A. and {Madhavacheril}, M. and {Manera}, M. and {Meisner}, A. and {Miquel}, R. and {Moodley}, K. and {Moustakas}, J. and {Mroczkowski}, T. and {Naess}, S. and {Newman}, J. and {Niemack}, M.~D. and {Niz}, G. and {Page}, L. and {Palanque-Delabrouille}, N. and {Partridge}, B. and {Percival}, W.~J. and {Prada}, F. and {Qu}, F.~J. and {Rossi}, G. and {Sanchez}, E. and {Schlegel}, D. and {Schubnell}, M. and {Sehgal}, N. and {Seo}, H. and {Sif{\'o}n}, C. and {Spergel}, D. and {Sprayberry}, D. and {Staggs}, S. and {Tarl{\'e}}, G. and {Vargas}, C. and {Vavagiakis}, E.~M. and {Weaver}, B.~A. and {Wollack}, E.~J. and {Zhou}, R. and {Zou}, H.},
        title = "{Evidence for large baryonic feedback at low and intermediate redshifts from kinematic Sunyaev-Zel'dovich observations with ACT and DESI photometric galaxies}",
      journal = {arXiv e-prints},
         year = 2024,
        month = jul,
          eid = {arXiv:2407.07152},
        pages = {arXiv:2407.07152},
          doi = {10.48550/arXiv.2407.07152},
archivePrefix = {arXiv},
       eprint = {2407.07152},
 primaryClass = {astro-ph.CO},
       adsurl = {https://ui.adsabs.harvard.edu/abs/2024arXiv240707152H}
}

@ARTICLE{Munchmeyer_2019_kSZ,
       author = {{M{\"u}nchmeyer}, Moritz and {Madhavacheril}, Mathew S. and {Ferraro}, Simone and {Johnson}, Matthew C. and {Smith}, Kendrick M.},
        title = "{Constraining local non-Gaussianities with kinetic Sunyaev-Zel'dovich tomography}",
      journal = {\prd},
         year = 2019,
        month = oct,
       volume = {100},
       number = {8},
          eid = {083508},
        pages = {083508},
          doi = {10.1103/PhysRevD.100.083508},
archivePrefix = {arXiv},
       eprint = {1810.13424},
 primaryClass = {astro-ph.CO},
       adsurl = {https://ui.adsabs.harvard.edu/abs/2019PhRvD.100h3508M}
}

@ARTICLE{Cayuso_2023_kSZ,
       author = {{Cayuso}, Juan and {Bloch}, Richard and {Hotinli}, Selim C. and {Johnson}, Matthew C. and {McCarthy}, Fiona},
        title = "{Velocity reconstruction with the cosmic microwave background and galaxy surveys}",
      journal = {\jcap},
         year = 2023,
        month = feb,
       volume = {2023},
       number = {2},
          eid = {051},
        pages = {051},
          doi = {10.1088/1475-7516/2023/02/051},
archivePrefix = {arXiv},
       eprint = {2111.11526},
 primaryClass = {astro-ph.CO},
       adsurl = {https://ui.adsabs.harvard.edu/abs/2023JCAP...02..051C}
}

@ARTICLE{Bolliet_2023_kSZ,
       author = {{Bolliet}, Boris and {Colin Hill}, J. and {Ferraro}, Simone and {Kusiak}, Aleksandra and {Krolewski}, Alex},
        title = "{Projected-field kinetic Sunyaev-Zel'dovich Cross-correlations: halo model and forecasts}",
      journal = {\jcap},
         year = 2023,
        month = mar,
       volume = {2023},
       number = {3},
          eid = {039},
        pages = {039},
          doi = {10.1088/1475-7516/2023/03/039},
archivePrefix = {arXiv},
       eprint = {2208.07847},
 primaryClass = {astro-ph.CO},
       adsurl = {https://ui.adsabs.harvard.edu/abs/2023JCAP...03..039B}
}

@ARTICLE{Roy_2023_kSZ,
       author = {{Roy}, Anirban and {van Engelen}, Alexander and {Gluscevic}, Vera and {Battaglia}, Nicholas},
        title = "{Probing the Circumgalactic Medium with Cosmic Microwave Background Polarization Statistical Anisotropy}",
      journal = {\apj},
         year = 2023,
        month = jul,
       volume = {951},
       number = {1},
          eid = {50},
        pages = {50},
          doi = {10.3847/1538-4357/acd194},
archivePrefix = {arXiv},
       eprint = {2201.05076},
 primaryClass = {astro-ph.CO},
       adsurl = {https://ui.adsabs.harvard.edu/abs/2023ApJ...951...50R}
}

@ARTICLE{Gatti_2021_DES_Y3,
       author = {{Gatti}, M. and {Sheldon}, E. and {Amon}, A. and {Becker}, M. and {Troxel}, M. and {Choi}, A. and {Doux}, C. and {MacCrann}, N. and {Navarro-Alsina}, A. and {Harrison}, I. and {Gruen}, D. and {Bernstein}, G. and {Jarvis}, M. and {Secco}, L.~F. and {Fert{\'e}}, A. and {Shin}, T. and {McCullough}, J. and {Rollins}, R.~P. and {Chen}, R. and {Chang}, C. and {Pandey}, S. and {Tutusaus}, I. and {Prat}, J. and {Elvin-Poole}, J. and {Sanchez}, C. and {Plazas}, A.~A. and {Roodman}, A. and {Zuntz}, J. and {Abbott}, T.~M.~C. and {Aguena}, M. and {Allam}, S. and {Annis}, J. and {Avila}, S. and {Bacon}, D. and {Bertin}, E. and {Bhargava}, S. and {Brooks}, D. and {Burke}, D.~L. and {Carnero Rosell}, A. and {Carrasco Kind}, M. and {Carretero}, J. and {Castander}, F.~J. and {Conselice}, C. and {Costanzi}, M. and {Crocce}, M. and {da Costa}, L.~N. and {Davis}, T.~M. and {De Vicente}, J. and {Desai}, S. and {Diehl}, H.~T. and {Dietrich}, J.~P. and {Doel}, P. and {Drlica-Wagner}, A. and {Eckert}, K. and {Everett}, S. and {Ferrero}, I. and {Frieman}, J. and {Garc{\'\i}a-Bellido}, J. and {Gerdes}, D.~W. and {Giannantonio}, T. and {Gruendl}, R.~A. and {Gschwend}, J. and {Gutierrez}, G. and {Hartley}, W.~G. and {Hinton}, S.~R. and {Hollowood}, D.~L. and {Honscheid}, K. and {Hoyle}, B. and {Huff}, E.~M. and {Huterer}, D. and {Jain}, B. and {James}, D.~J. and {Jeltema}, T. and {Krause}, E. and {Kron}, R. and {Kuropatkin}, N. and {Lima}, M. and {Maia}, M.~A.~G. and {Marshall}, J.~L. and {Miquel}, R. and {Morgan}, R. and {Myles}, J. and {Palmese}, A. and {Paz-Chinch{\'o}n}, F. and {Rykoff}, E.~S. and {Samuroff}, S. and {Sanchez}, E. and {Scarpine}, V. and {Schubnell}, M. and {Serrano}, S. and {Sevilla-Noarbe}, I. and {Smith}, M. and {Soares-Santos}, M. and {Suchyta}, E. and {Swanson}, M.~E.~C. and {Tarle}, G. and {Thomas}, D. and {To}, C. and {Tucker}, D.~L. and {Varga}, T.~N. and {Wechsler}, R.~H. and {Weller}, J. and {Wester}, W. and {Wilkinson}, R.~D.},
        title = "{Dark energy survey year 3 results: weak lensing shape catalogue}",
      journal = {\mnras},
         year = 2021,
        month = jul,
       volume = {504},
       number = {3},
        pages = {4312-4336},
          doi = {10.1093/mnras/stab918},
archivePrefix = {arXiv},
       eprint = {2011.03408},
 primaryClass = {astro-ph.CO},
       adsurl = {https://ui.adsabs.harvard.edu/abs/2021MNRAS.504.4312G}
}

@article{DES_Y3_2022,
	title = {Dark {Energy} {Survey} {Year} 3 results: {Cosmological} constraints from galaxy clustering and weak lensing},
	volume = {105},
	issn = {1550-79980556-2821},
	shorttitle = {Dark {Energy} {Survey} {Year} 3 results},
	url = {https://ui.adsabs.harvard.edu/abs/2022PhRvD.105b3520A},
	doi = {10.1103/PhysRevD.105.023520},
	urldate = {2024-10-27},
	journal = {Physical Review D},
	author = {Abbott, T. M. C. and Aguena, M. and Alarcon, A. and Allam, S. and Alves, O. and Amon, A. and Andrade-Oliveira, F. and Annis, J. and Avila, S. and Bacon, D. and Baxter, E. and Bechtol, K. and Becker, M. R. and Bernstein, G. M. and Bhargava, S. and Birrer, S. and Blazek, J. and Brandao-Souza, A. and Bridle, S. L. and Brooks, D. and Buckley-Geer, E. and Burke, D. L. and Camacho, H. and Campos, A. and Carnero Rosell, A. and Carrasco Kind, M. and Carretero, J. and Castander, F. J. and Cawthon, R. and Chang, C. and Chen, A. and Chen, R. and Choi, A. and Conselice, C. and Cordero, J. and Costanzi, M. and Crocce, M. and da Costa, L. N. and da Silva Pereira, M. E. and Davis, C. and Davis, T. M. and De Vicente, J. and DeRose, J. and Desai, S. and Di Valentino, E. and Diehl, H. T. and Dietrich, J. P. and Dodelson, S. and Doel, P. and Doux, C. and Drlica-Wagner, A. and Eckert, K. and Eifler, T. F. and Elsner, F. and Elvin-Poole, J. and Everett, S. and Evrard, A. E. and Fang, X. and Farahi, A. and Fernandez, E. and Ferrero, I. and Ferté, A. and Fosalba, P. and Friedrich, O. and Frieman, J. and García-Bellido, J. and Gatti, M. and Gaztanaga, E. and Gerdes, D. W. and Giannantonio, T. and Giannini, G. and Gruen, D. and Gruendl, R. A. and Gschwend, J. and Gutierrez, G. and Harrison, I. and Hartley, W. G. and Herner, K. and Hinton, S. R. and Hollowood, D. L. and Honscheid, K. and Hoyle, B. and Huff, E. M. and Huterer, D. and Jain, B. and James, D. J. and Jarvis, M. and Jeffrey, N. and Jeltema, T. and Kovacs, A. and Krause, E. and Kron, R. and Kuehn, K. and Kuropatkin, N. and Lahav, O. and Leget, P. -F. and Lemos, P. and Liddle, A. R. and Lidman, C. and Lima, M. and Lin, H. and MacCrann, N. and Maia, M. A. G. and Marshall, J. L. and Martini, P. and McCullough, J. and Melchior, P. and Mena-Fernández, J. and Menanteau, F. and Miquel, R. and Mohr, J. J. and Morgan, R. and Muir, J. and Myles, J. and Nadathur, S. and Navarro-Alsina, A. and Nichol, R. C. and Ogando, R. L. C. and Omori, Y. and Palmese, A. and Pandey, S. and Park, Y. and Paz-Chinchón, F. and Petravick, D. and Pieres, A. and Plazas Malagón, A. A. and Porredon, A. and Prat, J. and Raveri, M. and Rodriguez-Monroy, M. and Rollins, R. P. and Romer, A. K. and Roodman, A. and Rosenfeld, R. and Ross, A. J. and Rykoff, E. S. and Samuroff, S. and Sánchez, C. and Sanchez, E. and Sanchez, J. and Sanchez Cid, D. and Scarpine, V. and Schubnell, M. and Scolnic, D. and Secco, L. F. and Serrano, S. and Sevilla-Noarbe, I. and Sheldon, E. and Shin, T. and Smith, M. and Soares-Santos, M. and Suchyta, E. and Swanson, M. E. C. and Tabbutt, M. and Tarle, G. and Thomas, D. and To, C. and Troja, A. and Troxel, M. A. and Tucker, D. L. and Tutusaus, I. and Varga, T. N. and Walker, A. R. and Weaverdyck, N. and Wechsler, R. and Weller, J. and Yanny, B. and Yin, B. and Zhang, Y. and Zuntz, J. and {DES Collaboration}},
	month = jan,
	year = {2022},
	note = {Publisher: APS
ADS Bibcode: 2022PhRvD.105b3520A},
	pages = {023520},
}

@article{huang_modelling_2019,
	title = {Modelling baryonic physics in future weak lensing surveys},
	volume = {488},
	issn = {0035-8711},
	url = {https://doi.org/10.1093/mnras/stz1714},
	doi = {10.1093/mnras/stz1714},
	number = {2},
	journal = {Monthly Notices of the Royal Astronomical Society},
	author = {Huang, Hung-Jin and Eifler, Tim and Mandelbaum, Rachel and Dodelson, Scott},
	month = jun,
	year = {2019},
	note = {\_eprint: https://academic.oup.com/mnras/article-pdf/488/2/1652/28954328/stz1714.pdf},
	pages = {1652--1678},
}

@ARTICLE{Bigwood_2025_kSZ_GGL_benchmark,
       author = {{Bigwood}, Leah and {Yamamoto}, Masaya and {Siegel}, Jared and {Amon}, Alexandra and {McCarthy}, Ian G. and {Dave}, Romeel and {Salcido}, Jaime and {Schaller}, Matthieu and {Schaye}, Joop and {Yang}, Tianyi},
        title = "{The kinetic Sunyaev Zeldovich effect as a benchmark for AGN feedback models in hydrodynamical simulations: insights from DESI + ACT}",
      journal = {arXiv e-prints},
         year = 2025,
        month = oct,
          eid = {arXiv:2510.15822},
        pages = {arXiv:2510.15822},
          doi = {10.48550/arXiv.2510.15822},
archivePrefix = {arXiv},
       eprint = {2510.15822},
 primaryClass = {astro-ph.CO},
       adsurl = {https://ui.adsabs.harvard.edu/abs/2025arXiv251015822B}
}

@ARTICLE{Medlock_2025_FRB_Pk,
       author = {{Medlock}, Isabel and {Nagai}, Daisuke and {Angl{\'e}s-Alc{\'a}zar}, Daniel and {Gebhardt}, Matthew},
        title = "{Constraining Baryonic Feedback Effects on the Matter Power Spectrum with Fast Radio Bursts}",
      journal = {\apj},
         year = 2025,
        month = apr,
       volume = {983},
       number = {1},
          eid = {46},
        pages = {46},
          doi = {10.3847/1538-4357/adbc9c},
archivePrefix = {arXiv},
       eprint = {2501.17922},
 primaryClass = {astro-ph.CO},
       adsurl = {https://ui.adsabs.harvard.edu/abs/2025ApJ...983...46M}
}

@ARTICLE{Siegel_2025_kSZ_GGL,
       author = {{Siegel}, Jared and {Amon}, Alexandra and {McCarthy}, Ian G. and {Bigwood}, Leah and {Yamamoto}, Masaya and {Bulbul}, Esra and {Greene}, Jenny E. and {McCullough}, Jamie and {Schaller}, Matthieu and {Schaye}, Joop},
        title = "{Joint X-ray, kinetic Sunyaev-Zeldovich, and weak lensing measurements: toward a consensus picture of efficient gas expulsion from groups and clusters}",
      journal = {arXiv e-prints},
         year = 2025,
        month = sep,
          eid = {arXiv:2509.10455},
        pages = {arXiv:2509.10455},
          doi = {10.48550/arXiv.2509.10455},
archivePrefix = {arXiv},
       eprint = {2509.10455},
 primaryClass = {astro-ph.CO},
       adsurl = {https://ui.adsabs.harvard.edu/abs/2025arXiv250910455S}
}

@ARTICLE{Popik_tSZ_HOD_Model,
       author = {{Popik}, Chad and {Battaglia}, Nicholas and {Kusiak}, Aleksandra and {Bolliet}, Boris and {Hill}, J. Colin},
        title = "{On the Impacts of Halo Model Implementations in Sunyaev-Zeldovich Cross-Correlation Analyses}",
      journal = {arXiv e-prints},
         year = 2025,
        month = feb,
          eid = {arXiv:2502.13291},
        pages = {arXiv:2502.13291},
          doi = {10.48550/arXiv.2502.13291},
archivePrefix = {arXiv},
       eprint = {2502.13291},
 primaryClass = {astro-ph.CO},
       adsurl = {https://ui.adsabs.harvard.edu/abs/2025arXiv250213291P}
}

@ARTICLE{Val_Daalen_2011_BR,
       author = {{van Daalen}, Marcel P. and {Schaye}, Joop and {Booth}, C.~M. and {Dalla Vecchia}, Claudio},
        title = "{The effects of galaxy formation on the matter power spectrum: a challenge for precision cosmology}",
      journal = {\mnras},
         year = 2011,
        month = aug,
       volume = {415},
       number = {4},
        pages = {3649-3665},
          doi = {10.1111/j.1365-2966.2011.18981.x},
archivePrefix = {arXiv},
       eprint = {1104.1174},
 primaryClass = {astro-ph.CO},
       adsurl = {https://ui.adsabs.harvard.edu/abs/2011MNRAS.415.3649V}
}

@ARTICLE{Nishimichi_2019_DarkEmulator,
       author = {{Nishimichi}, Takahiro and {Takada}, Masahiro and {Takahashi}, Ryuichi and {Osato}, Ken and {Shirasaki}, Masato and {Oogi}, Taira and {Miyatake}, Hironao and {Oguri}, Masamune and {Murata}, Ryoma and {Kobayashi}, Yosuke and {Yoshida}, Naoki},
        title = "{Dark Quest. I. Fast and Accurate Emulation of Halo Clustering Statistics and Its Application to Galaxy Clustering}",
      journal = {\apj},
         year = 2019,
        month = oct,
       volume = {884},
       number = {1},
          eid = {29},
        pages = {29},
          doi = {10.3847/1538-4357/ab3719},
archivePrefix = {arXiv},
       eprint = {1811.09504},
 primaryClass = {astro-ph.CO},
       adsurl = {https://ui.adsabs.harvard.edu/abs/2019ApJ...884...29N}
}

@ARTICLE{Miyatake_2022_DarkEmulator,
       author = {{Miyatake}, Hironao and {Kobayashi}, Yosuke and {Takada}, Masahiro and {Nishimichi}, Takahiro and {Shirasaki}, Masato and {Sugiyama}, Sunao and {Takahashi}, Ryuichi and {Osato}, Ken and {More}, Surhud and {Park}, Youngsoo},
        title = "{Cosmological inference from an emulator based halo model. I. Validation tests with HSC and SDSS mock catalogs}",
      journal = {\prd},
         year = 2022,
        month = oct,
       volume = {106},
       number = {8},
          eid = {083519},
        pages = {083519},
          doi = {10.1103/PhysRevD.106.083519},
archivePrefix = {arXiv},
       eprint = {2101.00113},
 primaryClass = {astro-ph.CO},
       adsurl = {https://ui.adsabs.harvard.edu/abs/2022PhRvD.106h3519M}
}

@ARTICLE{Nelson_2018_Color_Bimodality,
       author = {{Nelson}, Dylan and {Pillepich}, Annalisa and {Springel}, Volker and {Weinberger}, Rainer and {Hernquist}, Lars and {Pakmor}, R{\"u}diger and {Genel}, Shy and {Torrey}, Paul and {Vogelsberger}, Mark and {Kauffmann}, Guinevere and {Marinacci}, Federico and {Naiman}, Jill},
        title = "{First results from the IllustrisTNG simulations: the galaxy colour bimodality}",
      journal = {\mnras},
         year = 2018,
        month = mar,
       volume = {475},
       number = {1},
        pages = {624-647},
          doi = {10.1093/mnras/stx3040},
archivePrefix = {arXiv},
       eprint = {1707.03395},
 primaryClass = {astro-ph.GA},
       adsurl = {https://ui.adsabs.harvard.edu/abs/2018MNRAS.475..624N}
}

@ARTICLE{Dvornik_2023_Halo_Model,
       author = {{Dvornik}, Andrej and {Heymans}, Catherine and {Asgari}, Marika and {Mahony}, Constance and {Joachimi}, Benjamin and {Bilicki}, Maciej and {Chisari}, Elisa and {Hildebrandt}, Hendrik and {Hoekstra}, Henk and {Johnston}, Harry and {Kuijken}, Konrad and {Mead}, Alexander and {Miyatake}, Hironao and {Nishimichi}, Takahiro and {Reischke}, Robert and {Unruh}, Sandra and {Wright}, Angus H.},
        title = "{KiDS-1000: Combined halo-model cosmology constraints from galaxy abundance, galaxy clustering, and galaxy-galaxy lensing}",
      journal = {\aap},
         year = 2023,
        month = jul,
       volume = {675},
          eid = {A189},
        pages = {A189},
          doi = {10.1051/0004-6361/202245158},
archivePrefix = {arXiv},
       eprint = {2210.03110},
 primaryClass = {astro-ph.CO},
       adsurl = {https://ui.adsabs.harvard.edu/abs/2023A&A...675A.189D}
}

@ARTICLE{RG_2025_kSZ_ACT_DESI_Spec,
       author = {{Ried Guachalla}, Bernardita and {Schaan}, Emmanuel and {Hadzhiyska}, Boryana and {Ferraro}, Simone and {Aguilar}, Jessica N. and {Ahlen}, Steven and {Battaglia}, Nicholas and {Bianchi}, Davide and {Bond}, Richard and {Brooks}, David and {Claybaugh}, Todd and {Coulton}, William R. and {de la Macorra}, Axel and {Devlin}, Mark J. and {Dey}, Arjun and {Doel}, Peter and {Dunkley}, Jo and {Fanning}, Kevin and {Forero-Romero}, Jaime and {Gazta\textbackslash\raisebox{-0.5ex}\textasciitildenaga}, Enrique and {Gontcho}, Satya Gontcho A and {Gutierrez}, Gaston and {Guy}, Julien and {Hill}, J. Colin and {Honscheid}, Klaus and {Juneau}, Stephanie and {Kisner}, Theodore and {Kremin}, Anthony and {Lambert}, Andrew and {Landriau}, Martin and {Le Guillou}, Laurent and {MacCrann}, Niall and {Manera}, Marc and {Meisner}, Aaron and {Miquel}, Ramon and {Moodley}, Kavilan and {Moustakas}, John and {Mroczkowski}, Tony and {Myers}, Adam D. and {Niemack}, Michael D. and {Niz}, Gustavo and {Palanque-Delabrouille}, Nathalie and {Percival}, Will and {P\textbackslash'erez-R\textbackslash`afols}, Ignasi and {Poppett}, Claire and {Prada}, Francisco and {Qu}, Frank J. and {Rossi}, Graziano and {Sanchez}, Eusebio and {Schlegel}, David and {Schubnell}, Michael and {Seo}, Hee-Jong and {Sif\textbackslash'on}, Crist\textbackslash'obal and {Spergel}, David N. and {Sprayberry}, David and {Tarl\textbackslash'e}, Gregory and {Vargas-Maga\textbackslash\raisebox{-0.5ex}\textasciitildena}, Mariana and {Vavagiakis}, Eve M. and {Weaver}, Benjamin A. and {Wollack}, Edward J. and {Zarrouk}, Pauline},
        title = "{Backlighting extended gas halos around luminous red galaxies: kinematic Sunyaev-Zel'dovich effect from DESI Y1 x ACT}",
      journal = {arXiv e-prints},
         year = 2025,
        month = mar,
          eid = {arXiv:2503.19870},
        pages = {arXiv:2503.19870},
archivePrefix = {arXiv},
       eprint = {2503.19870},
 primaryClass = {astro-ph.GA},
       adsurl = {https://ui.adsabs.harvard.edu/abs/2025arXiv250319870R}
}

@ARTICLE{Torres_2016_CMASS_Clustering,
       author = {{Rodr{\'\i}guez-Torres}, Sergio A. and {Chuang}, Chia-Hsun and {Prada}, Francisco and {Guo}, Hong and {Klypin}, Anatoly and {Behroozi}, Peter and {Hahn}, Chang Hoon and {Comparat}, Johan and {Yepes}, Gustavo and {Montero-Dorta}, Antonio D. and {Brownstein}, Joel R. and {Maraston}, Claudia and {McBride}, Cameron K. and {Tinker}, Jeremy and {Gottl{\"o}ber}, Stefan and {Favole}, Ginevra and {Shu}, Yiping and {Kitaura}, Francisco-Shu and {Bolton}, Adam and {Scoccimarro}, Rom{\'a}n and {Samushia}, Lado and {Schlegel}, David and {Schneider}, Donald P. and {Thomas}, Daniel},
        title = "{The clustering of galaxies in the SDSS-III Baryon Oscillation Spectroscopic Survey: modelling the clustering and halo occupation distribution of BOSS CMASS galaxies in the Final Data Release}",
      journal = {\mnras},
         year = 2016,
        month = aug,
       volume = {460},
       number = {2},
        pages = {1173-1187},
          doi = {10.1093/mnras/stw1014},
archivePrefix = {arXiv},
       eprint = {1509.06404},
 primaryClass = {astro-ph.CO},
       adsurl = {https://ui.adsabs.harvard.edu/abs/2016MNRAS.460.1173R}
}

@ARTICLE{Springel_2018_BR_TNG,
       author = {{Springel}, Volker and {Pakmor}, R{\"u}diger and {Pillepich}, Annalisa and {Weinberger}, Rainer and {Nelson}, Dylan and {Hernquist}, Lars and {Vogelsberger}, Mark and {Genel}, Shy and {Torrey}, Paul and {Marinacci}, Federico and {Naiman}, Jill},
        title = "{First results from the IllustrisTNG simulations: matter and galaxy clustering}",
      journal = {\mnras},
         year = 2018,
        month = mar,
       volume = {475},
       number = {1},
        pages = {676-698},
          doi = {10.1093/mnras/stx3304},
archivePrefix = {arXiv},
       eprint = {1707.03397},
 primaryClass = {astro-ph.GA},
       adsurl = {https://ui.adsabs.harvard.edu/abs/2018MNRAS.475..676S}
}

@ARTICLE{Zennaro_2024_GGL_w_Baryons,
       author = {{Zennaro}, Matteo and {Aric{\`o}}, Giovanni and {Garc{\'\i}a-Garc{\'\i}a}, Carlos and {Angulo}, Ra{\'u}l E. and {Ondaro-Mallea}, Lurdes and {Contreras}, Sergio and {Nicola}, Andrina and {Schaller}, Matthieu and {Schaye}, Joop},
        title = "{A 1\% accurate method to include baryonic effects in galaxy-galaxy lensing models}",
      journal = {arXiv e-prints},
         year = 2024,
        month = dec,
          eid = {arXiv:2412.08623},
        pages = {arXiv:2412.08623},
          doi = {10.48550/arXiv.2412.08623},
archivePrefix = {arXiv},
       eprint = {2412.08623},
 primaryClass = {astro-ph.CO},
       adsurl = {https://ui.adsabs.harvard.edu/abs/2024arXiv241208623Z}
}

@ARTICLE{Arico_2023_DESY3_Baryons,
       author = {{Aric{\`o}}, Giovanni and {Angulo}, Raul E. and {Zennaro}, Matteo and {Contreras}, Sergio and {Chen}, Angela and {Hern{\'a}ndez-Monteagudo}, Carlos},
        title = "{DES Y3 cosmic shear down to small scales: Constraints on cosmology and baryons}",
      journal = {\aap},
         year = 2023,
        month = oct,
       volume = {678},
          eid = {A109},
        pages = {A109},
          doi = {10.1051/0004-6361/202346539},
archivePrefix = {arXiv},
       eprint = {2303.05537},
 primaryClass = {astro-ph.CO},
       adsurl = {https://ui.adsabs.harvard.edu/abs/2023A&A...678A.109A}
}

@ARTICLE{Schneider_2022_Baryons_w_Cosmology,
       author = {{Schneider}, Aurel and {Giri}, Sambit K. and {Amodeo}, Stefania and {Refregier}, Alexandre},
        title = "{Constraining baryonic feedback and cosmology with weak-lensing, X-ray, and kinematic Sunyaev-Zeldovich observations}",
      journal = {\mnras},
         year = 2022,
        month = aug,
       volume = {514},
       number = {3},
        pages = {3802-3814},
          doi = {10.1093/mnras/stac1493},
archivePrefix = {arXiv},
       eprint = {2110.02228},
 primaryClass = {astro-ph.CO},
       adsurl = {https://ui.adsabs.harvard.edu/abs/2022MNRAS.514.3802S}
}

@ARTICLE{Troster_2022_Joint_Cosmo_Baryons,
       author = {{Tr{\"o}ster}, Tilman and {Mead}, Alexander J. and {Heymans}, Catherine and {Yan}, Ziang and {Alonso}, David and {Asgari}, Marika and {Bilicki}, Maciej and {Dvornik}, Andrej and {Hildebrandt}, Hendrik and {Joachimi}, Benjamin and {Kannawadi}, Arun and {Kuijken}, Konrad and {Schneider}, Peter and {Shan}, Huan Yuan and {van Waerbeke}, Ludovic and {Wright}, Angus H.},
        title = "{Joint constraints on cosmology and the impact of baryon feedback: Combining KiDS-1000 lensing with the thermal Sunyaev-Zeldovich effect from Planck and ACT}",
      journal = {\aap},
         year = 2022,
        month = apr,
       volume = {660},
          eid = {A27},
        pages = {A27},
          doi = {10.1051/0004-6361/202142197},
archivePrefix = {arXiv},
       eprint = {2109.04458},
 primaryClass = {astro-ph.CO},
       adsurl = {https://ui.adsabs.harvard.edu/abs/2022A&A...660A..27T}
}

@ARTICLE{Chisari_2019_Baryonic_Feedback_Cosmology,
       author = {{Chisari}, Nora Elisa and {Mead}, Alexander J. and {Joudaki}, Shahab and {Ferreira}, Pedro G. and {Schneider}, Aurel and {Mohr}, Joseph and {Tr{\"o}ster}, Tilman and {Alonso}, David and {McCarthy}, Ian G. and {Martin-Alvarez}, Sergio and {Devriendt}, Julien and {Slyz}, Adrianne and {van Daalen}, Marcel P.},
        title = "{Modelling baryonic feedback for survey cosmology}",
      journal = {The Open Journal of Astrophysics},
         year = 2019,
        month = jun,
       volume = {2},
       number = {1},
          eid = {4},
        pages = {4},
          doi = {10.21105/astro.1905.06082},
archivePrefix = {arXiv},
       eprint = {1905.06082},
 primaryClass = {astro-ph.CO},
       adsurl = {https://ui.adsabs.harvard.edu/abs/2019OJAp....2E...4C}
}

@ARTICLE{Steinborn_2015_AGN_Feedback_Subgrid,
       author = {{Steinborn}, Lisa K. and {Dolag}, Klaus and {Hirschmann}, Michaela and {Prieto}, M. Almudena and {Remus}, Rhea-Silvia},
        title = "{A refined sub-grid model for black hole accretion and AGN feedback in large cosmological simulations}",
      journal = {\mnras},
         year = 2015,
        month = apr,
       volume = {448},
       number = {2},
        pages = {1504-1525},
          doi = {10.1093/mnras/stv072},
archivePrefix = {arXiv},
       eprint = {1409.3221},
 primaryClass = {astro-ph.GA},
       adsurl = {https://ui.adsabs.harvard.edu/abs/2015MNRAS.448.1504S}
}

@ARTICLE{Pakmor_2023_MTNG_Clusters,
       author = {{Pakmor}, R{\"u}diger and {Springel}, Volker and {Coles}, Jonathan P. and {Guillet}, Thomas and {Pfrommer}, Christoph and {Bose}, Sownak and {Barrera}, Monica and {Delgado}, Ana Maria and {Ferlito}, Fulvio and {Frenk}, Carlos and {Hadzhiyska}, Boryana and {Hern{\'a}ndez-Aguayo}, C{\'e}sar and {Hernquist}, Lars and {Kannan}, Rahul and {White}, Simon D.~M.},
        title = "{The MillenniumTNG Project: the hydrodynamical full physics simulation and a first look at its galaxy clusters}",
      journal = {\mnras},
         year = 2023,
        month = sep,
       volume = {524},
       number = {2},
        pages = {2539-2555},
          doi = {10.1093/mnras/stac3620},
archivePrefix = {arXiv},
       eprint = {2210.10060},
 primaryClass = {astro-ph.CO},
       adsurl = {https://ui.adsabs.harvard.edu/abs/2023MNRAS.524.2539P}
}

@ARTICLE{Schaye_2023_FLAMINGO,
       author = {{Schaye}, Joop and {Kugel}, Roi and {Schaller}, Matthieu and {Helly}, John C. and {Braspenning}, Joey and {Elbers}, Willem and {McCarthy}, Ian G. and {van Daalen}, Marcel P. and {Vandenbroucke}, Bert and {Frenk}, Carlos S. and {Kwan}, Juliana and {Salcido}, Jaime and {Bah{\'e}}, Yannick M. and {Borrow}, Josh and {Chaikin}, Evgenii and {Hahn}, Oliver and {Hu{\v{s}}ko}, Filip and {Jenkins}, Adrian and {Lacey}, Cedric G. and {Nobels}, Folkert S.~J.},
        title = "{The FLAMINGO project: cosmological hydrodynamical simulations for large-scale structure and galaxy cluster surveys}",
      journal = {\mnras},
         year = 2023,
        month = dec,
       volume = {526},
       number = {4},
        pages = {4978-5020},
          doi = {10.1093/mnras/stad2419},
archivePrefix = {arXiv},
       eprint = {2306.04024},
 primaryClass = {astro-ph.CO},
       adsurl = {https://ui.adsabs.harvard.edu/abs/2023MNRAS.526.4978S}
}

@ARTICLE{Crain_2023_Galaxy_Population_Review,
       author = {{Crain}, Robert A. and {van de Voort}, Freeke},
        title = "{Hydrodynamical Simulations of the Galaxy Population: Enduring Successes and Outstanding Challenges}",
      journal = {\araa},
         year = 2023,
        month = aug,
       volume = {61},
        pages = {473-515},
          doi = {10.1146/annurev-astro-041923-043618},
archivePrefix = {arXiv},
       eprint = {2309.17075},
 primaryClass = {astro-ph.GA},
       adsurl = {https://ui.adsabs.harvard.edu/abs/2023ARA&A..61..473C}
}

@ARTICLE{Grandis_2024_Baryonic_Feedback,
       author = {{Grandis}, Sebastian and {Aric{\`o}}, Giovanni and {Schneider}, Aurel and {Linke}, Laila},
        title = "{Determining the baryon impact on the matter power spectrum with galaxy clusters}",
      journal = {\mnras},
         year = 2024,
        month = mar,
       volume = {528},
       number = {3},
        pages = {4379-4392},
          doi = {10.1093/mnras/stae259},
archivePrefix = {arXiv},
       eprint = {2309.02920},
 primaryClass = {astro-ph.CO},
       adsurl = {https://ui.adsabs.harvard.edu/abs/2024MNRAS.528.4379G}
}

@ARTICLE{Sunseri_2023_Baryonic_Feedback,
       author = {{Sunseri}, James and {Li}, Zack and {Liu}, Jia},
        title = "{Effects of baryonic feedback on the cosmic web}",
      journal = {\prd},
         year = 2023,
        month = jan,
       volume = {107},
       number = {2},
          eid = {023514},
        pages = {023514},
          doi = {10.1103/PhysRevD.107.023514},
archivePrefix = {arXiv},
       eprint = {2212.05927},
 primaryClass = {astro-ph.CO},
       adsurl = {https://ui.adsabs.harvard.edu/abs/2023PhRvD.107b3514S}
}

@ARTICLE{VanDaalen_2011_Baryonic_Feedback,
       author = {{van Daalen}, Marcel P. and {Schaye}, Joop and {Booth}, C.~M. and {Dalla Vecchia}, Claudio},
        title = "{The effects of galaxy formation on the matter power spectrum: a challenge for precision cosmology}",
      journal = {\mnras},
         year = 2011,
        month = aug,
       volume = {415},
       number = {4},
        pages = {3649-3665},
          doi = {10.1111/j.1365-2966.2011.18981.x},
archivePrefix = {arXiv},
       eprint = {1104.1174},
 primaryClass = {astro-ph.CO},
       adsurl = {https://ui.adsabs.harvard.edu/abs/2011MNRAS.415.3649V}
}

@ARTICLE{Paco_2021_CAMELS,
       author = {{Villaescusa-Navarro}, Francisco and {Angl{\'e}s-Alc{\'a}zar}, Daniel and {Genel}, Shy and {Spergel}, David N. and {Somerville}, Rachel S. and {Dave}, Romeel and {Pillepich}, Annalisa and {Hernquist}, Lars and {Nelson}, Dylan and {Torrey}, Paul and {Narayanan}, Desika and {Li}, Yin and {Philcox}, Oliver and {La Torre}, Valentina and {Maria Delgado}, Ana and {Ho}, Shirley and {Hassan}, Sultan and {Burkhart}, Blakesley and {Wadekar}, Digvijay and {Battaglia}, Nicholas and {Contardo}, Gabriella and {Bryan}, Greg L.},
        title = "{The CAMELS Project: Cosmology and Astrophysics with Machine-learning Simulations}",
      journal = {\apj},
         year = 2021,
        month = jul,
       volume = {915},
       number = {1},
          eid = {71},
        pages = {71},
          doi = {10.3847/1538-4357/abf7ba},
archivePrefix = {arXiv},
       eprint = {2010.00619},
 primaryClass = {astro-ph.CO},
       adsurl = {https://ui.adsabs.harvard.edu/abs/2021ApJ...915...71V}
}

@ARTICLE{SZ_effect_OG,
       author = {{Zeldovich}, Ya. B. and {Sunyaev}, R.~A.},
        title = "{The Interaction of Matter and Radiation in a Hot-Model Universe}",
      journal = {\apss},
         year = 1969,
        month = jul,
       volume = {4},
       number = {3},
        pages = {301-316},
          doi = {10.1007/BF00661821},
       adsurl = {https://ui.adsabs.harvard.edu/abs/1969Ap&SS...4..301Z}
}

@ARTICLE{Schaan_Ferarro_2016_kSZ_detection,
       author = {{Schaan}, Emmanuel and {Ferraro}, Simone and {Vargas-Maga{\~n}a}, Mariana and {Smith}, Kendrick M. and {Ho}, Shirley and {Aiola}, Simone and {Battaglia}, Nicholas and {Bond}, J. Richard and {De Bernardis}, Francesco and {Calabrese}, Erminia and {Cho}, Hsiao-Mei and {Devlin}, Mark J. and {Dunkley}, Joanna and {Gallardo}, Patricio A. and {Hasselfield}, Matthew and {Henderson}, Shawn and {Hill}, J. Colin and {Hincks}, Adam D. and {Hlozek}, Ren{\'e}e and {Hubmayr}, Johannes and {Hughes}, John P. and {Irwin}, Kent D. and {Koopman}, Brian and {Kosowsky}, Arthur and {Li}, Dale and {Louis}, Thibaut and {Lungu}, Marius and {Madhavacheril}, Mathew and {Maurin}, Lo{\"\i}c and {McMahon}, Jeffrey John and {Moodley}, Kavilan and {Naess}, Sigurd and {Nati}, Federico and {Newburgh}, Laura and {Niemack}, Michael D. and {Page}, Lyman A. and {Pappas}, Christine G. and {Partridge}, Bruce and {Schmitt}, Benjamin L. and {Sehgal}, Neelima and {Sherwin}, Blake D. and {Sievers}, Jonathan L. and {Spergel}, David N. and {Staggs}, Suzanne T. and {van Engelen}, Alexander and {Wollack}, Edward J. and {ACTPol Collaboration}},
        title = "{Evidence for the kinematic Sunyaev-Zel'dovich effect with the Atacama Cosmology Telescope and velocity reconstruction from the Baryon Oscillation Spectroscopic Survey}",
      journal = {\prd},
         year = 2016,
        month = apr,
       volume = {93},
       number = {8},
          eid = {082002},
        pages = {082002},
          doi = {10.1103/PhysRevD.93.082002},
archivePrefix = {arXiv},
       eprint = {1510.06442},
 primaryClass = {astro-ph.CO},
       adsurl = {https://ui.adsabs.harvard.edu/abs/2016PhRvD..93h2002S}
}

@ARTICLE{Mroczkowski_2019_kSZ_review,
       author = {{Mroczkowski}, Tony and {Nagai}, Daisuke and {Basu}, Kaustuv and {Chluba}, Jens and {Sayers}, Jack and {Adam}, R{\'e}mi and {Churazov}, Eugene and {Crites}, Abigail and {Di Mascolo}, Luca and {Eckert}, Dominique and {Macias-Perez}, Juan and {Mayet}, Fr{\'e}d{\'e}ric and {Perotto}, Laurence and {Pointecouteau}, Etienne and {Romero}, Charles and {Ruppin}, Florian and {Scannapieco}, Evan and {ZuHone}, John},
        title = "{Astrophysics with the Spatially and Spectrally Resolved Sunyaev-Zeldovich Effects. A Millimetre/Submillimetre Probe of the Warm and Hot Universe}",
      journal = {\ssr},
         year = 2019,
        month = feb,
       volume = {215},
       number = {1},
          eid = {17},
        pages = {17},
          doi = {10.1007/s11214-019-0581-2},
archivePrefix = {arXiv},
       eprint = {1811.02310},
 primaryClass = {astro-ph.CO},
       adsurl = {https://ui.adsabs.harvard.edu/abs/2019SSRv..215...17M}
}

@article{Cromer_2022_Baryons_WL_Mass,
	title = {Towards 1\% accurate galaxy cluster masses: including baryons in weak-lensing mass inference},
	volume = {2022},
	issn = {1475-7516},
	shorttitle = {Towards 1\% accurate galaxy cluster masses},
	url = {https://ui.adsabs.harvard.edu/abs/2022JCAP...10..034C},
	doi = {10.1088/1475-7516/2022/10/034},
	urldate = {2024-10-27},
	journal = {Journal of Cosmology and Astroparticle Physics},
	author = {Cromer, Dylan and Battaglia, Nicholas and Miyatake, Hironao and Simet, Melanie},
	month = oct,
	year = {2022},
	note = {Publisher: IOP
ADS Bibcode: 2022JCAP...10..034C},
	pages = {034},
}

@ARTICLE{Hadzhiyska_2023_SZ_sims,
       author = {{Hadzhiyska}, Boryana and {Ferraro}, Simone and {Pakmor}, R{\"u}diger and {Bose}, Sownak and {Delgado}, Ana Maria and {Hern{\'a}ndez-Aguayo}, C{\'e}sar and {Kannan}, Rahul and {Springel}, Volker and {White}, Simon D.~M. and {Hernquist}, Lars},
        title = "{Interpreting Sunyaev-Zel'dovich observations with MillenniumTNG: mass and environment scaling relations}",
      journal = {\mnras},
         year = 2023,
        month = nov,
       volume = {526},
       number = {1},
        pages = {369-382},
          doi = {10.1093/mnras/stad2751},
archivePrefix = {arXiv},
       eprint = {2305.00992},
 primaryClass = {astro-ph.CO},
       adsurl = {https://ui.adsabs.harvard.edu/abs/2023MNRAS.526..369H}
}

@ARTICLE{Padmanabhan_2012_VR,
       author = {{Padmanabhan}, Nikhil and {Xu}, Xiaoying and {Eisenstein}, Daniel J. and {Scalzo}, Richard and {Cuesta}, Antonio J. and {Mehta}, Kushal T. and {Kazin}, Eyal},
        title = "{A 2 per cent distance to z = 0.35 by reconstructing baryon acoustic oscillations - I. Methods and application to the Sloan Digital Sky Survey}",
      journal = {\mnras},
         year = 2012,
        month = dec,
       volume = {427},
       number = {3},
        pages = {2132-2145},
          doi = {10.1111/j.1365-2966.2012.21888.x},
archivePrefix = {arXiv},
       eprint = {1202.0090},
 primaryClass = {astro-ph.CO},
       adsurl = {https://ui.adsabs.harvard.edu/abs/2012MNRAS.427.2132P}
}

@ARTICLE{Ried_Guachalla_2024_VR,
       author = {{Ried Guachalla}, Bernardita and {Schaan}, Emmanuel and {Hadzhiyska}, Boryana and {Ferraro}, Simone},
        title = "{Velocity reconstruction in the era of DESI and Rubin/LSST. I. Exploring spectroscopic, photometric, and hybrid samples}",
      journal = {\prd},
         year = 2024,
        month = may,
       volume = {109},
       number = {10},
          eid = {103533},
        pages = {103533},
          doi = {10.1103/PhysRevD.109.103533},
archivePrefix = {arXiv},
       eprint = {2312.12435},
 primaryClass = {astro-ph.CO},
       adsurl = {https://ui.adsabs.harvard.edu/abs/2024PhRvD.109j3533R}
}

@ARTICLE{Hadzhiyska_2024_VR,
       author = {{Hadzhiyska}, Boryana and {Ferraro}, Simone and {Ried Guachalla}, Bernardita and {Schaan}, Emmanuel},
        title = "{Velocity reconstruction in the era of DESI and Rubin/LSST. II. Realistic samples on the light cone}",
      journal = {\prd},
         year = 2024,
        month = may,
       volume = {109},
       number = {10},
          eid = {103534},
        pages = {103534},
          doi = {10.1103/PhysRevD.109.103534},
archivePrefix = {arXiv},
       eprint = {2312.12434},
 primaryClass = {astro-ph.CO},
       adsurl = {https://ui.adsabs.harvard.edu/abs/2024PhRvD.109j3534H}
}

@ARTICLE{Fowler_2007_ACT,
       author = {{Fowler}, J.~W. and {Niemack}, M.~D. and {Dicker}, S.~R. and {Aboobaker}, A.~M. and {Ade}, P.~A.~R. and {Battistelli}, E.~S. and {Devlin}, M.~J. and {Fisher}, R.~P. and {Halpern}, M. and {Hargrave}, P.~C. and {Hincks}, A.~D. and {Kaul}, M. and {Klein}, J. and {Lau}, J.~M. and {Limon}, M. and {Marriage}, T.~A. and {Mauskopf}, P.~D. and {Page}, L. and {Staggs}, S.~T. and {Swetz}, D.~S. and {Switzer}, E.~R. and {Thornton}, R.~J. and {Tucker}, C.~E.},
        title = "{Optical design of the Atacama Cosmology Telescope and the Millimeter Bolometric Array Camera}",
      journal = {\ao},
         year = 2007,
        month = jun,
       volume = {46},
       number = {17},
        pages = {3444-3454},
          doi = {10.1364/AO.46.003444},
archivePrefix = {arXiv},
       eprint = {astro-ph/0701020},
 primaryClass = {astro-ph},
       adsurl = {https://ui.adsabs.harvard.edu/abs/2007ApOpt..46.3444F}
}

@ARTICLE{Swetz_2011_ACT,
       author = {{Swetz}, D.~S. and {Ade}, P.~A.~R. and {Amiri}, M. and {Appel}, J.~W. and {Battistelli}, E.~S. and {Burger}, B. and {Chervenak}, J. and {Devlin}, M.~J. and {Dicker}, S.~R. and {Doriese}, W.~B. and {D{\"u}nner}, R. and {Essinger-Hileman}, T. and {Fisher}, R.~P. and {Fowler}, J.~W. and {Halpern}, M. and {Hasselfield}, M. and {Hilton}, G.~C. and {Hincks}, A.~D. and {Irwin}, K.~D. and {Jarosik}, N. and {Kaul}, M. and {Klein}, J. and {Lau}, J.~M. and {Limon}, M. and {Marriage}, T.~A. and {Marsden}, D. and {Martocci}, K. and {Mauskopf}, P. and {Moseley}, H. and {Netterfield}, C.~B. and {Niemack}, M.~D. and {Nolta}, M.~R. and {Page}, L.~A. and {Parker}, L. and {Staggs}, S.~T. and {Stryzak}, O. and {Switzer}, E.~R. and {Thornton}, R. and {Tucker}, C. and {Wollack}, E. and {Zhao}, Y.},
        title = "{Overview of the Atacama Cosmology Telescope: Receiver, Instrumentation, and Telescope Systems}",
      journal = {\apjs},
         year = 2011,
        month = jun,
       volume = {194},
       number = {2},
          eid = {41},
        pages = {41},
          doi = {10.1088/0067-0049/194/2/41},
archivePrefix = {arXiv},
       eprint = {1007.0290},
 primaryClass = {astro-ph.IM},
       adsurl = {https://ui.adsabs.harvard.edu/abs/2011ApJS..194...41S}
}

@ARTICLE{Thornton_2016_ACT,
       author = {{Thornton}, R.~J. and {Ade}, P.~A.~R. and {Aiola}, S. and {Angil{\`e}}, F.~E. and {Amiri}, M. and {Beall}, J.~A. and {Becker}, D.~T. and {Cho}, H. -M. and {Choi}, S.~K. and {Corlies}, P. and {Coughlin}, K.~P. and {Datta}, R. and {Devlin}, M.~J. and {Dicker}, S.~R. and {D{\"u}nner}, R. and {Fowler}, J.~W. and {Fox}, A.~E. and {Gallardo}, P.~A. and {Gao}, J. and {Grace}, E. and {Halpern}, M. and {Hasselfield}, M. and {Henderson}, S.~W. and {Hilton}, G.~C. and {Hincks}, A.~D. and {Ho}, S.~P. and {Hubmayr}, J. and {Irwin}, K.~D. and {Klein}, J. and {Koopman}, B. and {Li}, Dale and {Louis}, T. and {Lungu}, M. and {Maurin}, L. and {McMahon}, J. and {Munson}, C.~D. and {Naess}, S. and {Nati}, F. and {Newburgh}, L. and {Nibarger}, J. and {Niemack}, M.~D. and {Niraula}, P. and {Nolta}, M.~R. and {Page}, L.~A. and {Pappas}, C.~G. and {Schillaci}, A. and {Schmitt}, B.~L. and {Sehgal}, N. and {Sievers}, J.~L. and {Simon}, S.~M. and {Staggs}, S.~T. and {Tucker}, C. and {Uehara}, M. and {van Lanen}, J. and {Ward}, J.~T. and {Wollack}, E.~J.},
        title = "{The Atacama Cosmology Telescope: The Polarization-sensitive ACTPol Instrument}",
      journal = {\apjs},
         year = 2016,
        month = dec,
       volume = {227},
       number = {2},
          eid = {21},
        pages = {21},
          doi = {10.3847/1538-4365/227/2/21},
archivePrefix = {arXiv},
       eprint = {1605.06569},
 primaryClass = {astro-ph.IM},
       adsurl = {https://ui.adsabs.harvard.edu/abs/2016ApJS..227...21T}
}

@ARTICLE{Henderson_2016_ACT,
       author = {{Henderson}, S.~W. and {Allison}, R. and {Austermann}, J. and {Baildon}, T. and {Battaglia}, N. and {Beall}, J.~A. and {Becker}, D. and {De Bernardis}, F. and {Bond}, J.~R. and {Calabrese}, E. and {Choi}, S.~K. and {Coughlin}, K.~P. and {Crowley}, K.~T. and {Datta}, R. and {Devlin}, M.~J. and {Duff}, S.~M. and {Dunkley}, J. and {D{\"u}nner}, R. and {van Engelen}, A. and {Gallardo}, P.~A. and {Grace}, E. and {Hasselfield}, M. and {Hills}, F. and {Hilton}, G.~C. and {Hincks}, A.~D. and {Hloẑek}, R. and {Ho}, S.~P. and {Hubmayr}, J. and {Huffenberger}, K. and {Hughes}, J.~P. and {Irwin}, K.~D. and {Koopman}, B.~J. and {Kosowsky}, A.~B. and {Li}, D. and {McMahon}, J. and {Munson}, C. and {Nati}, F. and {Newburgh}, L. and {Niemack}, M.~D. and {Niraula}, P. and {Page}, L.~A. and {Pappas}, C.~G. and {Salatino}, M. and {Schillaci}, A. and {Schmitt}, B.~L. and {Sehgal}, N. and {Sherwin}, B.~D. and {Sievers}, J.~L. and {Simon}, S.~M. and {Spergel}, D.~N. and {Staggs}, S.~T. and {Stevens}, J.~R. and {Thornton}, R. and {Van Lanen}, J. and {Vavagiakis}, E.~M. and {Ward}, J.~T. and {Wollack}, E.~J.},
        title = "{Advanced ACTPol Cryogenic Detector Arrays and Readout}",
      journal = {Journal of Low Temperature Physics},
         year = 2016,
        month = aug,
       volume = {184},
       number = {3-4},
        pages = {772-779},
          doi = {10.1007/s10909-016-1575-z},
archivePrefix = {arXiv},
       eprint = {1510.02809},
 primaryClass = {astro-ph.IM},
       adsurl = {https://ui.adsabs.harvard.edu/abs/2016JLTP..184..772H}
}

@ARTICLE{Planck_2020_Maps,
       author = {{Planck Collaboration} and {Aghanim}, N. and {Akrami}, Y. and {Arroja}, F. and {Ashdown}, M. and {Aumont}, J. and {Baccigalupi}, C. and {Ballardini}, M. and {Banday}, A.~J. and {Barreiro}, R.~B. and {Bartolo}, N. and {Basak}, S. and {Battye}, R. and {Benabed}, K. and {Bernard}, J. -P. and {Bersanelli}, M. and {Bielewicz}, P. and {Bock}, J.~J. and {Bond}, J.~R. and {Borrill}, J. and {Bouchet}, F.~R. and {Boulanger}, F. and {Bucher}, M. and {Burigana}, C. and {Butler}, R.~C. and {Calabrese}, E. and {Cardoso}, J. -F. and {Carron}, J. and {Casaponsa}, B. and {Challinor}, A. and {Chiang}, H.~C. and {Colombo}, L.~P.~L. and {Combet}, C. and {Contreras}, D. and {Crill}, B.~P. and {Cuttaia}, F. and {de Bernardis}, P. and {de Zotti}, G. and {Delabrouille}, J. and {Delouis}, J. -M. and {D{\'e}sert}, F. -X. and {Di Valentino}, E. and {Dickinson}, C. and {Diego}, J.~M. and {Donzelli}, S. and {Dor{\'e}}, O. and {Douspis}, M. and {Ducout}, A. and {Dupac}, X. and {Efstathiou}, G. and {Elsner}, F. and {En{\ss}lin}, T.~A. and {Eriksen}, H.~K. and {Falgarone}, E. and {Fantaye}, Y. and {Fergusson}, J. and {Fernandez-Cobos}, R. and {Finelli}, F. and {Forastieri}, F. and {Frailis}, M. and {Franceschi}, E. and {Frolov}, A. and {Galeotta}, S. and {Galli}, S. and {Ganga}, K. and {G{\'e}nova-Santos}, R.~T. and {Gerbino}, M. and {Ghosh}, T. and {Gonz{\'a}lez-Nuevo}, J. and {G{\'o}rski}, K.~M. and {Gratton}, S. and {Gruppuso}, A. and {Gudmundsson}, J.~E. and {Hamann}, J. and {Handley}, W. and {Hansen}, F.~K. and {Helou}, G. and {Herranz}, D. and {Hildebrandt}, S.~R. and {Hivon}, E. and {Huang}, Z. and {Jaffe}, A.~H. and {Jones}, W.~C. and {Karakci}, A. and {Keih{\"a}nen}, E. and {Keskitalo}, R. and {Kiiveri}, K. and {Kim}, J. and {Kisner}, T.~S. and {Knox}, L. and {Krachmalnicoff}, N. and {Kunz}, M. and {Kurki-Suonio}, H. and {Lagache}, G. and {Lamarre}, J. -M. and {Langer}, M. and {Lasenby}, A. and {Lattanzi}, M. and {Lawrence}, C.~R. and {Le Jeune}, M. and {Leahy}, J.~P. and {Lesgourgues}, J. and {Levrier}, F. and {Lewis}, A. and {Liguori}, M. and {Lilje}, P.~B. and {Lilley}, M. and {Lindholm}, V. and {L{\'o}pez-Caniego}, M. and {Lubin}, P.~M. and {Ma}, Y. -Z. and {Mac{\'\i}as-P{\'e}rez}, J.~F. and {Maggio}, G. and {Maino}, D. and {Mandolesi}, N. and {Mangilli}, A. and {Marcos-Caballero}, A. and {Maris}, M. and {Martin}, P.~G. and {Martinelli}, M. and {Mart{\'\i}nez-Gonz{\'a}lez}, E. and {Matarrese}, S. and {Mauri}, N. and {McEwen}, J.~D. and {Meerburg}, P.~D. and {Meinhold}, P.~R. and {Melchiorri}, A. and {Mennella}, A. and {Migliaccio}, M. and {Millea}, M. and {Mitra}, S. and {Miville-Desch{\^e}nes}, M. -A. and {Molinari}, D. and {Moneti}, A. and {Montier}, L. and {Morgante}, G. and {Moss}, A. and {Mottet}, S. and {M{\"u}nchmeyer}, M. and {Natoli}, P. and {N{\o}rgaard-Nielsen}, H.~U. and {Oxborrow}, C.~A. and {Pagano}, L. and {Paoletti}, D. and {Partridge}, B. and {Patanchon}, G. and {Pearson}, T.~J. and {Peel}, M. and {Peiris}, H.~V. and {Perrotta}, F. and {Pettorino}, V. and {Piacentini}, F. and {Polastri}, L. and {Polenta}, G. and {Puget}, J. -L. and {Rachen}, J.~P. and {Reinecke}, M. and {Remazeilles}, M. and {Renault}, C. and {Renzi}, A. and {Rocha}, G. and {Rosset}, C. and {Roudier}, G. and {Rubi{\~n}o-Mart{\'\i}n}, J.~A. and {Ruiz-Granados}, B. and {Salvati}, L. and {Sandri}, M. and {Savelainen}, M. and {Scott}, D. and {Shellard}, E.~P.~S. and {Shiraishi}, M. and {Sirignano}, C. and {Sirri}, G. and {Spencer}, L.~D. and {Sunyaev}, R. and {Suur-Uski}, A. -S. and {Tauber}, J.~A. and {Tavagnacco}, D. and {Tenti}, M. and {Terenzi}, L. and {Toffolatti}, L. and {Tomasi}, M. and {Trombetti}, T. and {Valiviita}, J. and {Van Tent}, B. and {Vibert}, L. and {Vielva}, P. and {Villa}, F. and {Vittorio}, N. and {Wandelt}, B.~D. and {Wehus}, I.~K. and {White}, M. and {White}, S.~D.~M. and {Zacchei}, A. and {Zonca}, A.},
        title = "{Planck 2018 results. I. Overview and the cosmological legacy of Planck}",
      journal = {\aap},
         year = 2020,
        month = sep,
       volume = {641},
          eid = {A1},
        pages = {A1},
          doi = {10.1051/0004-6361/201833880},
archivePrefix = {arXiv},
       eprint = {1807.06205},
 primaryClass = {astro-ph.CO},
       adsurl = {https://ui.adsabs.harvard.edu/abs/2020A&A...641A...1P}
}

@article{Buchs_2019_DES,
	title = {Phenotypic redshifts with self-organizing maps: {A} novel method to characterize redshift distributions of source galaxies for weak lensing},
	volume = {489},
	issn = {0035-8711},
	shorttitle = {Phenotypic redshifts with self-organizing maps},
	url = {https://ui.adsabs.harvard.edu/abs/2019MNRAS.489..820B},
	doi = {10.1093/mnras/stz2162},
	urldate = {2024-11-08},
	journal = {Monthly Notices of the Royal Astronomical Society},
	author = {Buchs, R. and Davis, C. and Gruen, D. and DeRose, J. and Alarcon, A. and Bernstein, G. M. and Sánchez, C. and Myles, J. and Roodman, A. and Allen, S. and Amon, A. and Choi, A. and Masters, D. C. and Miquel, R. and Troxel, M. A. and Wechsler, R. H. and Abbott, T. M. C. and Annis, J. and Avila, S. and Bechtol, K. and Bridle, S. L. and Brooks, D. and Buckley-Geer, E. and Burke, D. L. and Carnero Rosell, A. and Carrasco Kind, M. and Carretero, J. and Castander, F. J. and Cawthon, R. and D'Andrea, C. B. and da Costa, L. N. and De Vicente, J. and Desai, S. and Diehl, H. T. and Doel, P. and Drlica-Wagner, A. and Eifler, T. F. and Evrard, A. E. and Flaugher, B. and Fosalba, P. and Frieman, J. and García-Bellido, J. and Gaztanaga, E. and Gruendl, R. A. and Gschwend, J. and Gutierrez, G. and Hartley, W. G. and Hollowood, D. L. and Honscheid, K. and James, D. J. and Kuehn, K. and Kuropatkin, N. and Lima, M. and Lin, H. and Maia, M. A. G. and March, M. and Marshall, J. L. and Melchior, P. and Menanteau, F. and Ogando, R. L. C. and Plazas, A. A. and Rykoff, E. S. and Sanchez, E. and Scarpine, V. and Serrano, S. and Sevilla-Noarbe, I. and Smith, M. and Soares-Santos, M. and Sobreira, F. and Suchyta, E. and Swanson, M. E. C. and Tarle, G. and Thomas, D. and Vikram, V. and {DES Collaboration}},
	month = oct,
	year = {2019},
	note = {Publisher: OUP
ADS Bibcode: 2019MNRAS.489..820B},
	pages = {820--841},
}
